%% file: ms.tex
\documentclass[apj,numberedappendix,revtex4,twocolappendix]{emulateapj}
\usepackage{colordvi}
\usepackage{epsfig}
\usepackage{graphicx}
\usepackage{amssymb}
\usepackage{threeparttable}

\newcommand{\Ha}{\hbox{{\rm H}\kern 0.1em$\alpha$}}
\newcommand{\Hb}{\hbox{{\rm H}\kern 0.1em$\beta$}}
\newcommand{\Hd}{\hbox{{\rm H}\kern 0.1em$\delta$}}

\newcommand{\MgII}{\hbox{{\rm Mg}\kern 0.1em{\sc ii}}}
\newcommand{\CIV}{\hbox{{\rm C}\kern 0.1em{\sc iv}}}
\newcommand{\NeV}{\hbox{[{\rm Ne}\kern 0.1em{\sc v}]}}
\newcommand{\OII}{\hbox{[{\rm O}\kern 0.1em{\sc ii}]}}
\newcommand{\NeIII}{\hbox{[{\rm Ne}\kern 0.1em{\sc iii}]}}
\newcommand{\OIII}{\hbox{[{\rm O}\kern 0.1em{\sc iii}]}}
\newcommand{\NII}{\hbox{[{\rm N}\kern 0.1em{\sc ii}]}}
\newcommand{\SII}{\hbox{[{\rm S}\kern 0.1em{\sc ii}]}}

\newcommand{\lmass}{log(M/M$_{\odot}$)}

\newcommand{\sige}{$\Sigma_{\rm e}$}

\newcommand{\sigone}{$\Sigma_{1}$}
\newcommand{\sigoneq}{$\Sigma_{1}^{\rm {\scriptscriptstyle Q}}$}
\newcommand{\sigonesf}{$\Sigma_{1}^{\rm {\scriptscriptstyle SF}}$}

\newcommand{\sigeq}{$\Sigma_{\rm e}^{\rm {\scriptscriptstyle Q}}$}
\newcommand{\sigesf}{$\Sigma_{\rm e}^{\rm {\scriptscriptstyle SF}}$}

\newcommand{\sigonem}{$\log\Sigma_{1}-\log M_{\star}$}
\newcommand{\sigm}{$\log\Sigma-\log M_{\star}$}

\begin{document}

\title{A universal structural and star-forming relation since
  $z\sim3$: connecting compact star-forming and quiescent galaxies }

\author{Guillermo Barro\altaffilmark{1,2},
Sandra M. Faber\altaffilmark{1},
David C.~Koo\altaffilmark{1}, 
Avishai Dekel\altaffilmark{3},
Jerome J. Fang\altaffilmark{1}, 
Jonathan R.~Trump\altaffilmark{4,5},
Pablo G.~P\'{e}rez-Gonz\'{a}lez\altaffilmark{6},
Camilla Pacifici\altaffilmark{7},
Joel R. Primack\altaffilmark{8},
Rachel S. Somerville \altaffilmark{9}, 
Haojing Yan\altaffilmark{10}, 
Yicheng Guo\altaffilmark{1},
Fengshan Liu\altaffilmark{11},
Daniel Ceverino\altaffilmark{12},
Dale D.~Kocevski\altaffilmark{13}, 
Elizabeth McGrath\altaffilmark{13}}
\altaffiltext{1}{University of California, Santa Cruz}
\altaffiltext{2}{University of California, Berkeley}
\altaffiltext{3}{The Hebrew University}
\altaffiltext{4}{Pennsylvania State University}
\altaffiltext{5}{Hubble Fellow}
\altaffiltext{6}{Universidad Complutense de Madrid}
\altaffiltext{7}{Space Telescope Science Institute}
\altaffiltext{8}{Santa Cruz Institute for Particle Physics}
\altaffiltext{9}{Rutgers University}
\altaffiltext{10}{University of Missouri}
\altaffiltext{11}{Shenyang Normal University}
\altaffiltext{12}{Centro de Astrobiologia, CSIC-INTA}
\altaffiltext{13}{Colby College}

\slugcomment{Submitted to the Astrophysical Journal} 
\slugcomment{Last edited: \today}
\label{firstpage}
\begin{abstract} 

We study the evolution of the core ($r<1$~kpc) and effective
($r<r_{\rm e}$) stellar-mass surface densities (\sigone~and \sige) in
star-forming (SFGs) and quiescent galaxies. As early as $z=3$, both
populations occupy distinct linear relations in $\log\Sigma_{\rm e}$
and $\log\Sigma_{\rm 1}$ vs. $\log{\rm M}_{\star}$.  These structural
relations exhibit almost constant slopes and scatter while their
normalizations decline with time. For SFGs, the normalization in
\sige~and \sigone~declines by a factor of $\sim$2 since $z=3$. Such
mild declines suggest that SFGs build dense cores by moving {\it
  along} these relations. We define this evolution as the {\it
  structural main sequence} ($\Sigma$-MS), analogous to the
star-formation rate main sequence (SFR-MS). Quiescent galaxies follow
different relations (\sigeq, \sigoneq) off the $\Sigma$-MS by having
higher densities than SFGs of the same mass and redshift. The
normalization of \sigeq~declines by a factor of 10 since $z=3$, while
only a factor of $\sim$2 in \sigoneq. Thus, a dense stellar core is
present in quiescent galaxies at all redshifts and the formation of
such core in SFGs is the main requirement for quenching
star-formation. Expressed in 2D as deviations off the SFR-MS and off
\sigoneq~at each redshift, the distribution of massive
(\lmass~$>10.3$) galaxies forms a {\it universal}, L-shaped track that
relates two fundamental physical processes: compaction and quenching.
Compaction is a process of substantial core-growth in SFGs relative to
the evolution in the $\Sigma$-MS. This process increases the
core-to-total mass and S\'ersic index, thereby, making ``compact''
SFGs structurally similar to quiescent galaxies.  Quenching occurs
once compact SFGs reach a maximum central density above
$\Sigma_{1}^{\rm
  Q}\equiv\Sigma_{1}-0.65\log(M_{\star}-10.5)\gtrsim9.5$~M$_{\odot}$/kpc$^{2}$. This
threshold provides the most effective selection criterion to identify
the star-forming progenitors of quiescent galaxies at all redshifts.

\end{abstract}

\section{Introduction}\label{intro}

Studies of galaxy evolution from the peak of cosmic star formation to
the present day have matured tremendously over the past two decades.
The advent of large multi-wavelength photometric surveys have enabled
inferences of the global stellar population properties such as stellar
mass, age, and star formation rate (SFR). Large-area surveys such us
Sloan Digital Sky Survey (SDSS), NMBS, zCOSMOS, UltraVISTA, and
zFOURGE have robustly established the shape and evolution of the mass
function of star-forming galaxies (SFGs) and quiescent galaxies since
$z\sim4$, cementing our understanding of galaxy build up and shutdown
(\citealt{peng10}; \citealt{brammer11}; Ilbert et al. 2013;
\citealt{muzzin13smf}; Woo et al.  2013; \citealt{tomczak13}).
Furthermore, the sensitivity and high spatial resolution of {\it
  Hubble Space Telescope} (HST) have extended those mass functions
further back in cosmic time (\citealt{bouwens10}; Finkelstein et
al. 2013; Oesch et al. 2014), and have made a pivotal contribution to
the study of galaxy sizes and morphologies (e.g., \citealt{vdw12};
Shibuya et al. 2015). Deep multi-band surveys with HST, such as GOODS
(\citealt{goods}) and CANDELS (\citealt{candelsgro};
\citealt{candelskoe}) have thus provided an exquisite dataset to
quantify the simultaneous evolution of the galaxy stellar populations
and structural properties across cosmic time.

The consensus is that strong correlations between structure and
stellar populations (i.e., a Hubble sequence) exist up to $z=4$
(\citealt{franx08}; \citealt{kriek09}; \citealt{wuyts11a}).  One such
relation is the star formation rate (SFR) main sequence (SFR-MS)
(\citealt{mainseq}; \citealt{elbaz07}; \citealt{salim07};
\citealt{pannella09}; \citealt{magdis10}; \citealt{wuyts11a};
\citealt{elbaz11}; \citealt{rodi10b}; \citealt{whitaker12b};
\citealt{pannella14}), which is thought to describe a relatively
smooth mode of galaxy growth (\citealt{elbaz07}; \citealt{rodi10b}) in
which gas inflow and SFR have reached a steady-state phase (e.g.,
\citealt{dekel13a}). SFGs on the SFR-MS typically have larger sizes
and exponential disk profiles, while quiescent galaxies of the same
mass have more concentrated mass profiles (higher S\'ersic indices)
and smaller sizes. A dichotomy is also present in the size-mass
relations, where quiescent galaxies exhibit a much steeper slope than
SFGs, and a lower normalization, i.e., higher densities
(\citealt{williams10}; \citealt{newman12}; \citealt{vdw14}). While
both the SFR-MS and the size-mass scaling relations evolve with time,
the fundamental structural differences in SFGs and quiescent galaxies
are always present, suggesting that having concentrated (denser)
surface density profiles is a requisite for quenching
(\citealt{kauffmann03}, 2006; \citealt{schiminovich07};
\citealt{bell08}; \citealt{cheung12}; \citealt{fang13};
\citealt{lang14}). In other words, SFGs must grow dense stellar cores
before quenching.

There is increasing observational evidence that SFGs with dense cores
exist at every redshift.  At $z\gtrsim2$, SFGs with the highest
central densities are remarkably compact and have high S\'ersic
indices and spheroidal morphologies, lacking any signature of an
underlying disk (\citealt{wuyts11b}; \citealt{barro13, barro14a,
  barro14b}; \citealt{patel13}; \citealt{stefanon13};
\citealt{williams13}; \citealt{nelson14}). These galaxies resemble the
quiescent population at the same redshift but are radically different
from other SFGs that have irregular and clumpy appearances
(\citealt{elmegreen04}; \citealt{genzel08}; \citealt{guo15}). This
suggests that compact SFGs are formed by strongly dissipational
processes. Some of these processes, like mergers and disk
instabilities, are indeed expected to be more frequent at earlier
times due to the higher gas-to-total mass ratios in SFGs
(\citealt{tacconi10,tacconi13}; \citealt{daddi10a}).  The increased
gas mass relative to the SFR makes such systems prone to gravitational
collapse on scales of $\sim$1~kpc, causing substantial core-growth
resulting from a gas-fed central starburst and/or an inward migration
of clumps (\citealt{dekel09b}; \citealt{ceverino10};
\citealt{genel14}; \citealt{wellons14}; \citealt{ceverino15};
\citealt{zolotov14}).  At lower redshifts, SFGs with dense stellar
cores have clearly recognizable disk structures, but their profiles
are dominated by a central bulge (\citealt{wuyts12};
\citealt{bruce12}; \citealt{bruce14b};
\citealt{lang14}). Interestingly, quiescent galaxies at low-z also
seem to have bulge+disk morphologies (\citealt{mcgrath08};
\citealt{bundy10}; \citealt{vdw11a}), suggesting that quenching takes
place among galaxies with similar morphologies.

The common denominator in the evolution of massive galaxies described
above is the growth of a dense stellar core. This suggests that it is
possible to describe the general processes of structural growth and
star-formation quenching using a unique quantity tracing central
stellar mass density. Here we study the evolution since $z\sim3$ of
the stellar mass surface density within a radius of 1~kpc, \sigone. We
build upon previous results at lower redshift which show that
\sigone~is closely related with quiescence, and follows a much tighter
correlation with stellar mass than the effective radius or the
effective surface density, \sige (e.g., \citealt{cheung12},
\citealt{fang13}; \citealt{dokkum14}; \citealt{woo15}). We aim to
answer whether this relation holds at high redshift, if it is a more
fundamental quenching predictor, and if the global build up and
quenching of SFGs can be described in simple terms using \sigone,
i.e., if it can be used to track galaxies in transit from the
star-forming to quiescence phase.

The paper is structured in three parts: In \S~\ref{data}, we give an
overview of the observational dataset and the estimated properties of
the galaxy sample. In \S~\ref{relations} we analyze the correlations
in effective and central mass surface densities, \sige~and~\sigone,
vs. stellar mass for SFGs and quiescent galaxies from $z=3-0.5$. In
\S~\ref{profiles} we study how the best-fit relations in
\sige~and~\sigone~relate to each other for galaxies with a S\'ersic
mass profile. In \S~\ref{sigmapaths} we analyze galaxy evolutionary
paths based on the best-fit $\Sigma_{\rm e, 1}$ structural relations
and predictions of hydro-dynamical simulations.  We find that the
structural evolution of SFGs can be described in terms of 2 phases: A
steady phase of size and core growth that we call the $\Sigma$ ``main
sequence'', and a ``compaction'' phase of rapid core growth.  In
\S~\ref{2dsequence} we study the relative distance of massive SFGs and
quiescent galaxies from the SFR-MS and the quiescent structural
relations as a function of redshift. We find that their distribution
defines a universal (redshift-independent) L-shape sequence that
traces 2 fundamental transformations: the compaction of SFGs to form a
dense stellar core, and the quenching of star-formation in those
galaxies at maximum central density. Lastly, in \S~\ref{cqprops} we
discuss the evolution in the structural properties and visual
morphologies of galaxies within the compaction-quenching sequence as a
function of time.

Throughout this paper, we quote magnitudes in the AB system, assume a
\citep{chabrier} initial mass function (IMF) and adopt the following
cosmological parameters: ($\Omega_{M}$,$\Omega_{\Lambda}$, h) = (0.3,
0.7, 0.7).

\begin{figure*}[t]
\includegraphics[width=18.5cm,angle=0.]{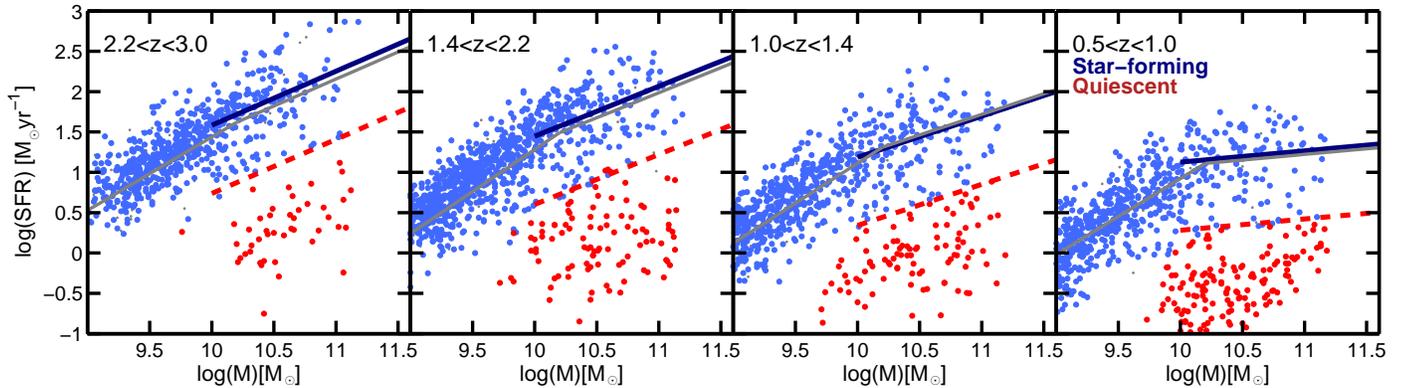}
\caption{\label{sfrms} Star-formation rate vs. stellar mass for
  galaxies in the CANDELS/GOODS-S field at $0.5<z<3.0$.  The blue
  lines show our fits to the SFR-MS at each redshift, which agree with
  previous results above \lmass$>10$. The gray lines show the results
  of \citet{whitaker13}, which highlight the change in the slope of
  the SFR-MS at lower masses.  SFGs are indicated by the blue
  points. Quiescent galaxies (red points) are selected to lie 0.7 dex
  below the SFR-MS, i.e., below the red dashed lines.}
\end{figure*}

\section{Data}\label{data}

This paper is based on a sample of massive galaxies built from the
{\it HST}/WFC3 F160W selected catalog for the CANDELS GOODS-S field \citep{guo13}.
Consistent, multi-wavelength photometry was measured using TFIT
\citep{tfit}, implemented as described by \citet{guo12}.  Photometric
redshifts were computed using EAZY \citep{eazy} and yielded errors of
$\Delta z/(1+z)=3$\% . Stellar masses were derived using FAST
\citep{fast} and based on a grid of \citet{bc03} models that assume a
\citet{chabrier} IMF, solar metallicity, exponentially declining star
formation histories, and the \citet{calzetti} dust extinction law (see
\citealt{santini15} for a detailed description). 

Star formation rates (SFRs) were computed by combining IR and
rest-frame UV (uncorrected for extinction) luminosities
(\citealt{ken98} and \citealt{bell05}) and adopting a \citet{chabrier}
IMF (see \citealt{barro11b} for more details):
$SFR_{\mathrm{UV+IR}}=1.09\times10^{-10}(L_{\mathrm{IR}}+3.3L_{2800})$. Total
IR luminosities ($L_{\mathrm{IR}}$$\equiv L$[8-1000$\mu$m]) were
derived from \citet{ce01} templates fitting MIPS 24$\mu$m fluxes,
applying a {\it Herschel}-based re-calibration \citep{elbaz11}. For
galaxies undetected by MIPS below a 2$\sigma$ level (20$\mu$Jy) SFRs
come from rest-frame UV luminosities that are corrected for extinction
as derived from SED fits \citep{wuyts11a}. 

The half-light radii, measured along the major axis, and S\`ersic
indices were determined from {\it HST}/WFC3 $H$ images using GALFIT
\citep{galfit} with PSFs created and processed to replicate the
conditions of the observed data \citep{vdw12}.  The stellar mass
profiles were computed by fitting multi-band SEDs derived from surface
brightness profiles in 9 HST bands measured with IRAF/ellipse (see
\citealt{liu13} and Liu et al. in prep for more details).  Following
\citet{wuyts12}, we impose an additional constraint on the spatially
resolved SED-fit by requiring that the integrated profile matches the
observed flux in IRAC ch1 and ch2. We apply this constraint by
assuming that the integrated IRAC-F160W color is the same at all
radii.

The multi-band HST mosaics were PSF-matched to the resolution of
F160W, that has a half width at half maximum of HWHM=0.''09. The
profiles have an intrinsic spatial resolution ranging from
$r=0.6-0.7$~kpc within the redshift range of the sample and thus
resolve the inner 1~kpc of the galaxies.  However, part of the light
can be smeared to larger radius. To correct for this effect, we
estimate a S\'ersic dependent correction to the mass profile within
1~kpc. The correction is computed from a grid S\'ersic profiles with
$n=0-4$ and $r_{e}=0.1--10$~kpc degraded to F160W resolution at
different redshift. The correction ranges from $\sim0.4$~dex at
$n\gtrsim2$ to $\sim0.2-0$~dex at $n=1-0$.

Our goal is to analyze the distribution of massive SFGs and quiescent
galaxies at $0.5<z<3$ in the structural scaling relations in order to
understand their differential evolution. The deep CANDELS photometry
enables the selection of stellar mass complete samples down to
\lmass$\sim10.3$ at $z=3$ (e.g., Tal et al. 2014). Thus, we use this
threshold to limit our sample selection for the analysis in \S~3.4 and
3.5. Nonetheless, in the next section, we use a larger sample of
galaxies down to \lmass~$=9$ ($\sim$70\% complete) to characterize the
SFR and structural correlations over a wider dynamical range.

\section{Results}\label{results}

\subsection{SFR and structural scaling relations}\label{relations}

\begin{figure*}[t]
\includegraphics[width=18.5cm,angle=0.]{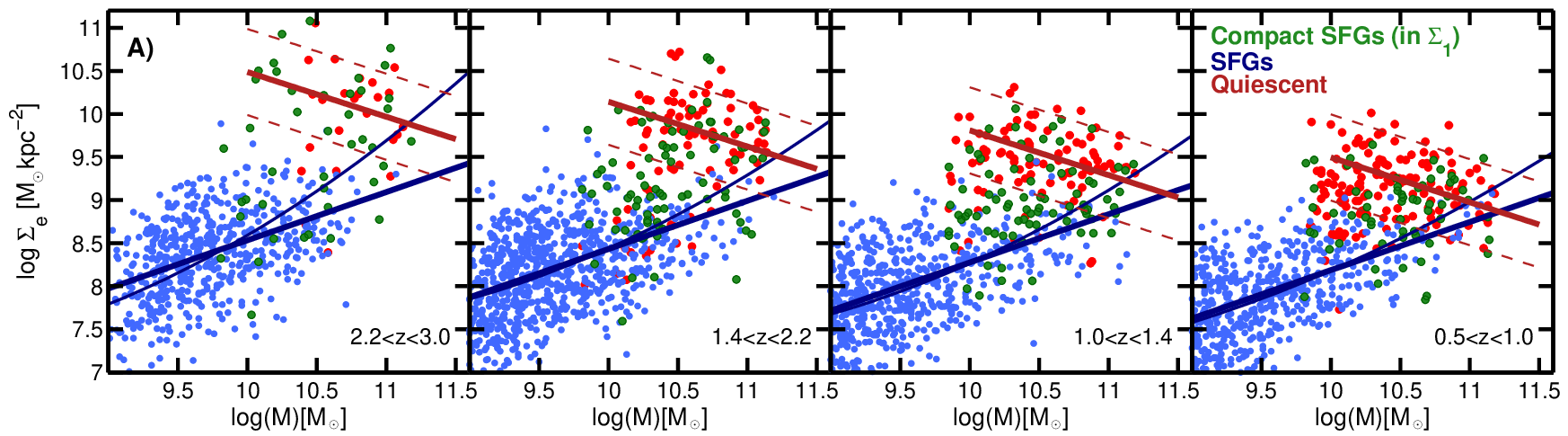}\\
\includegraphics[width=18.5cm,angle=0.]{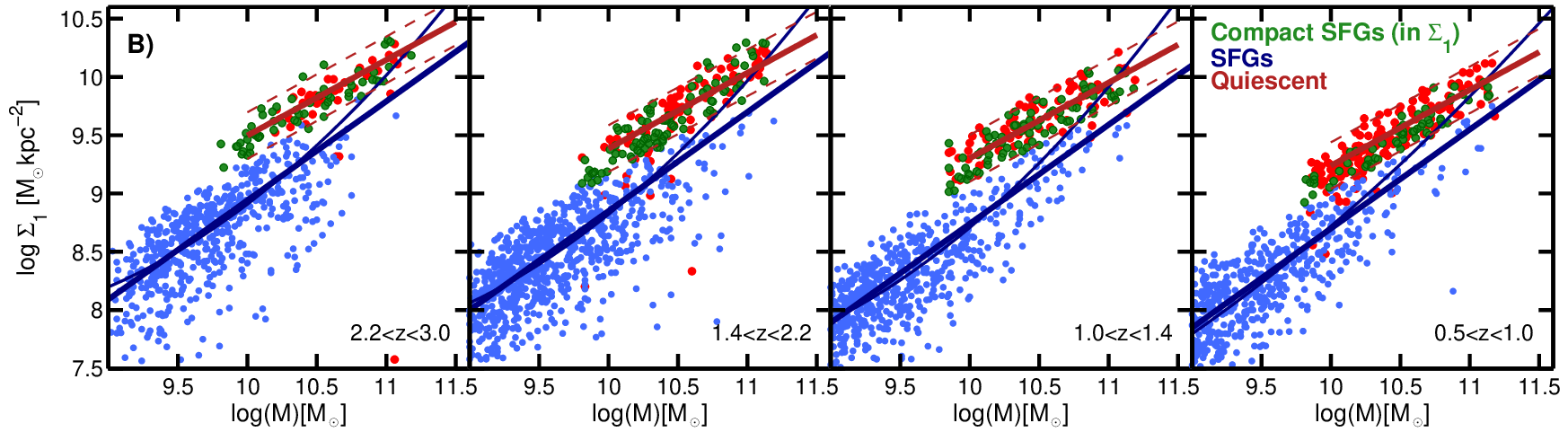}
\caption{\label{sequences} Surface density vs. stellar mass for
  galaxies in the CANDELS/GOODS-S field at $0.5<z<3.0$. Panels A and B
  show the surface density within the effective radius, \sige, and
  within the inner 1~kpc, \sigone, respectively. The blue and red
  circles show SFGs and quiescent galaxies selected using the SFR
  criterion of Figure~\ref{sfrms}. The thick blue and red lines depict
  the best-fit \sigm~ relations for the two populations. The dashed
  red lines show the 2$\sigma$ scatter around the quiescent
  relation. SFGs and quiescent galaxies exhibit clear and distinct
  scaling relations since $z\sim3$, which are well-described by single
  power-laws. The scatter in the \sigone~relations is a factor of
  $\sim2$ smaller than that of \sige. The slopes of the scaling
  relations remains approximately constant with time. The zero-points
  of the star-forming relations decline slowly with time (see
  Figure~\ref{zeropt}). Based on this smooth evolution, we speculate
  that SFGs follow, on average, evolutionary paths along that
  relation. We define this track as a structural ``main sequence''
  ($\Sigma$-MS) . At \lmass$\gtrsim10$ we find an increasing number of
  compact SFGs with high surface densities, similar to those of
  quiescent galaxies (green circles selected within the quiescent
  relation in \sigone). We capture this trend by fitting a
  second-order polynomial to \sigm, which shows a steeper slope at the
  high-mass end (thin blue line). These galaxies may deviate upwards
  from the $\Sigma$-MS due to dissipational ``compaction'' events that
  cause a rapid core growth (see \S~\ref{hydropaths}). Fit parameters
  are given in Tables~\ref{powerlaw} and \ref{2ndorder}}
\end{figure*} 

Figure~\ref{sfrms} show the distribution of SFR vs. stellar mass for
the galaxies in GOODS-S.  The majority of star-forming galaxies follow
a relatively tight relation between SFR and stellar mass. The observed
relation suggests that galaxy star formation histories are
predominantly regular and smooth, i.e., galaxies grow in a secular
mode which is usually referred to as the SFR main sequence (SFR-MS;
\citealt{mainseq}; \citealt{elbaz07}; \citealt{salim07};
\citealt{pannella09}; \citealt{magdis10}; \citealt{wuyts11a};
\citealt{elbaz11}; \citealt{rodi10b}; \citealt{whitaker12b};
\citealt{pannella14}). Following previous works, we characterize the
SFR-MS as a single power-law at \lmass~$\gtrsim$10.
\begin{equation}
\log {\rm SFR} =\mu \Big[\log\Big(\frac{M_{\star}}{M_{\odot}}\Big)-10.5\Big]+ \log {\rm C}
\end{equation}
Note that at lower masses, the SFR-MS exhibits a steeper slope (e.g.,
\citet{whitaker14}; \citet{coren15}). The best-fit SFR-MS at every
redshift is in excellent agreement with the results of
\citet{whitaker14} and \citet{coren15} at the high-mass end. The slope
and the normalization are consistent with their values within the
errors (Table~\ref{powerlaw}).  We select star-forming galaxies above
and quiescent galaxies below a threshold of $\Delta{\rm SFR}=-0.7$~dex
(dashed red line), where $\Delta {\rm SFR}\equiv\log{\rm SFR}-\log{\rm
  SFR^{MS}}$. This classification line is $\sim2\sigma$ below the
SFR-MS, which has a typical observational scatter of 0.3~dex
(\citealt{whitaker14}; \citealt{speagle14}; \citealt{coren15}).

Panel A of Figure~\ref{sequences} shows the distribution in effective
surface mass density, $\Sigma_{\rm e}=0.5 M_{\star}/\pi r_{\rm
  e}^{2}$, vs. mass for the galaxies in Figure~\ref{sfrms}. SFGs and
quiescent galaxies follow well-defined size-mass relations, which are
characterized by low-linear relations, $\log r_{\rm e}\propto a\log M$
(e.g., \citealt{law12a}; \citealt{mosleh12}; \citealt{szo12};
\citealt{newman12}; \citealt{vdw14}). Those relations can be expressed
in terms of \sige:

\begin{equation}\label{eq:sige}
\log \Sigma_{\rm e} = \alpha\Big[\log\Big(\frac{M_{\star}}{M_{\odot}}\Big)-10.5\Big]+\log{\rm A}(z)
\end{equation}

\noindent
where $\alpha$ is related to the slope of the size-mass relation $a$
as $\alpha=1-2a$.  The red and blue lines show the best-fit
\sige~relations for quiescent and star-forming galaxies. The best-fit
slopes are relatively constant with time, $\alpha^{\rm SF}\sim-0.5$,
$\alpha^{\rm Q}\sim0.6$ and agree well with the results of
\citet{vdw14} for the size-mass relations of both populations ($a^{\rm
  SF}\sim0.2$ and $a^{\rm Q}\sim0.8$). There are too few quiescent
galaxies at $2.2<z<3.0$ to accurately fit the slope, and so we fix the
slope (but not the normalization) in this bin to match the value at
$1.4<z<2.2$. Note that quiescent galaxies have a steep slope in the
size-mass relation, which leads to an anti-correlation in \sigeq.
Meanwhile, SFGs have a shallow slope in size-mass ($a^{\rm SF}<0.5$)
and a positive correlation between \sigesf~and mass. The scatter in
\sige~is consistent with $\sim$2$\times$ that of the size-mass
relations, $\sigma(\log\Sigma_{e})\sim0.5$~dex and 0.3~dex, for SFGs
and quiescent galaxies, as expected from $\Delta\log\Sigma_{\rm
  e}\propto2\Delta\log r_{\rm e}$. The redshift-dependent
normalizations decline from $z=3$ to $z=0.5$. Such decline is much
steeper for quiescent galaxies than for SFGs (1~dex vs. 0.3~dex) as
noted in previous works (\citealt{buitrago08}; \citealt{newman12};
\citealt{vdw14}).

\input{etab1} 
\input{etab2}

The bottom row (B panels) of Figure~\ref{sequences} shows the redshift
evolution of the central surface mass density within 1~kpc,
$\Sigma_{1\rm kpc}=M(<1~\rm kpc)/\pi (1~{\rm kpc})^2$, versus the
stellar mass. Similarly to \sige, we characterize the observed
correlation in \sigone~as a log-linear relation:
\begin{equation}\label{eq:sigone}
\log \Sigma_{1} = \beta\Big[\log\Big(\frac{M_{\star}}{M_{\odot}}\Big)-10.5\Big]+\log{\rm B}(z)
\end{equation}
Again, we find clear correlations for both SFGs and quiescent galaxies
at every redshift since $z\sim3$. The slopes of these relations are
positive and relatively constant with time, $\beta^{\rm SF}=0.9$ and
$\beta^{\rm Q}=0.7$. By comparison with \sige, the dispersion is
$\sim2\times$ tighter, $\sigma(\log\Sigma_{1})\sim0.25$~dex and
0.14~dex, for SFGs and quiescent galaxies, in good agreement with the
results of \citet{fang13} at $z=0$. The normalization of the
star-forming \sigonesf~relation declines by $\sim$0.3~dex from $z=3$
to $z=0.5$ similar to the evolution in \sigesf. Interestingly, for
quiescent galaxies, \sigoneq~declines by a similar amount, in stark
contrast with the strong decline of $\sim1$~dex in \sigeq.  The lower
scatter and weaker redshift evolution indicates that \sigone~is a more
robust and reliable structural parameter than \sige. Note also that,
by measuring the mass inside a fixed physical core aperture,
\sigone~is closer to the concept of a cosmic {\it clock}, i.e., it
only increases with a stellar mass growth, unlike the effective size
($r_{e}$), which can also decrease (e.g., due to a substantial mass
growth closer to galaxy center, or to fading of a extended
star-forming region).

Two main conclusions arise from the distribution of SFGs and quiescent
galaxies in Figure~\ref{sequences}: 1) the slopes of the structural
relations are almost constant with time, and the normalizations
decline for all of them; 2) at any redshift, quiescent galaxies are
denser than SFGs of the same mass, although this difference declines
with time in \sige~due to fast evolution of \sigeq. A reasonable
assumption based on the first conclusion is that the evolutionary
paths of individual SFGs approximately follow the best-fit \sigesf
relation which, by analogy with the SFR-MS, defines a structural main
sequence, $\Sigma$-MS, that could be interpreted as phase of smooth
structural growth.  The second conclusion indicates that increasing
the surface density by forming a dense stellar core is a requisite for
quenching the star-formation, as similarly suggested in previous
works. Together, these conclusions suggest the need for an
intermediate phase to bridge the $\Sigma$-MS and quiescent sequence,
in which SFGs become more compact and centrally concentrated before
quenching. We further discuss the possible evolutionary paths of SFGs
and quiescent galaxies in \S~\ref{sigmapaths}.

\subsection{Surface mass profiles: relating \sige~and \sigone}\label{profiles}

In this section we analyze the relation between the \sige~and
\sigone~structural relations for SFGs and quiescent galaxies in terms
of their surface density profiles. Galaxies's surface brightness and
stellar mass profiles are typically described by \citet{sersic}
profiles, in which case, the central and effective densities are
related to one another by the profile parameters. Assuming the
following characterization of the mass profile:
\begin{equation}\label{eq:massprof}
M(r)=M_{e}~\rm exp\Big(-b_{n}\Big[\Big(r/r_{e}\Big)^{1/n}-1\Big]\Big)
\end{equation}
where $n$ is the sersic index and $M_{e}$ is the effective mass at
$r_{e}$, \sigone~and \sige~can be determined from one of them for a
given value of $n$ and $r_{e}$.  Note that in doing so we assume that
the mass and light profiles are the same, $r_{e,{\rm mass}}=r_{e,{\rm
    light}}$. This is a better approximation for quiescent galaxies,
which exhibit only weak color gradients (e.g., \citealt{guo12};
\citealt{szo12}; \citealt{wuyts12}). Nonetheless, the parametrization
is useful to discuss the expected differences in \sigone~and
\sige~when the assumption fails.

\noindent
By integrating Equation~\ref{eq:massprof} we obtain:
\begin{equation}\label{gammaeq}
\log \Sigma_{1\rm kpc} - \log M_{\star} = -\log\pi - \log\gamma(2n,b_{n}r_{e}^{-1/n}) 
\end{equation}
where $\gamma$ is the incomplete gamma function, which depends on $n$
and $r_{e}$ (see e.g., \citealt{graham05}). As shown in
Figure~\ref{gammaplot}, the $\gamma$ function can be approximated as a
second order polynomial for a given $n$.
\begin{equation}\label{gammapol}
\log\gamma(2n,b_{n}r_{e}^{-1/n}) = c_{0} + c_{1}\log r_{e} + c_{2}(\log r_{e})^{2}
\end{equation}
where the $c_{i}$ coefficients depend on the S\'ersic index. Combining
the linear term of this equation with the best-fit relations for
\sigone~and \sige~as a function of mass, and the definition of the
latter in terms of $r_{e}$ ($\log r_{e}=0.5(\log
M_{\star}-\log\Sigma_{e})$), we obtain the following relations for the
slope and the scatter of the \sigonem~relation in terms of the slope
in the \sige~and size-mass relations.
\begin{eqnarray}\label{gammaslopes}
\beta&\sim&1+0.5c_{1}(\alpha-1);\;\;\;\; \sigma(\log\Sigma_{1})\sim0.5c_{1}\sigma(\log\Sigma_{\rm e})\\\nonumber
\beta&\sim&1-c_{1}a; \;\;\;\;\;\;\;\;\;\;\;\;\;\;\;\;\; \sigma(\log\Sigma_{1})\sim c_{1}\sigma(\log r_{\rm e})
\end{eqnarray}
Quiescent galaxies typically have high S\'ersic values, $n=3-4$
(\citealt{williams10}; \citealt{bell12}). Therefore, for a value of
$c_{1}(4)=0.55$, we obtain
$\sigma(\log\Sigma_{1})\sim0.3\sigma(\log\Sigma_{\rm e}$), which
explains the smaller scatter in \sigone~with respect to
\sige. Furthermore, for a value of $\alpha\sim-0.5$, the expected
slope of the \sigoneq~relation is $\beta\sim0.6$, which is consistent
with the best-fit value. Thus, a single S\'ersic model is good
approximation to explain the differences in the \sigoneq~and
\sigeq~relations.

\begin{figure}[t]
\centering
\includegraphics[width=8.3cm,angle=0.]{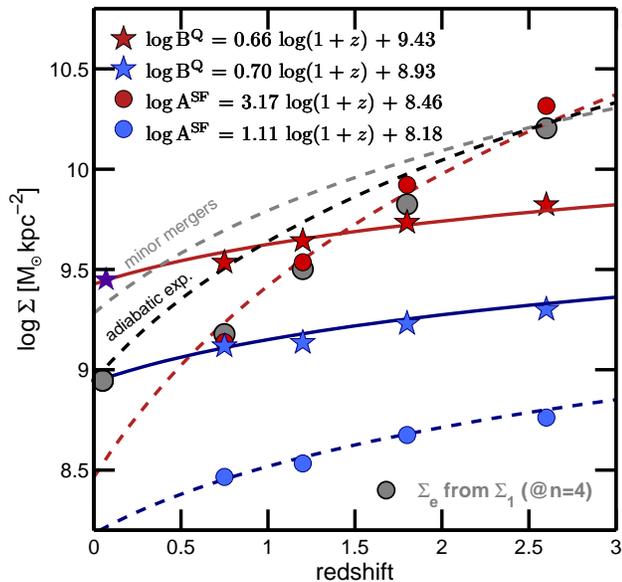}
\caption{\label{zeropt} Redshift evolution in the normalization of the
  \sigone~(stars; $\log{\rm B}$) and \sige~(circles; $\log{\rm A}$)
  relations for SFGs (blue) and quiescent galaxies (red). The solid
  and dashed lines show the best-fit relations to the normalizations
  in \sigone~and \sige~as a function of redshift.  The purple star
  shows the zero-point of the quiescent \sigone~relation at $z=0$ from
  \citet{fang13}. The dashed gray line shows the predicted evolution
  in \sigeq~due to minor mergers, assuming that \sigoneq~evolves only
  due to mass growth outside 1~kpc. The dashed black line shows the
  predicted evolution in \sigeq~assuming that \sigoneq~evolves only
  due to mass loss and adiabatic expansion. The gray circles show the
  predicted normalization in \sigeq~inferred from \sigoneq~using
  Equation~\ref{gammaeq} and a S\'ersic of $n=4$.}
\end{figure} 

\begin{figure}[t]
\includegraphics[width=8cm,angle=0.]{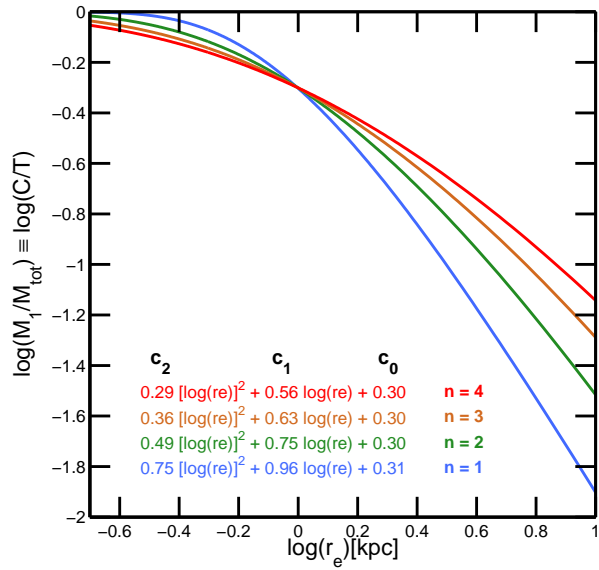}
\caption{\label{gammaplot} Ratio of the mass in the central 1~kpc to
  the total mass (defined as C/T) as a function of the effective
  radius, $r_{e}$, and $n$ for a single-S\'ersic mass profile. The C/T
  ratio follows a gamma function (Equation~\ref{gammaeq}), which can
  be expressed as a second-order polynomial in $r_{e}$ for different
  values of $n$ (bottom-left corner; Equation~\ref{gammapol}). The
  linear coefficient of those relations, $c_{1}$, determines the
  relative values of the slope and typical scatter of the \sige,
  \sigone and $r_{e}$ scaling relations with stellar mass
  (Equation~\ref{gammaslopes}). The best-fit structural relations for
  quiescent galaxies, \sigeq~and \sigoneq~agree well with one another
  for a S\'ersic profile of $n=4$ (see gray and red lines in
  Figure~\ref{sigma1sersic}).}
\end{figure} 

\begin{figure}[t]
\includegraphics[width=9.5cm,angle=0.]{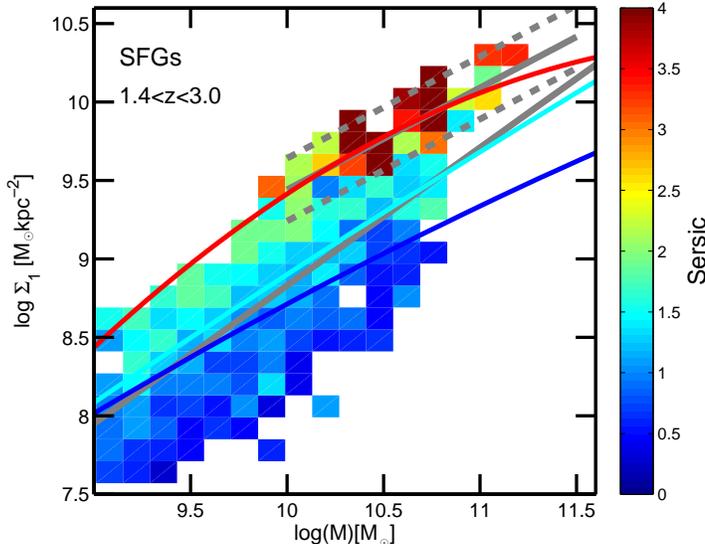}
\caption{\label{sigma1sersic} \sigone~vs. stellar mass for only SFGs
  at $1.4<z<3.0$ color coded by median S\'ersic index. The gray lines
  show the best-fit \sigonem~relations for SFGs and quiescent galaxies
  at $z=2$ depicted in Figure~\ref{sequences}. The SFGs in the
  overlapping region with the quiescent \sigone~relation (dashed gray
  lines) exhibit higher S\'ersic index. This could indicate that those
  SFGs arrive on the quiescent sequence as a result of a structural
  transformation that makes their mass profiles more centrally
  concentrated. The red line shows the \sigoneq~relation inferred from
  the observed \sigeq~relation for quiescent galaxies using
  Equations~\ref{gammaeq} and ~\ref{gammapol} and S\'ersic value of
  $n=4$. The observed and predicted \sigone~relations show an
  excellent agreement. The blue and cyan lines show the predicted
  \sigonesf~relation for SFGs derived from \sigesf~assuming $n=1$ and
  $n=2$, respectively. Qualitatively, the slopes agree with the
  observed relation, but the systematically lower normalizations
  suggest that the zero-point of \sigesf~depends on the S\'ersic.}
\end{figure}

SFGs exhibit a broad range in S\'ersic values, and their profiles can
often deviate from a single S\'ersic profile, e.g., in a disk with a
denser bulge (\citealt{wuyts11a}; \citealt{bruce12}).  The color code
in Figure~\ref{sigma1sersic} illustrates the spread in S\'ersic values
for SFGs in the \sigone~vs. mass diagram.  At a given stellar mass,
the S\'ersic increases with \sigone. This correlation suggest that as
SFGs grow a dense stellar core, their mass profiles become more
centrally concentrated. I.e., the width of the \sigone~distribution
for SFGs depends on their structural properties. The SFGs in the
overlapping region with \sigoneq, in particular, exhibit S\'ersic
values similar to those of quiescent galaxies. The blue and cyan lines
show the predicted \sigonesf~relation based on \sigesf~for $n=1$ and
$n=2$. Qualitatively, the prediction for $n=2$ matches the slope and
normalization of the best-fit \sigonesf~relation (gray). However, the
relations determined from \sigesf~ with $n\lesssim2$ have lower
normalizations than the observed distribution of SFGs with similar
S\'ersic values. This suggests that the normalization of the
\sige~distribution for SFGs also increases with $n$.

\subsection{Redshift evolution of the normalization of the quiescent structural relations}\label{quievol}

Using equation~\ref{gammaeq} and the relation between \sige~and
r$_{e}$, we quantify the evolution in the normalization of the
structural relations ($\log{\rm A}$ and $\log{\rm B}$) due to
different evolutionary processes that preserve the S\'ersic profile
but cause a relative change in size and stellar mass characterized as
$\Delta\log r_{e}=\eta\Delta\log {\rm M}$. If the slope of the
structural relations remain constant in the process, the
normalizations change as $\Delta\log\Sigma_{\rm
  e,1}-(\alpha,\beta)\Delta\log$M, which leads to:
\begin{eqnarray}\label{eq:zeropt}
\Delta \log{\rm A} &=&  (1-2\eta-\alpha)\Delta\log M \\\nonumber
\Delta \log{\rm B} &\sim&  (1-c_{1}\eta-\beta)\Delta\log M
\end{eqnarray} 
These relations are useful to test whether some of the proposed
evolutionary processes are consistent with the simultaneous evolution
in the normalizations of both structural relations
(Table~\ref{powerlaw}). For quiescent galaxies, where the relative
evolution in the normalization of \sige~vs.  \sigone~is largest, there
are three main evolutionary channels that can explain such large
difference. The leading explanation is that quiescent galaxies
experience a large increase in size compared to the mass growth due to
minor mergers (i.e., accretion of smaller satellite galaxies;
\citealt{bezanson09}; \citealt{hopkins09cores}; \citealt{oser12}).
Alternatively quiescent galaxies can have a large increase in size
({\it puff up}) due to feedback or stellar mass loss associated with
passive evolution (i.e., death of old stars) which cause adiabatic
expansion (\citealt{damjanov11}; \citealt{poggianti13};
\citealt{porter14}; \citealt{dokkum14}). Lastly, the strong decline in
\sigeq~could be caused by the arrival of new quiescent with
progressively larger sizes at lower redshifts (e.g.,
\citealt{poggianti13}; \citealt{carollo13}; \citealt{porter14}).  Each
of these scenarios leads to a different evolution in the normalization
of \sigeq~and \sigoneq. Attending to the best-fit relations in the
upper-left corner of Figure~\ref{zeropt}, the expected evolution in
the ratio of normalizations is $\Delta\log{\rm A}$/$\Delta\log{\rm
  B}\sim5$.

If the evolution of quiescent galaxies is driven by minor mergers, the
expectation is that the core mass remains relatively unchanged in the
process. Thus the normalization in \sigoneq~changes only due to the
total mass growth as $\Delta \log{\rm B}\sim-0.65\Delta\log M$, while
the normalization in \sigeq~follows Equation~8 as $\Delta \log{\rm
  A}\sim-1.7\Delta\log M$, for conservative value of $\eta\sim1.6$
(e.g., \citealt{newman12}). The predicted relative change in such case
is $\Delta\log{\rm A}$/$\Delta\log{\rm B}\sim2.6$, which is
substantially lower than the observed ratio (gray line in
Figure~\ref{zeropt}).  Therefore a simple minor merger scenario is
inconsistent with the observed results. This tension can be partially
alleviated if mergers cause a larger size growth ($\eta\gg1.6$ ) or if
they also cause an increase in \sigone, e.g., due to projection
effects in the 2D surface density. The latter, however, is
inconsistent with the results of \citet{dokkum14}, who showed that the
number density of the most massive cores decreases with time.

An alternative, or most likely complementary, scenario to explain the
fast decline in the normalization of \sigeq~is that newly quenched
galaxies at lower redshifts have larger sizes and lower density cores
(e.g., \citealt{poggianti13}; \citealt{carollo13}). Assuming that all
those quiescent galaxies have also high S\'ersic indices ($n\sim4$),
we use equations 5 and 6 estimate the redshift dependent normalization
in \sigeq~based on the observed values of the the normalization in
\sigoneq~(gray circles in Figure~\ref{zeropt}). This prediction
exhibits a much better agreement with observations down to $z=0.75$,
suggesting that a single S\'ersic model is a good approximation at
those redshifts. The main difference with a minor merger scenario is
that the latter increases the size and preserves the core density at
the expense of breaking the single S\'ersic mass profile. Note however
that at $z=0$, the predicted value of \sigeq~based on the
\sigoneq~value of \citep{fang13} deviates from the expected trend
towards larger sizes for a given central density. This could signal a
more prominent role of minor mergers at lower redshifts or a change in
the ratio of wet-to-dry mergers (e.g, \citealt{carlos12};
\citealt{porter14}).

\begin{figure*}[t]
\centering
\includegraphics[width=8.3cm,angle=0.]{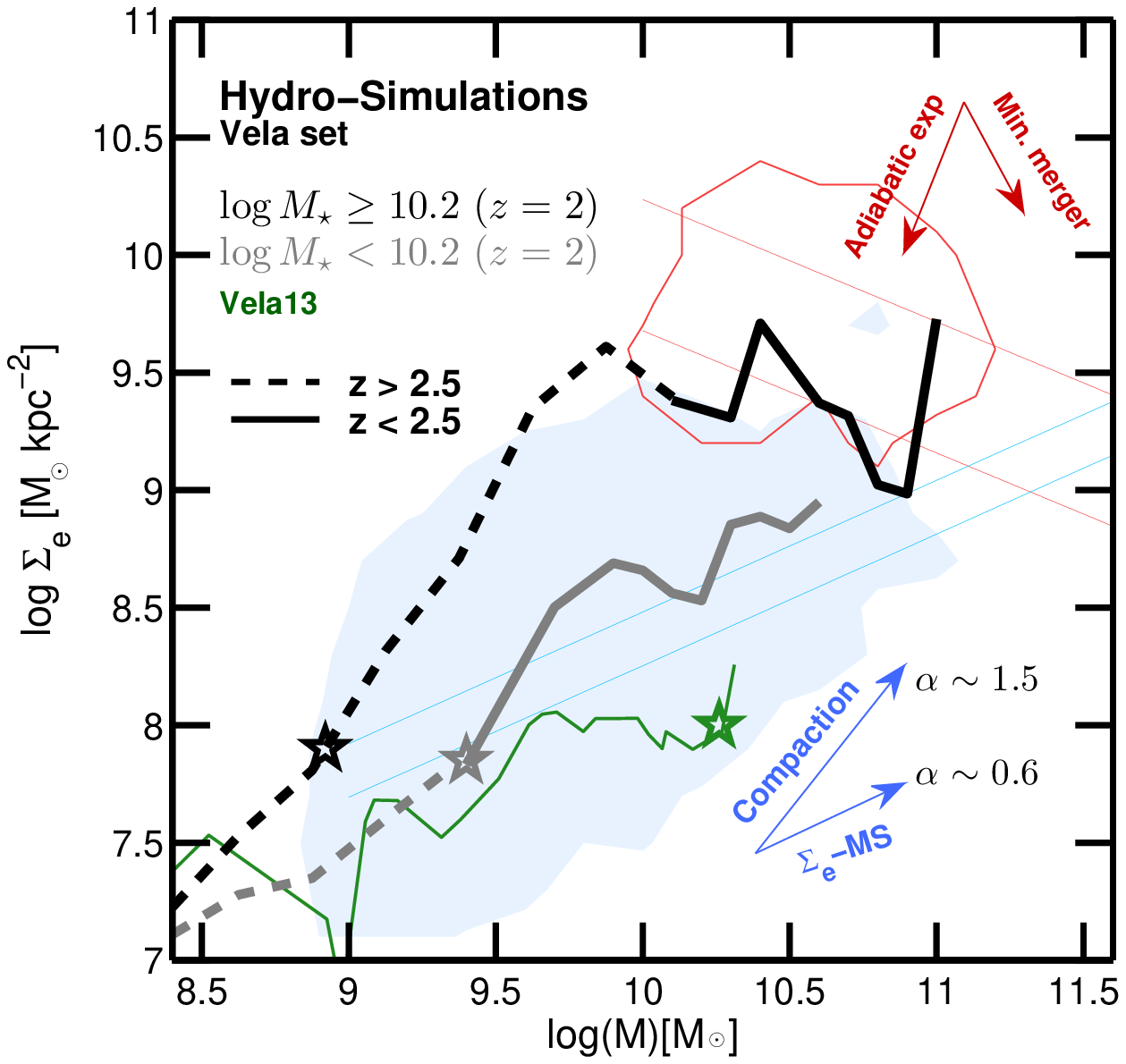}
\includegraphics[width=8.3cm,angle=0.]{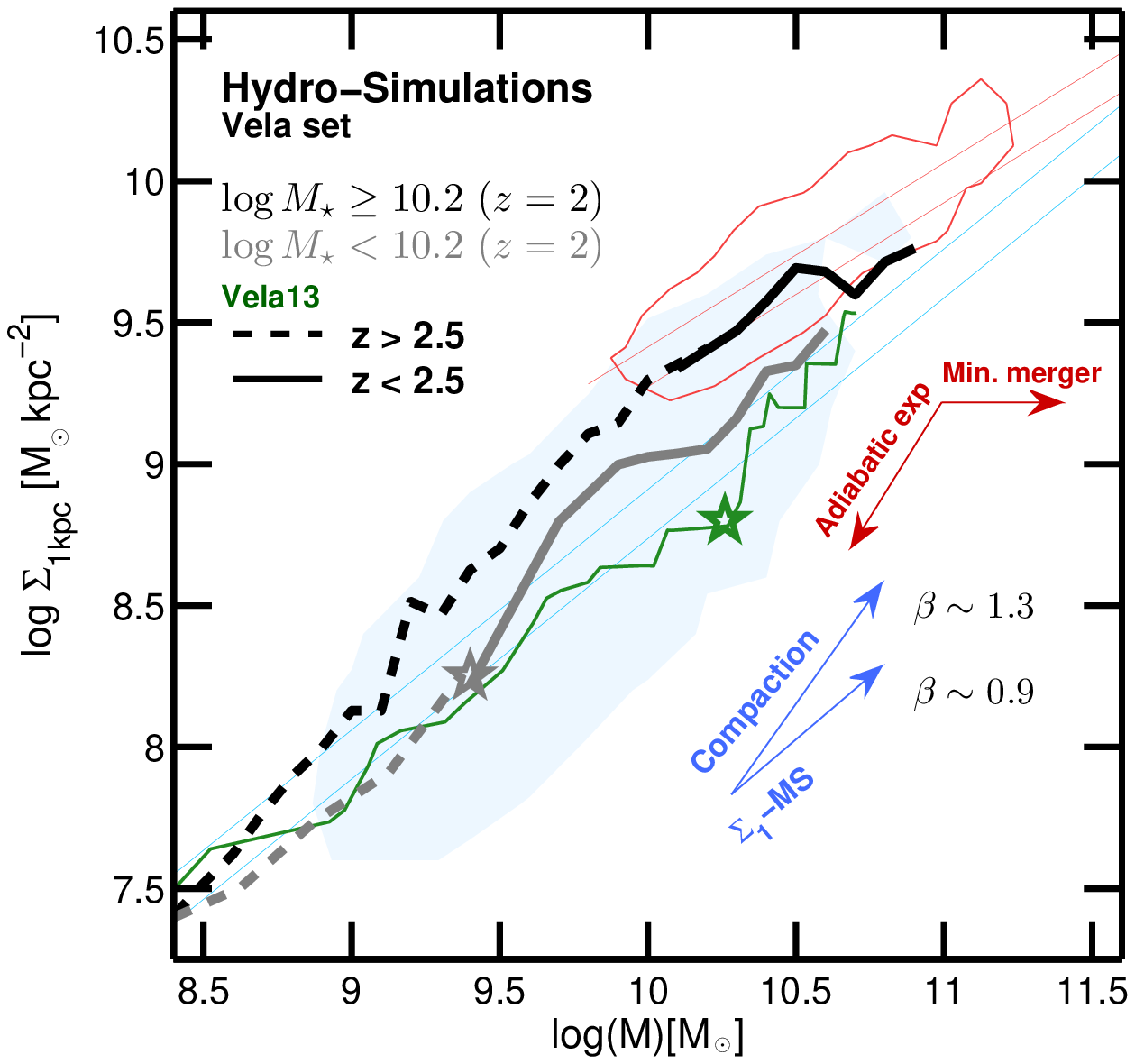}
\caption{\label{simul} Galaxy evolutionary tracks in \sige~(left) and
  \sigone~(right) vs. stellar mass as a function of time for the Vela
  set of 35 hydrodynamical simulations described in \citet{ceverino14}
  and \citet{zolotov14}. The black and gray lines show the median
  evolutionary tracks for galaxies above and below a threshold of
  \lmass$=10.2$ at $z=2$. The dashed and solid parts of the tracks
  indicate the evolution at redshift higher and lower than $z=2.5$,
  respectively.  The blue and red contours show the distribution of
  SFGs and quiescent galaxies at $z=1-2$. The blue and red lines show
  the best-fit \sige~and \sigone~relations at $z\sim1$ and $z\sim2$
  from Figure~\ref{sequences}. The model tracks are in excellent
  agreement with the observed distributions in \sige~and \sigone~(see
  also Figure 12 of Ceverino et al. 2015). Moreover, the tracks
  exhibit two fundamental phases with different slopes (blue arrows):
  (1) the $\Sigma$-MS, a phase of smooth structural growth that
  follows the best-fit $\Sigma_{\rm e, 1}$ relations from the data
  ($\alpha\sim0.6$, $\beta\sim0.9$), and (2) compaction, a period of
  steep core-growth ($\alpha\sim1.5$, $\beta\sim1.3$), usually
  triggered by a strongly dissipational event. The simulations exhibit
  a {\it downsizing} trend, such that the most massive galaxies evolve
  earlier and experience a stronger compaction event (open star) due
  to higher gas fraction at high-z.  The different loci of SFGs and
  quiescent galaxies is caused by the downsizing trend: massive
  galaxies form the backbone of the sequence by quenching first (and
  rapidly) at high-z, while low mass galaxies arrive on the sequence
  from below in a late compaction event. The green line shows the
  track of Vela13, which exhibits the latest compaction at $z\sim1$,
  remaining the longest in the $\Sigma$-MS. The red arrows indicate
  possible evolutionary tracks for quiescent galaxies for a minor
  merger or adiabatic expansion scenario (see \S~\ref{quievol}).}
\end{figure*} 

Another evolutionary scenario that preserves the shape (S\'ersic
index) of the mass profile is adiabatic expansion. In this case,
galaxies puff up due to a decline in the gravitational potential
caused by the death old stars. For a value of $\eta=-1$ (e.g.,
\citealt{damjanov09}), \sigeq decreases as $\Delta \log{\rm
  A}\sim3.5\Delta\log M_{\star}$, while \sigoneq~decreases as $\Delta
\log{\rm B}\sim\Delta\log M_{\star}$. Thus, the ratio between the two
is $\Delta\log{\rm A}$/$\Delta\log{\rm B}\sim3.5$, which is larger
than the prediction from minor mergers, but still underestimates the
observed trend (dashed black line).  The nearly unity relation in
$\log{\rm B}$ with stellar mass indicates that the mass loss required
to reproduce the observed evolution in \sigoneq~from adiabatic
expansion is $\Delta\log M_{\star}\sim(1+z)^{0.66}$
(Figure~\ref{zeropt}). However, as noted by \citet{dokkum14}, a simple
formation model in which quiescent galaxies quenched as early as
$z\sim5$ implies a much slower mass decline ($\sim(1+z)^{0.06}$).

As pointed out in previous works, the most likely scenario is that all
the evolutionary channels above play a role in the evolution of the
normalizations. Nonetheless, quantifying their relative contribution
as a function of time requires precise estimates of the ages and
quenching times of quiescent galaxies (e.g., \citealt{belli14b}), as
well as large enough samples to characterize the number densities at
the extremes of the distributions (i.e., the smallest or densest
galaxies) whose disappearance indicates the need for a size growth
\citealt{vdw14}) or mass loss (\citealt{dokkum14}).

In passing, we note that the decreasing trend with time in \sigone~has
also implications for the galaxy dynamics. \citet{fang13} found an
excellent correlation in \sigone~and the velocity dispersion in the
central 1~kpc, which closely followed the virial theorem,
$\Sigma_{1}\propto\sigma^{2}$. If this correlation holds at high-z,
the decrease in $\Delta\log B(z)\sim0.3$~dex from $z=3$ to $z=0$
implies that the Faber-Jackson \citep{faber76} relation at $z\gtrsim2$
should be 0.15~dex higher than the local value, as suggested in
\citet{belli14a,belli14b}.

\subsection{Galaxy evolutionary paths in $\Sigma_{\rm }$, $\Sigma_{1}$ vs. $M_{\star}$: the structural ``main sequence'' and compaction:}\label{sigmapaths}

In this section we study possible evolutionary paths for SFGs and
quiescent galaxies in \sige~and \sigone~vs. mass based on the
evolution in the normalization of the structural relations computed in
\S~\ref{relations} and predictions of theoretical simulations.

\subsubsection{Observational trends}

Based on the conclusions in \S~\ref{relations}, we speculate that the
relatively constant slope and weak evolution of the normalization in
\sigesf~and \sigone~suggest that SFGs follow evolutionary paths {\it
  along} these sequences, which we define as the $\Sigma$-MS. In the
following discussion we adopt this assumption, and thus we refer to
the $\Sigma$-MS as log-linear tracks in $\Delta\log\Sigma_{(\rm e,
  1)}=(\alpha,\beta)\Delta\log M$ with slopes of $\alpha\sim0.7$ and
$\beta\sim0.9$.

SFGs in the $\Sigma$-MS increase their surface density with
time. However, since the slope of the \sigonesf~relation is $\beta<1$
the mass in the core grows more slowly than the total mass of galaxy,
i.e., the core-to-total mass ratio decreases with time.  This is
consistent with the notion of inside-out growth of an exponential
profile (disk), due to galactic-scale star-formation and accretion of
higher angular momentum material, which causes both $r_{e}$ and
\sigone~to increase proportionally to the stellar mass growth (e.g.,
\citealt{nelson12,nelson13}). This scenario also agrees with
Figure~\ref{sigma1sersic}, which shows that the $\Sigma$-MS (gray)
describes an evolution at approximately constant S\'ersic.

For quiescent galaxies, which lack in-situ star-formation, the
evolutionary paths are thought to be driven by either minor mergers or
adiabatic expansion due to mass loss (see previous section). In a
minor merger scenario, quiescent galaxies follow flat tracks in
\sigone~and steeply declining tracks in \sige~($\alpha\sim-2.2$). In
an adiabatic expansion scenario, quiescent galaxies lose stellar mass
and puff up, both within the $r_{e}$ and the central 1~kpc, which
causes a steep decline in \sige~($\alpha=-3$) and
\sigone~($\beta=-1$). The evolutionary tracks for quiescent galaxies
in these scenarios are shown as red arrows in Figure~\ref{simul}.

Next, we focus on the evolutionary paths that take SFGs to the
higher-density structural relations of quiescent galaxies. To first
order, SFGs growing along the $\Sigma$-MS intersect the quiescent
structural relations at \lmass~$\gtrsim11.5$. However, confining
quenching to this evolutionary track would overproduce the number of
massive quiescent galaxies (e.g., \citealt{dokkum15}). This unique
track also requires a population of extremely dense, low mass SFGs
that are not observed. An alternative scenario to explain the
emergence of quiescent galaxies is that some SFGs follow a steeper
path upwards from the $\Sigma$-MS as result of some period(s) of fast
core growth. These periods can be caused by strongly dissipational,
``compaction'' events, e.g., major mergers (\citealt{hopkins08a}) or
disk instabilities (\citealt{elme08}; \citealt{dekel09b}) which funnel
large amounts of gas to the center of the galaxy. Compaction events
enhance star-formation in the core increasing the central density and,
in the most extreme cases, collapsing the whole galaxy to a much
smaller radius.

The compaction scenario was discussed in \citet{wuyts11b} and
\citet{barro13} on the basis that some massive SFGs seem to lie on the
steep size-mass relation of quiescent galaxies. These compact SFGs
exhibit smaller sizes and higher S\'ersic indices than the typical
SFGs (see also Figure~\ref{sigma1sersic}) suggesting that they have
experienced a substantial structural transformation. To account for
the presence of these compact population on the \sige~and
\sigone~relations, we fit the \sigm~distribution using a second order
polynomial, which allows for a change in the slope at the high mass
end. The thin blue lines in Figure~\ref{sequences} indicate that the
relations become indeed steeper at \lmass$\gtrsim10.5$ in both
\sige~and \sigone~(see also Table~\ref{2ndorder}) due to the
increasing number of SFGs with higher surface densities and S\'ersic
index (green circles).

In the following, we refer to any evolutionary paths upwards from the
$\Sigma$-MS due to phase(s) of high-efficiency core growth as
compaction track(s). We emphasize that compaction is, primarily, a
core building process, and thus it is more efficient at increasing
\sigone~than \sige, as the latter depends also on the {\it overall}
galaxy size.  See for example the SFGs with dense cores selected
within the scatter of \sigoneq~(green circles in
Figure~\ref{sequences}) which have a larger scatter in \sige. We
further discuss this difference in \S~3.5.

\subsubsection{Trends in hydrodynamical simulations}\label{hydropaths}

In order to provide further support for the structural evolutionary
paths discussed above, and obtain some insight on the physical
mechanism(s) that cause them, we study the evolutionary paths of a
sample of high-resolution hydrodynamical simulations in the same
parameter space. Figure~\ref{simul} shows the evolutionary tracks in
\sigone~and \sige~vs. mass for the Vela set of 35 simulations
described in \citet{ceverino14} and \citet{zolotov14}.  In the latter,
the authors analyzed the SFR, structure and kinematics of the
simulated SFGs, showing that the driving force behind the most
significant changes in all these properties is a phase (or phases) of
dissipational ``compaction'' caused by, e.g., mergers, disk
instabilities, interactions, counter rotating streams or
tidal-compressions, which trigger intense episodes of gas inflow and
SFR, as mentioned in the previous section.

Following the approach in \citet{zolotov14}, we divide the sample in
two groups according to their mass (in Figure~6, high-mass in black,
low-mass in grey), and we show their median evolutionary tracks at
early (dashed; $z\geq2.5$) and late (solid; $z<2.5$) times. The late
evolution matches the redshift range of the galaxy sample, shown as
blue and red contours. Qualitatively, the tracks have a similar
behavior in both panels, and match the observed distributions. To
first order, the overall evolutionary paths can be described in terms
of two phases: an early phase of steady structural growth
approximately following the slope of the $\Sigma$-MS ($\alpha\sim0.6$,
$\beta\sim0.9$), and a compaction phase, which causes a steep increase
in \sige and ~\sigone~($\alpha\sim1.5$, $\beta\sim1.3$). The steep
increase in \sige~also implies a size shrinkage, particularly in the
high-mass simulations (see also Figure~9 in
\citealt{zolotov14}). Overall, the dependence on $r_{e}$, which can
change rapidly due to gas accretion and/or interactions, implies a
larger spread in \sige~with respect to \sigone~for a given stellar
mass.

The main differences in the black and gray tracks are: 1) a {\it
  downsizing} effect, i.e., the most massive galaxies evolve earlier
and faster; 2) the evolutionary tracks of massive galaxies have a
higher normalization; 3) the main compaction event is stronger and
happens earlier for the most massive galaxies. The decline in the
normalization is the result of a decline in the gas fractions, and
thus SFRs, as a function of time (\citealt{dekel13a};
\citealt{zolotov14}). 

The different loci of SFGs and quiescent galaxies (blue and red
contours) also results from the gradual decline in the gas reservoirs
and the differential evolution with stellar mass. The quiescent locus
consist, almost exclusively, of massive galaxies that experienced a
strong compaction event at $z>2.5$ (black lines in
Figure~\ref{simul}). These galaxies quenched star-formation in the
core, but continued to grow in stellar mass due to star-formation in a
re-grown disk \citep{zolotov14}. The new disk causes a sudden increase
in size, which leads to decreasing tracks with mass in \sige, but
nearly flat tracks in \sigone. The star-forming locus consist mostly
of low mass galaxies which populate the high-mass region at lower
redshifts $z<2.5$ (gray lines). These galaxies experience weaker
compaction events, after which the core resumes star-formation and
continues on an evolutionary path with a similar slope as the
\sigone-MS. Later, these SFGs can have a secondary compaction event
and arrive on the quiescent sequence from below, as illustrated by the
green track of Vela13. This galaxy has the latest compaction event and
thus spends most of its life in the $\Sigma$-MS. Interestingly, during
the $\Sigma$-MS phase, the gas mass in the central 1~kpc remains
almost constant, as expected in the case of a simple ``bathtub'' model
evolution (e.g., \citealt{dekel13a}). This ``bathtub'' phase
strengthens the notion that the structural main sequence describes a
phase of smooth, steady-state evolution that coincides with the SFR
main sequence.

Lastly, note that despite the overall agreement in observations and
simulations, there are some differences between the simulated tracks
and the observed parameter space in $\Sigma$ vs. mass, particularly at
high-z. At $z>2.5$, all massive SFGs evolve directly into the
high-density, quiescent region (red contour) due to strong compaction
events, which also cause significant mass growth ($\Delta\log {\rm
  M_{\star}}>1$~dex). This evolution under-predicts the number of
massive SFGs (\lmass$>10$) with low \sige~and \sigone~at $z>2.5$,
which suggests that the $\Sigma$-MS evolutionary channel is
underrepresented, or happens only at low redshifts, in this sample
simulations.

\begin{figure*}[t]
\includegraphics[width=18.5cm,angle=0.]{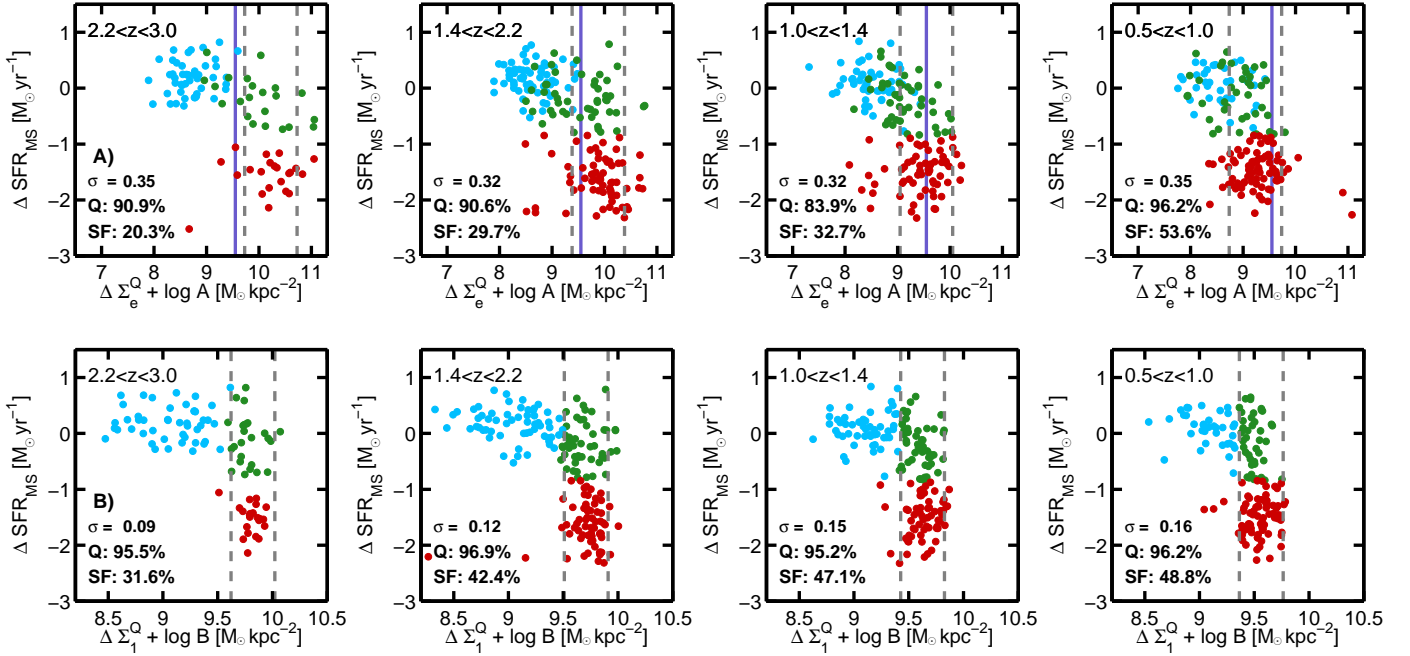}
\caption{\label{deltaseq} $\Delta$SFR vs. $\Delta\Sigma$ plots from
  the SFR-MS and the \sige~(top) and \sigone~(bottom) quiescent
  structural relations as a function of redshift for galaxies with
  \lmass~$>10.3$. The x-axis is normalized to the zero-point of the
  quiescent relations to illustrate the evolution with time.  Blue and
  green circles indicate SFGs outside and within the $2\sigma$ scatter
  of \sigoneq~(dashed lines). We refer to the latter as compact SFGs.
  The relative fractions of SFGs and quiescent galaxies found within
  the $2\sigma$ scatter of \sigeq~and \sigoneq~(dashed lines) are
  indicated in the bottom left. The L-shaped nature of the
  distributions indicate that SFGs become compact {\it before} they
  quench their star-formation. Moreover, the L-shape persist across
  redshifts, implying a {\it universal} process of compaction followed
  by quenching (see Figure~\ref{usr}). The distributions in the top
  and bottom panels are very similar; however, \sigoneq~exhibits a
  tighter scatter and a much slower evolution of the zero-point. The
  purple line shows the compactness threshold in \sigeq from B13 which
  is only efficient at $z\gtrsim2$. A fraction of the compact SFGs
  selected with $\Delta$\sigoneq~are not compact in \sigeq. This
  indicates that those SFGs have compact cores, but have larger
  $r_{e}$ than typical quiescent galaxies. Thus, a threshold in
  \sigoneq~is a more efficient selection criterion to identify compact
  SFGs and quiescent galaxies.}
\end{figure*}

\subsubsection{Summary of evolutionary paths}

Based on the excellent agreement in the observed galaxy distributions
and model evolutionary tracks in the \sigm~diagrams, we speculate that
the overall long-term structural evolution of SFGs can be expressed in
terms of 2 fundamental phases, namely, a $\Sigma$-MS and compaction.
These phases approximately follow linear tracks in
$\log\Sigma_{1,e}\propto[\alpha,\beta]\log M_{\star}$.  The $\Sigma$
main sequence ($\alpha\sim0.6$, $\beta\sim0.9$), is a relatively
smooth phase of structural growth, consistent with a period of
steady-state gas accretion and galactic scale star-formation. As a
result, the central and effective densities increase with time, as
does the galaxy size, i.e., similar to the typical inside-out growth
of a disk. Compaction, is a phase (or phases) of enhanced core-growth
($\alpha>1$, $\beta>1$) fueled by strong gas infall to the galaxy
center, as a result of gravitational instabilities. In this phase, the
central and effective densities increase steeply, and the galaxy's
effective radius can decrease due to the large increase in stellar
mass close to the center or due to a structural collapse.  Overall,
the intensity and duration of compaction events decreases with time
due to the similarly decreasing gas fractions.

For quiescent galaxies, the simulations reproduce the expected
evolutionary path of decreasing \sige~and relatively constant
\sigone~with increasing mass. However, this path is fueled by low
levels of in-situ star-formation, which appears to contradict the fast
quenching (\citealt{vandesande13}; \citealt{bezanson13};
\citealt{onodera14}) and the lack of diffuse star-forming components
(\citealt{szo11,szo12}) in quiescent galaxies at $z>1.5$. The leading
theory is that those quiescent galaxies grow due to minor mergers in
absence of further star-formation. Alternatively, it could be that
some of the SFGs with high surface densities are rejuvenated
quiescent galaxies with re-grown star-forming disks. Nonetheless,
given the rapidly increasing number of quiescent galaxies (see
\S~\ref{ndensitysect}) and the lack of apparent disk signatures in
those SFGs (see \S~\ref{cqprops}), rejuvenated quiescent galaxies
with extended star-forming disks can only make a small fraction of the
sample.


\subsection{Relative distances from the SFR-MS and the quiescent structural relations: compaction and quenching}\label{2dsequence}

In this section we study the relative distributions of galaxies with
respect to the SFR-MS and the $\Sigma_{\rm e, 1}^{\rm Q}$ structural
relations to identify candidate quenching galaxies as a function of
redshift. Qualitatively, this analysis is similar to studying the
distribution with respect to the SFR- and $\Sigma$- main
sequences. However, using the $\Sigma_{\rm e, 1}^{\rm Q}$
frames the selection around quiescent galaxies, searching for SFGs
with similar structural properties to them.

\begin{figure*}[t]
\centering
\includegraphics[width=19cm,angle=0.]{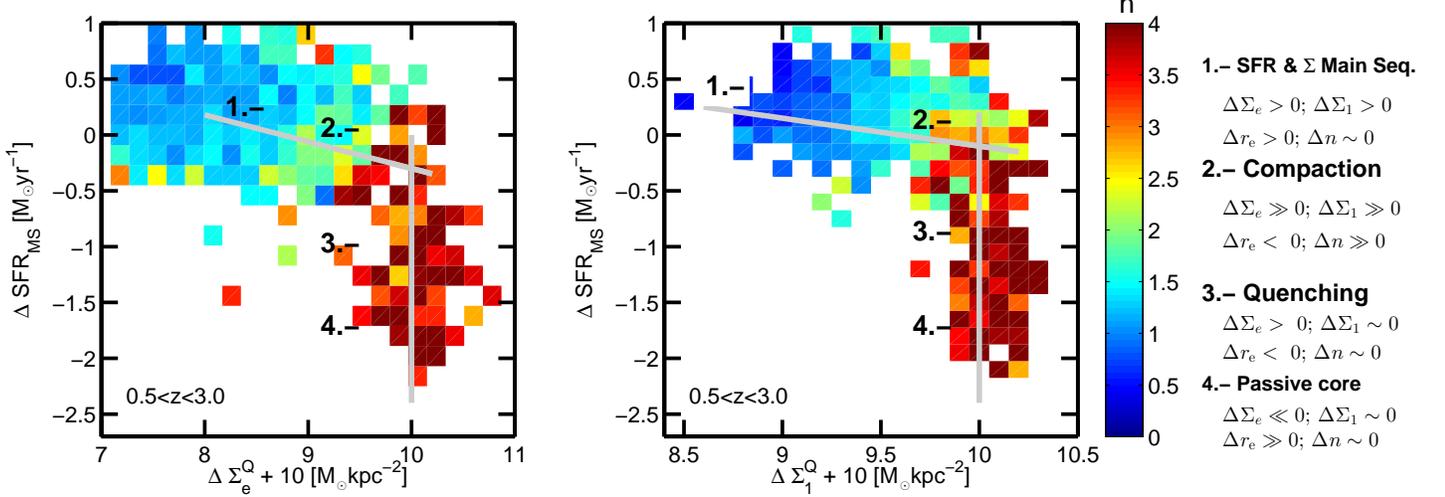}
\caption{\label{usr} Stacked $\Delta$SFR-MS vs. $\Delta\Sigma_{\rm
    e}^{\rm q}$ (left) and $\Delta\Sigma_{1}^{\rm q}$ (right) for all
  massive galaxies at $0.5<z<3$, color coded by S\'ersic index. The
  $\Delta\Sigma$ values are normalized at \sigeq$=$\sigoneq$=$10. The
  gray lines show the best-fit linear relations for SFGs and quiescent
  galaxies.  The L-shape and the intrinsic scatter of these diagrams
  is independent of redshift, which suggests that all massive galaxies
  follow the same universal evolutionary path from star-forming to
  quiescent. The numbers and the legend on the right side summarize
  the main phases of such evolution: 1) The SFR and $\Sigma$- MS phase
  is a nearly horizontal track in which galaxies increase both their
  core mass and effective radius at relatively constant S\'ersic
  (i.e., a disk growth); 2) Compaction is a phase of enhanced core
  growth, that increases the S\'ersic index. Both 1) and 2) follow
  horizontal tracks, however, only compaction causes a substantial
  increase in $n$, pushing SFGs galaxies towards the knee of the
  relation. 3) Quenching is a nearly vertical track that indicates
  that the full shut down of star-formation happens at maximum
  \sige~and \sigone. 4) In the passive core phase $\Sigma_{1}$ remains
  constant (or declines slowly due to dying stars), while $\Sigma_{\rm
    e}$ decreases faster due to size growth or new arrivals, causing a
  large evolution in the normalization of \sigeq
  (Figure~\ref{deltaseq}).  The slight tilt of the $\Sigma$-MS to
  compaction phase indicates that the onset of quenching starts begins
  before reaching the maximum stellar density.}
\end{figure*} 

\subsubsection{Compact SFGs as progenitors of quiescent galaxies}

As discussed in the previous section, the higher normalization of the
quiescent \sigeq~and \sigoneq~relations with respect to those of SFGs
indicates that quenching is preceded by an increase in the surface
density above a certain threshold. However, as shown in
Figure~\ref{sequences}, such characteristic density scales with
stellar mass, challenging the simple notion of a {\it fixed} quenching
threshold at a given surface density or stellar mass (e.g.,
\citealt{kauffmann03}; \citealt{franx08}). In turn, the most effective
quiescent criterion is a selection with respect to the \sigeq~or
\sigoneq~sequences with stellar mass, i.e., a relative offset from the
structural relations. Such relative selection includes fewer
SFGs. However, there is always overlap with quiescent galaxies at
every redshift. A possible interpretation, outlined in the previous
section, is that those overlapping SFGs acquire quiescent morphologies
while they are still star-forming as a result of a compaction process,
i.e., a structural transformation that increases the central density
and S\'ersic index, and reduces the size {\it before} quenching
star-formation.  Such an evolutionary sequence was confirmed in
\citet[][hereafter B13]{barro13,barro14a} for SFGs at $z\sim2$. In
B13, the authors used a selection on specific SFR and relative
distance to the quiescent size-mass relation to identify compact SFGs
at $z\sim2$, finding that those galaxies have similar sizes, S\'ersic
indices and spheroidal morphologies as the quiescent population.

Building on this idea, in this work we define {\it compact} SFGs as
those SFGs ($\Delta {\rm SFR_{\rm MS}}>-0.7$~dex) found within the
$\sim2\times$ the scatter of the quiescent structural scaling
relations at a given redshift, $\Delta \Sigma^{\rm
  Q}_{e,1}\equiv\log\Sigma_{e,1} - \log\Sigma^{Q}_{e,1}(z) >
-2\sigma(\log\Sigma^{Q})$, where $\Sigma_{e,1}$ is either the central
or effective mass density, and we use
$2\sigma(\log\Sigma^{Q}_{e})=0.5$~dex and
$2\sigma(\log\Sigma^{Q}_{1})=0.2$~dex, respectively. This definition
differs from previous works where {\it compact} is an absolute term to
identify the smallest galaxies at high-z ($r_{e}\lesssim1$~kpc; e.g.,
\citealt{cassata11,cassata13}). Here, compact is a {\it relative} term
referring to the densest/smallest SFGs at every redshift. Panels A and
B of Figure~\ref{deltaseq} illustrate the selection of compact SFGs in
\sige~and \sigone~(dashed line). In the x-axis we add the zero-point
at each redshift to illustrate the different time evolution in the
normalization of these relations. Panel A shows also the {\it
  compactness} threshold of B13 (purple line), which, by definition,
matches $\Delta \Sigma^{\rm Q}_{\rm e}<0.5$~dex at
$z\sim1.8$. However, as the normalization of \sigeq declines with
time, a fixed selection threshold gets progressively fewer
galaxies. In turn, $\Delta {\rm SFR_{\rm MS}}-\Delta\Sigma^{\rm Q}$ is
essentially a redshift-independent extension of the method to identify
star-forming progenitors of quiescent galaxies.

The relative distributions of SFGs and quiescent galaxies in both
panels of Figure~\ref{deltaseq} are very similar. By definition, both
$\Delta \Sigma^{\rm Q}_{1}$ and $\Delta \Sigma^{\rm Q}_{\rm e}$ select
all quiescent galaxies within the typical scatter of the structural
relations. A \sige~selection exhibits a few more catastrophic
outliers, most likely due to extreme values of the S\'ersic also
affecting the $r_{e}$ (see e.g., \citealt{vdw12}). Nonetheless, both
\sige~and \sigone~identify the bulk of the quiescent population. The
selection of compact SFGs is also largely consistent. However,
\sigone~selects $\sim10-15\%$ more compact galaxies at each
redshift. These compact SFGs in \sigone~scattered outside the
\sige~selection have both dense centers and extended star-forming
profiles, which leads an overall higher $r_{e}$ in a single S\'ersic
fit (e.g., \citealt{bruce12}). Nonetheless, as discussed for example
in \citet{fang13}, these galaxies will eventually increase their
\sige~as the extended star-forming disk fades. Thus, they are bonafide
quiescent progenitors. The lower efficiency of \sige~for selecting
compact SFGs is a result of $r_{e}$ not being monotonic with time
(i.e., $r_{e}$ both increases and decreases with redshift).

In summary, \sigone~is closer to a {\it clock} (it only increases),
and it also exhibits a tighter scatter and a slower evolution the
normalization, so we conclude that $\Delta$\sigoneq~is a more
efficient criterion to identify compact SFGs. In fact, as shown in
panel B of Figure~\ref{deltaseq}, a single threshold of
$\Sigma_{1}-0.65\log(M_{\star}-10.5)\gtrsim9.5$~M$_{\odot}$/kpc$^{2}$
identifies the majority of compact SFGs and quiescent galaxies at
redshifts $z>0.5$.  Hereafter, we will refer to compact SFGs as those
selected in \sigone.

\subsubsection{A universal compaction-quenching sequence}\label{cqsequence}

The remarkable similarity in the galaxy distributions of
Figure~\ref{deltaseq} as a function of time suggest that the relative
distance from the SFR-MS and $\Sigma_{\rm e, 1}^{\rm Q}$ relations
describes a universal evolutionary sequence for massive galaxies,
which is independent of redshift. Figure~\ref{usr} illustrates this
sequence showing the stacked distribution in $\Delta {\rm SFR_{\rm
    MS}}$ vs.  $\Delta\Sigma^{\rm Q}$ for the redshift range $0.5 < z
< 3.0$.  The upside-down, L-shaped sequence summarizes the notion that
forming a dense stellar core is a pre-requisite for quenching
star-formation (\citealt{cheung12}, \citealt{bell12};
\citealt{fang13}). The color code in S\'ersic index emphasizes that
the evolution requires {\it both} the growth of a dense stellar core,
and a structural transformation from an exponential (disk) profile to
more a concentrated ($n\gtrsim2$) S\'ersic profile. Incidentally,
Figure~\ref{usr} also illustrates why the S\'ersic index is a better
quiescent indicator than a constant threshold in \sige,~\sigone~ or
velocity dispersion, ~$\sigma$ (e.g., \citealt{bell12}).

In terms of the galaxy evolutionary paths discussed in
\S~\ref{sigmapaths} the two branches of the L-shaped sequence describe
two fundamental transitions from the $\Sigma$- and SFR- main
sequences, respectively: compaction and quenching. The horizontal
branch of SFGs represents constant evolution along the SFR-MS, but an
eventual departure from the $\Sigma$-MS due to a compaction event. The
$\Sigma$-MS describes the growth of an exponential disk ($n\sim1$) due
to in-situ star-formation, which increases \sigone, \sige~and the
overall size of the galaxy. Compaction involves a stronger increase in
\sigone~and \sige, leading to a steeper mass profile ($n\gtrsim2$) and
a smaller $r_{\rm e}$. The vertical branch shows that quenching
(defined as $\Delta {\rm SFR_{\rm MS}}>-0.7$~dex) takes places at
maximum central density in compact SFGs with similar morphologies to
the quiescent population, i.e., compaction precedes the {\it full}
shut down of star-formation.  The slight tilt in the horizontal
branch, $\Delta \rm SFR_{\rm MS}\sim-0.20\Delta\Sigma^{Q}_{1}$,
$\Delta \rm SFR_{\rm MS}\sim-0.25\Delta\Sigma^{Q}_{e}$ (gray lines),
however, suggests that while quenching is completed at maximum central
density it starts some time before.

Following the interpretation of B13, we characterize the evolutionary
pace along compaction-quenching path as declining from fast to slow as
a function of time. The gradient in quenching speed arises naturally
from the decline in sSFR with time (e.g., \citealt{speagle14}). Since
$\Delta\log\Sigma_{1,e}\propto\Delta\log M\propto{\rm sSFR(z)}$, the
central densities increase more slowly with time.  The larger SFR at
high-z is associated with higher gas fractions (\citealt{tacconi10};
\citealt{tacconi13}), which are likely related with larger accretion
rates from the dark matter haloes (\citealt{dekel13a,dekel13b}).
Simple stability arguments predict that such gas rich galaxies are
prone also to stronger gravitational instabilities, and thus favor
more expedite compaction and quenching processes. In such case, the
tilt in Figure~\ref{usr}, could be due to gas {\it starvation} after
an instability-induced starburst, i.e., a wet-inflow that causes a
peak in SFR at maximum gas density, and declines progressively with
the gas supply, while the core mass increases (e.g.,
\citealt{feldmann15}; \citealt{zolotov14}). The truncation of the gas
supply to the galaxy center can be caused by a combination of feedback
(e.g., outflows) and stabilization of the disk due to the increasing
central density (\citealt{martig09}).

\begin{figure}[t]
\includegraphics[width=8.5cm,angle=0.]{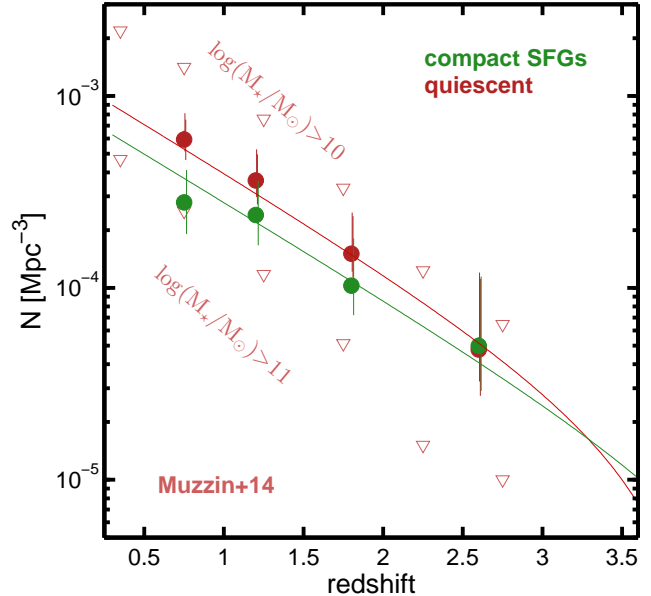}
\caption{\label{ndensity} Redshift evolution of the number density of
  massive (\lmass$>10.3$) compact SFGs and quiescent galaxies. Compact
  SFGs are selected using the \sigoneq~criterion. The red triangles
  indicate the evolution in the number density of quiescent galaxies
  in different mass ranges from \citet{muzzin13smf}. The green and red
  lines show the best-fit evolutionary model to the observed number of
  compact SFGs and quiescent galaxies. The model is based on the
  assumption that quiescent galaxies are descendants of compact SFGs
  that have a characteristic quenching timescale $\lambda^{\rm Q}$, but
  can also increase in number due to compaction events in more
  extended SFGs that have a frequency $\lambda^{\rm SFG}$. This simple
  model that assumes a continuous replenishment of compact SFGs at
  every redshift can account for the observed evolution in the two
  populations.}
\end{figure}

\begin{figure*}[t]
\centering
\includegraphics[width=8.5cm,angle=0.]{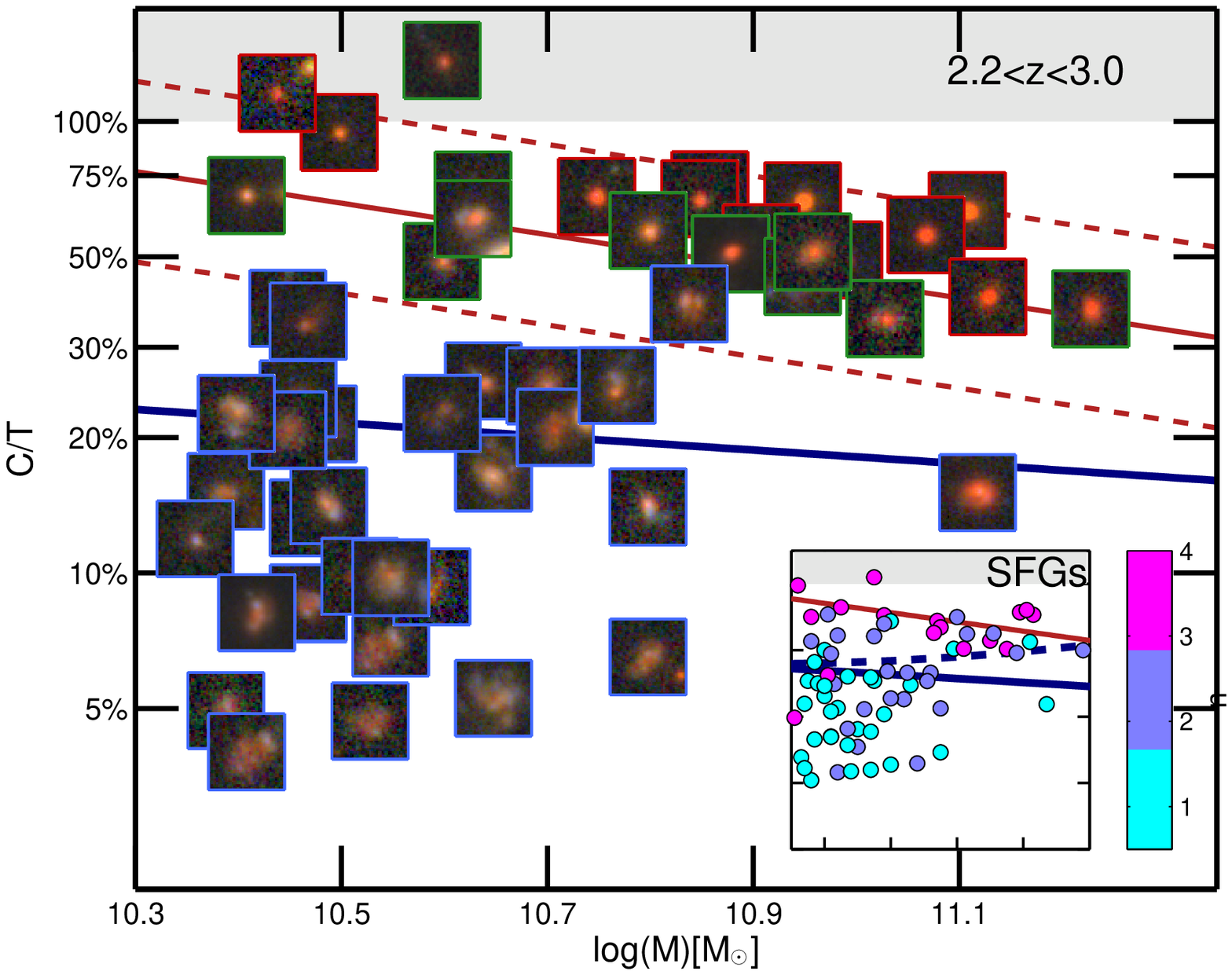}
\includegraphics[width=8.5cm,angle=0.]{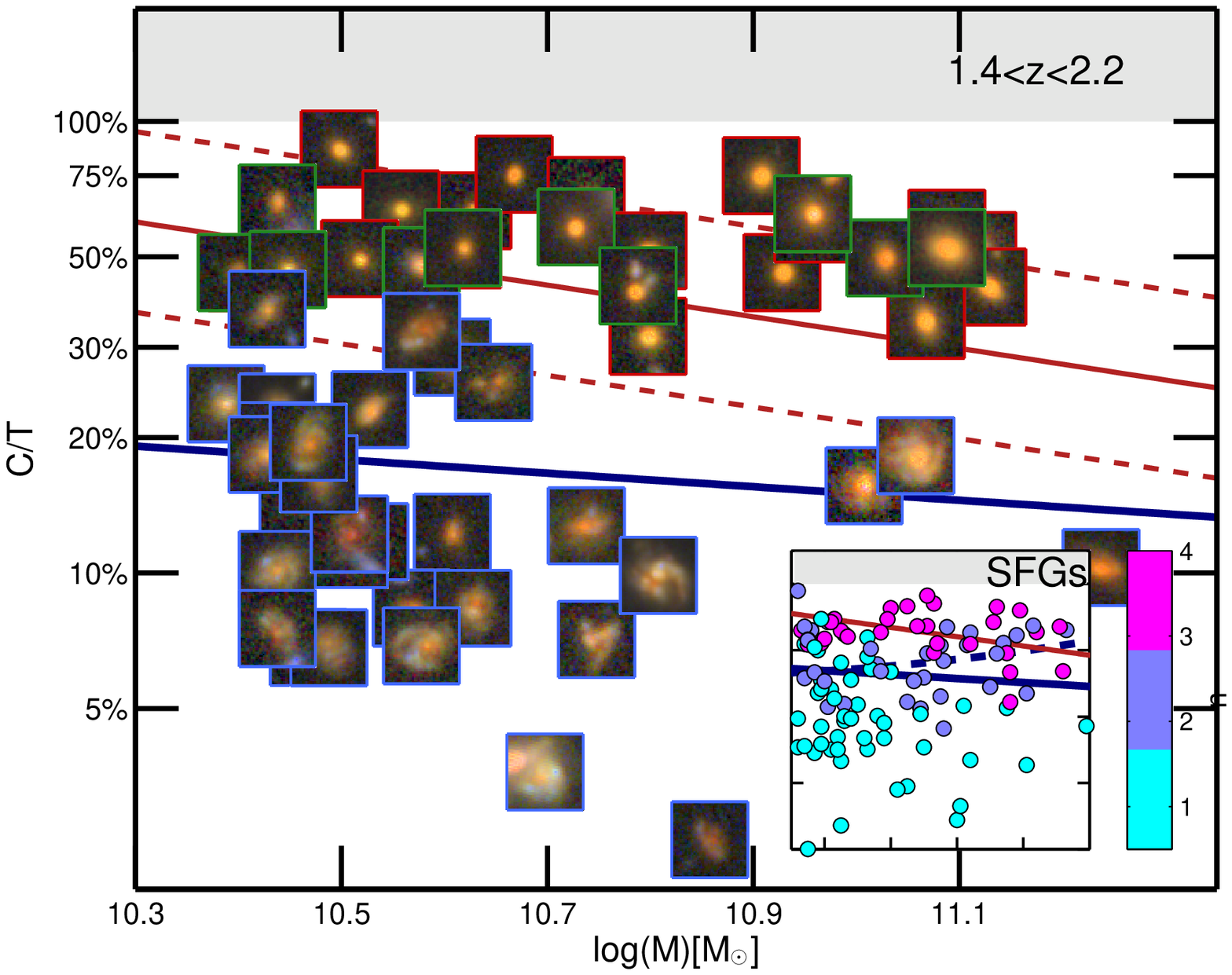}
\includegraphics[width=8.5cm,angle=0.]{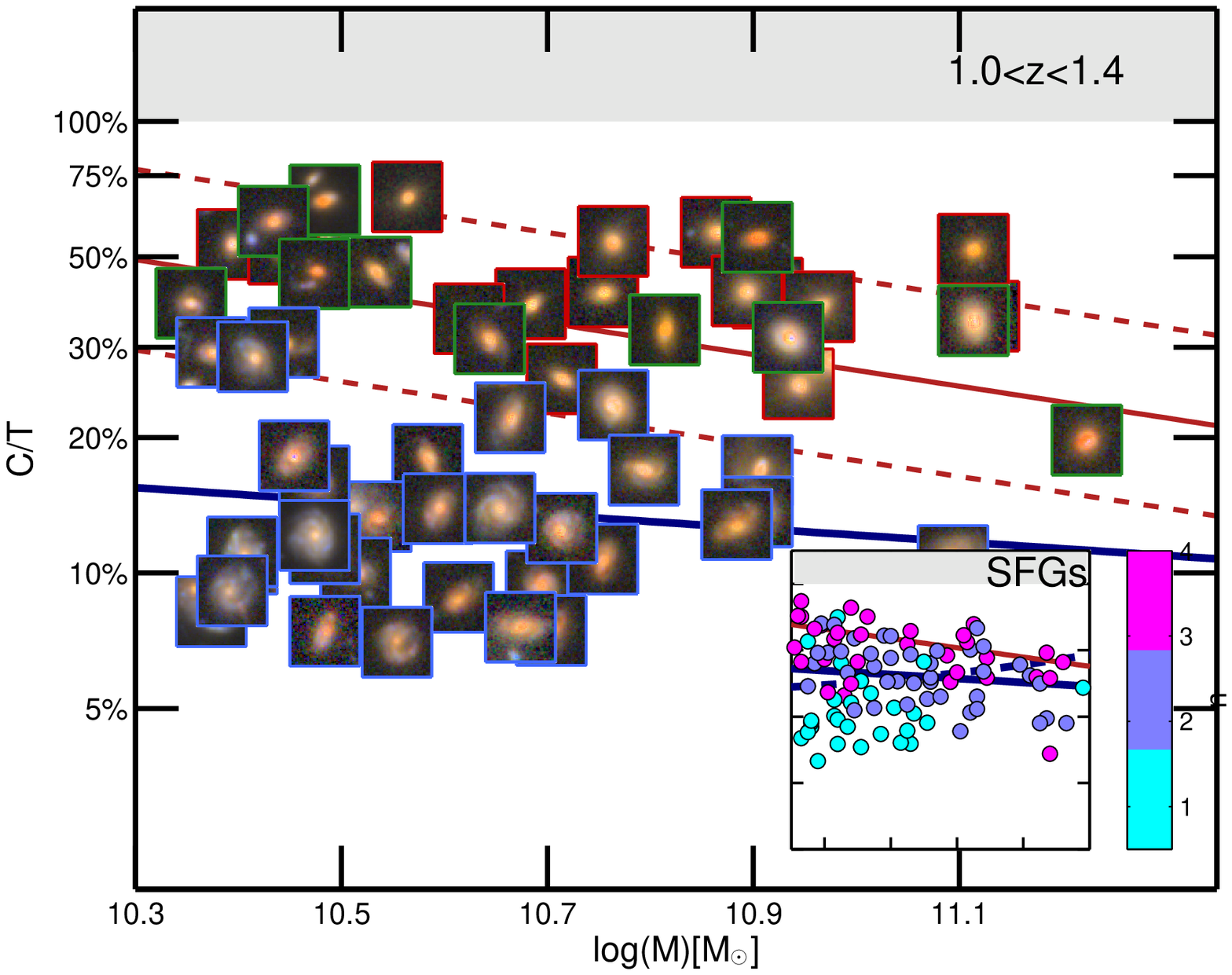}
\includegraphics[width=8.5cm,angle=0.]{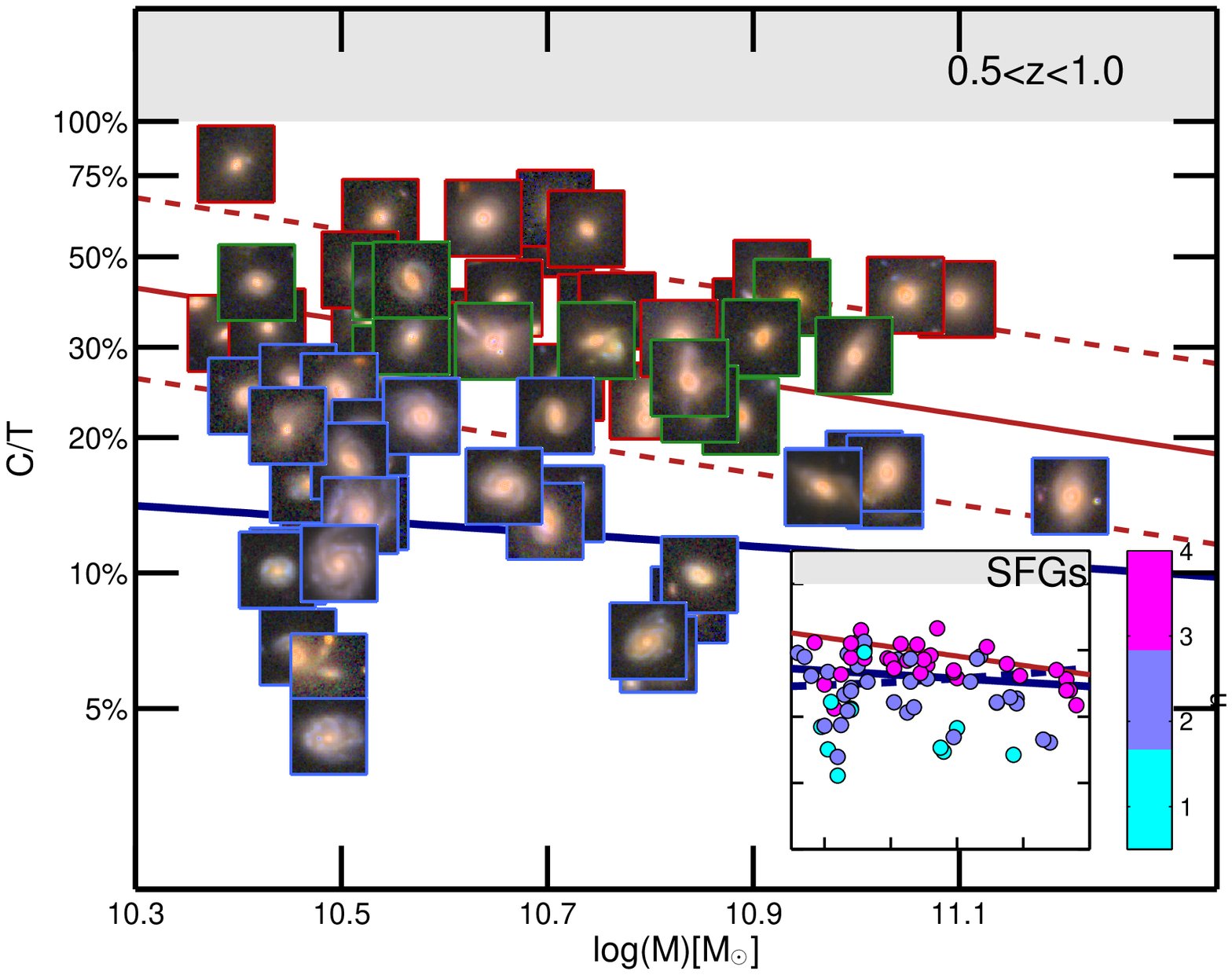}
\caption{\label{btdiagrams} Core to total mass ratio (C/T) vs. stellar
  mass for a representative sub-sample of massive galaxies at
  different redshifts. The color composite ACS/WFC3 $zJH$ images
  illustrate the evolution in the morphology of SFGs and quiescent
  galaxies as a function of time. The color of the border around the
  postage stamps indicate the location of the galaxy in the three
  regions of the compaction-quenching diagram (Figure~\ref{usr}). The
  solid blue and red lines indicate the best fit \sigone~relations for
  SFG and quiescent galaxies at each redshift. The dashed lines
  indicate the typical width of the \sigoneq~relation, which is the
  threshold used to select compact SFGs. The insets show the same
  diagram color coded by S\'ersic index for all massive SFGs. SFGs
  evolving in the $\Sigma$-MS decrease their C/T with time as the disk
  component becomes more prominent (i.e., the disk grows faster than
  the core). Similarly, quiescent galaxies decrease their C/T with
  time as they grow larger stellar envelopes.  However, SFGs that have
  a compaction event, experience a significant core growth relative to
  the total mass (increasing their C/T), moving upwards from the
  $\Sigma$-MS to the quiescent relation.  The visual appearances of
  compact SFGs (green) are very similar to those of quiescent galaxies
  at all redshifts, but are most different from other SFGs only at
  high-z (see also Figure~\ref{postages}).}
\end{figure*} 

At lower redshift, the smaller gas fractions and longer dynamical
timescales suggest that compaction mechanisms become a mixture of
weaker instabilities, which can still cause enhanced gas inflows or
inward migration of stellar clumps (e.g., \citealt{dekel09b};
\citealt{bournaud11a}; \citealt{genel12}), and {\it secular} processes
(\citealt{kor04}) associated with torques and dynamical friction in
the presence of bars and spiral arms. The latter play an important
role increasing the central density in SFGs that already exhibit
relatively quiescent centers, i.e., lacking enough star-formation to
sustain the core growth required to reach
\sigoneq~(\citealt{wuyts12,wuyts13}; \citealt{lang14};
\citealt{bruce14a,bruce14b}). In those SFGs, quenching is also
expected to be a slow process, related with gas consumption in the
star-forming disk (fading; e.g., \citealt{fang13};
\citealt{tacchella15}). The slow quenching process depends also on
additional mechanisms to prevent further gas accretion into the disk,
e.g., virial shock heating in massive haloes (\citealt{croton06};
\citealt{dekel06}) or AGN feedback (. As mentioned in the previous
section, some of these SFGs are compact in \sigone~but not in
\sige. Therefore, the tilt in $\Delta \rm SFR_{\rm MS}$
vs. $\Delta$\sigeq~can be caused by the size shrinkage ($r_{e}$)
associated with fading rather than with an increase in their core
mass.

\subsubsection{Number density of compact SFGs and quiescent galaxies}\label{ndensitysect}

\begin{figure*}[t]
\centering
\includegraphics[width=4.4cm,angle=0.]{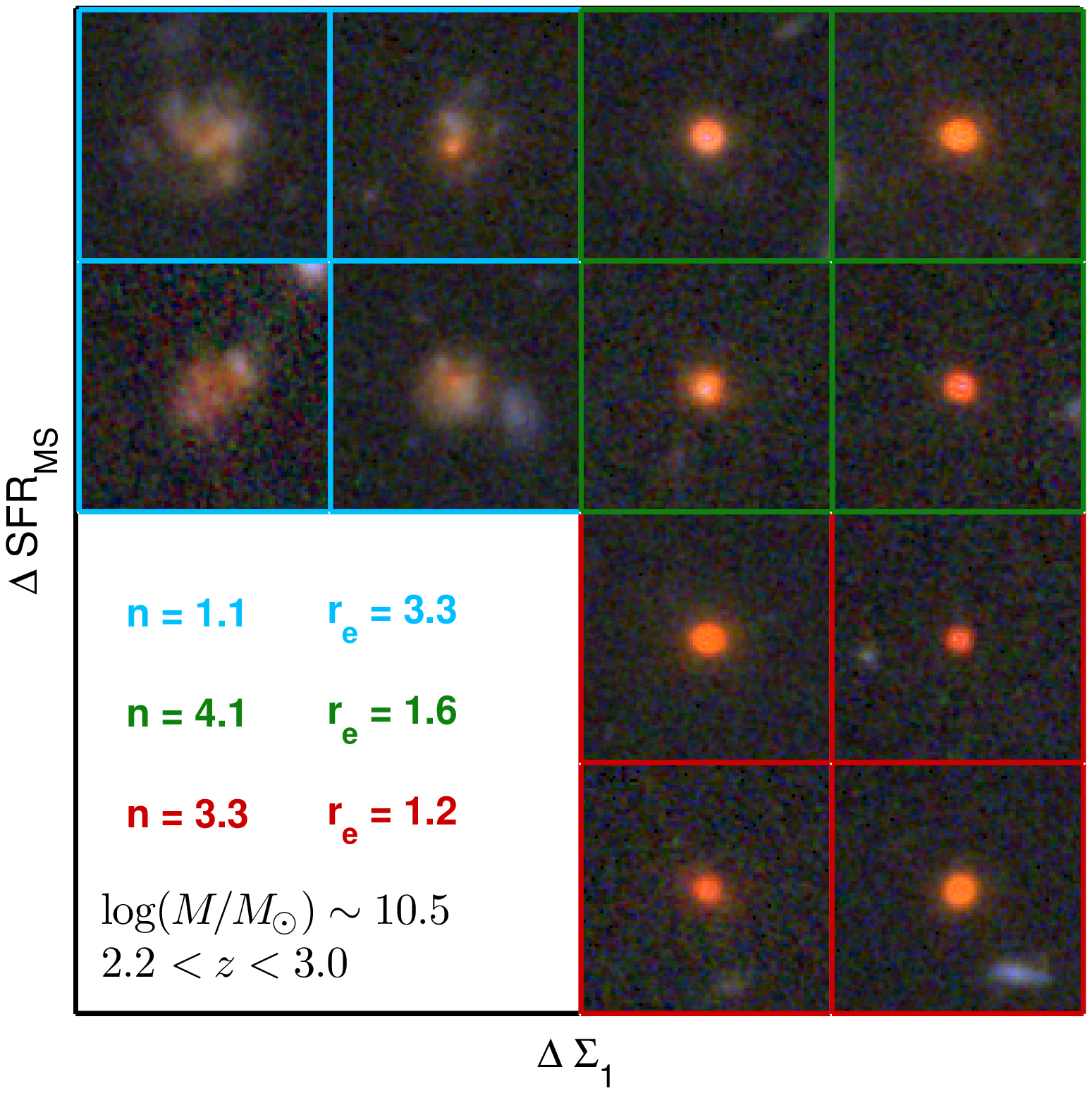}
\includegraphics[width=4.4cm,angle=0.]{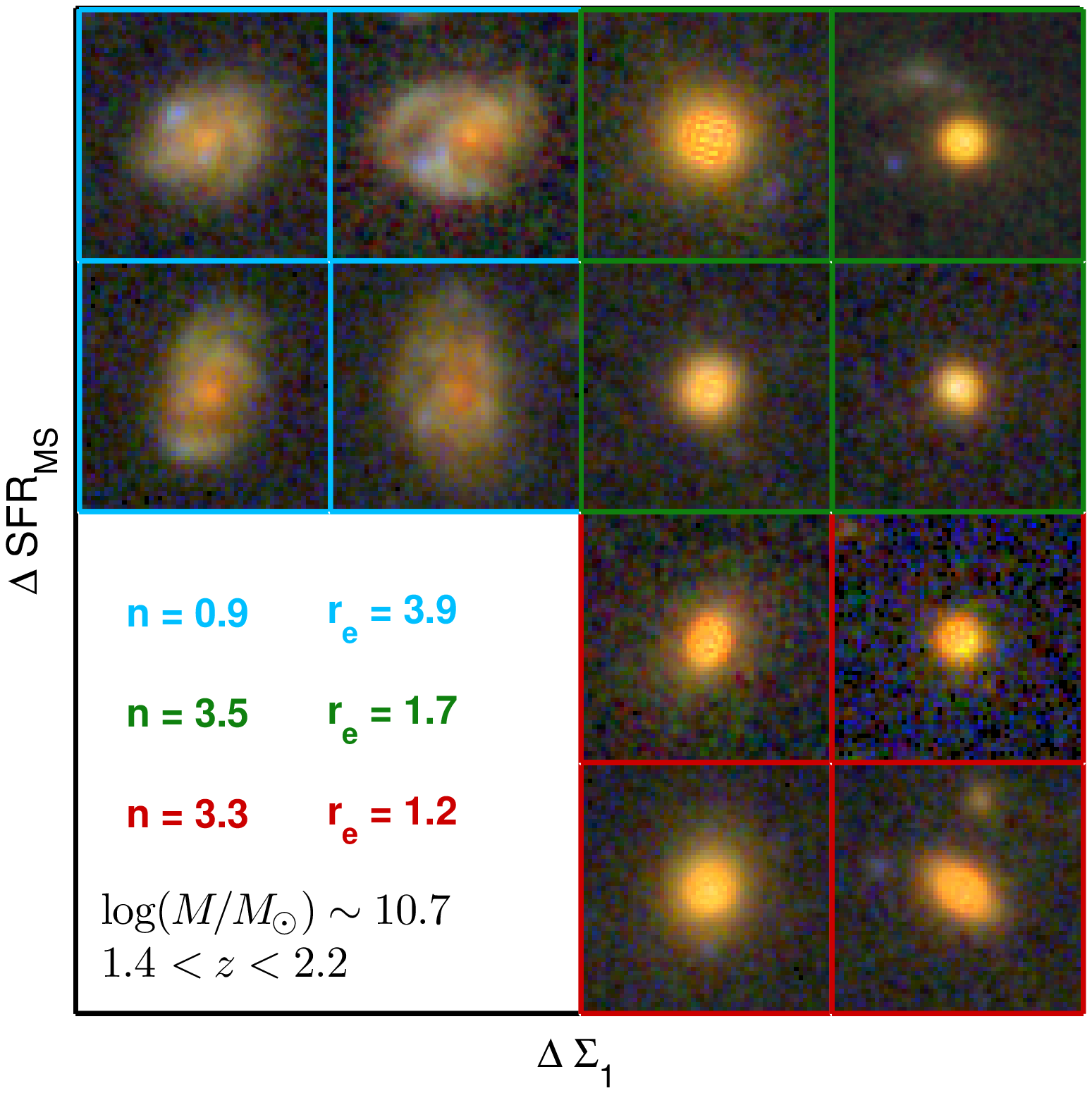}
\includegraphics[width=4.4cm,angle=0.]{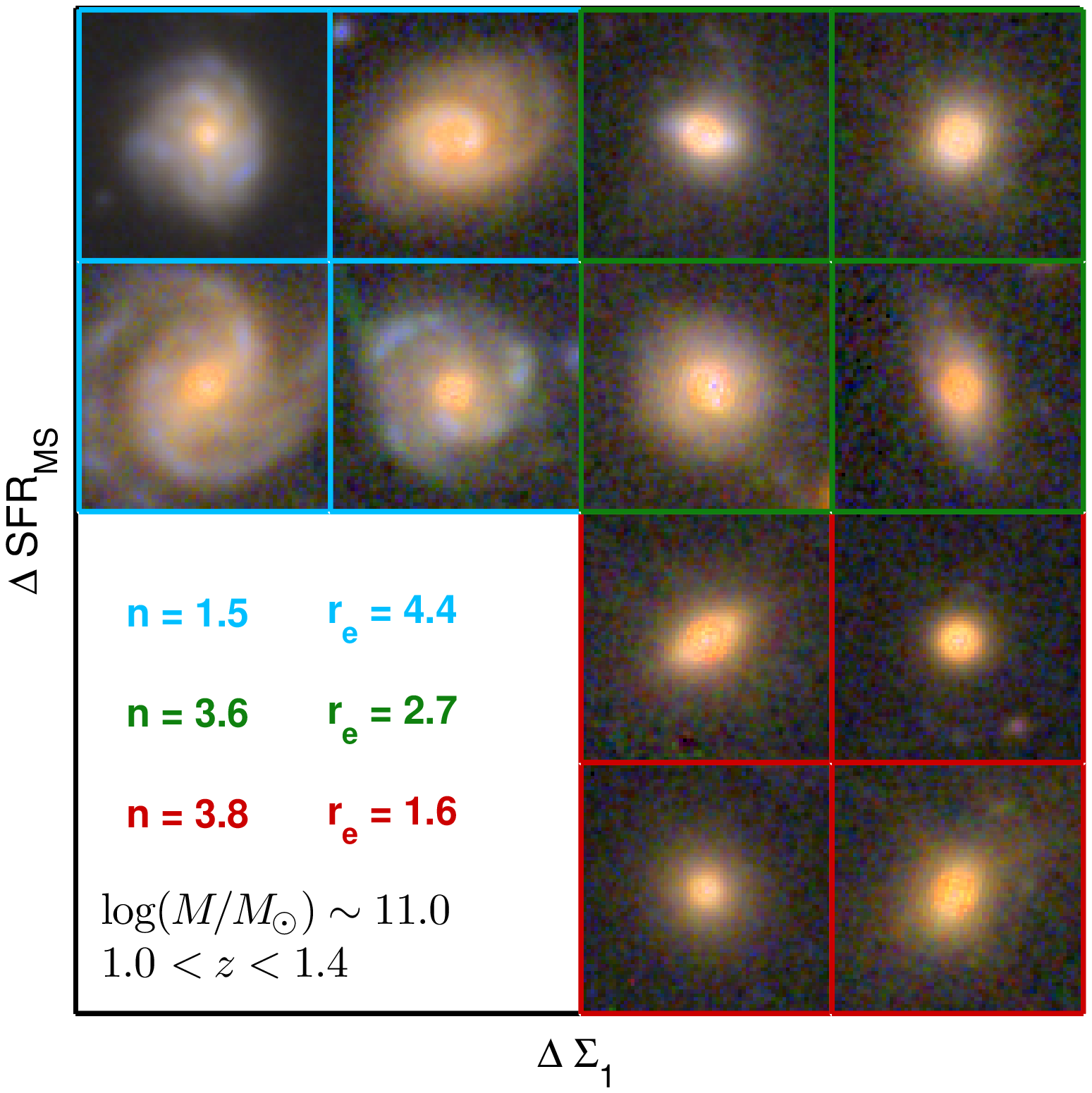}
\includegraphics[width=4.4cm,angle=0.]{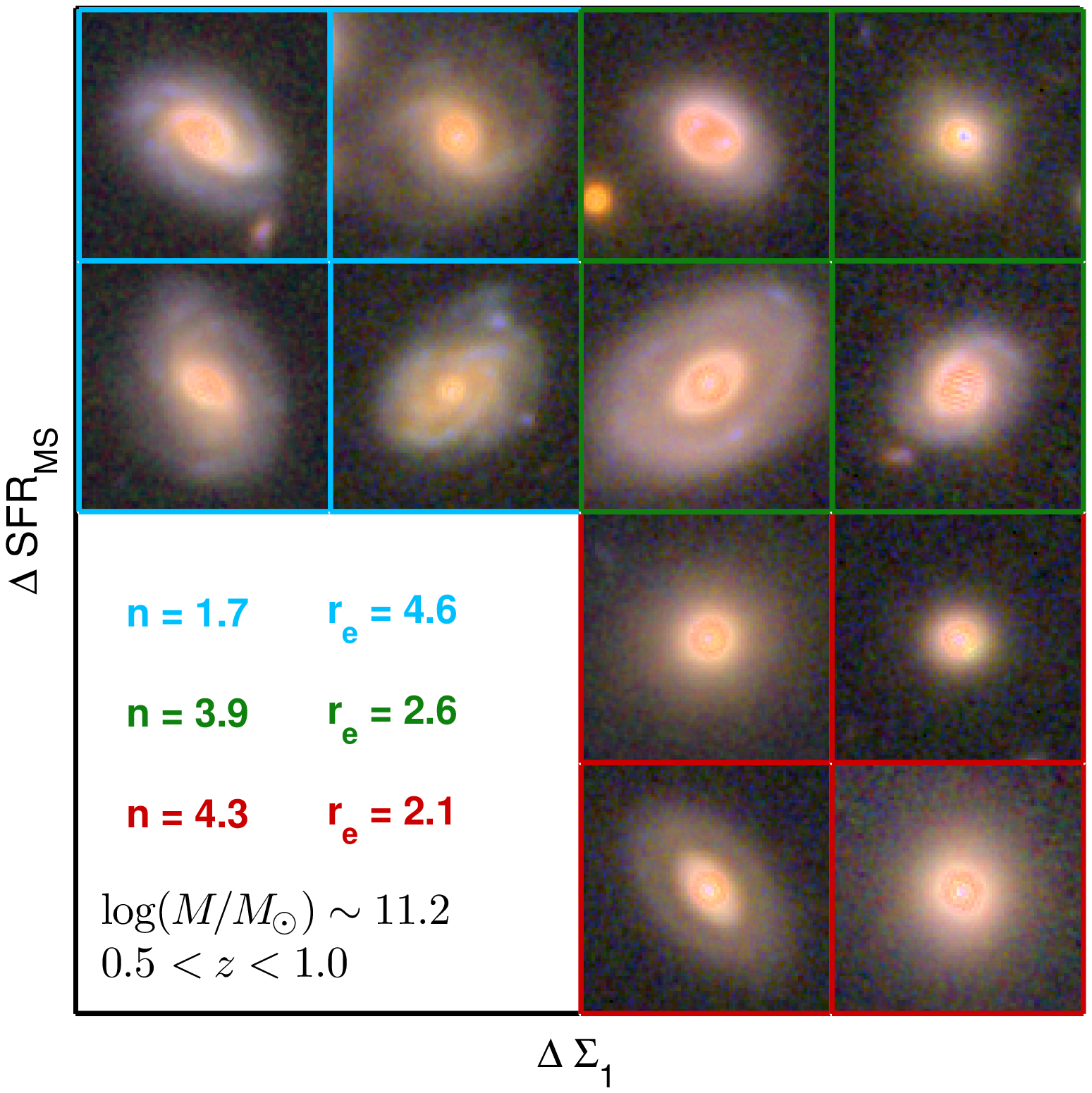}
\caption{\label{postages} Postage stamps in the
  $\Delta$SFR$-\Delta$\sigoneq~plane, showing galaxies in three
  different phases of the compaction-quenching sequence (see
  Figure~\ref{deltaseq}): extended SFGs (blue), compact SFGs (green)
  and quiescent galaxies (red). Each panel shows galaxies with
  increasing stellar masses as a function of time (left to right)
  following the expected mass growth from $z=3$ to $z=0.5$ predicted
  by semi-analytic models. The numbers on the inset indicate the
  median $n$ and $r_{e}$ of the galaxies in each group. The
  evolutionary sequence, discussed in \S~\ref{cqsequence}, suggests
  that some SFGs grow along the $\Sigma$-MS from $z=3$ to $z=0.5$
  (blue), while others experience a compaction event (green),
  departing from the $\Sigma$-MS.  Compaction produces a steeper mass
  profile, resulting in a strong increase in \sigone~and \sige. After
  compaction, galaxies quench (red), showing only minor changes in
  their structural and visual appearances relative to their compact
  SFG progenitors. The $\Sigma$-MS describes the growth of a disk,
  starting from a small, irregular morphology and becoming
  progressively more settled with time, increasing both its size and
  central density (bulge growth). The pre- and post- compaction
  appearances of SFGs at high-z are radically different, suggesting the
  action of a strongly dissipative process. However, at low-z, many
  compact SFGs exhibit disks, indicating that compaction affects
  mostly the central region of the galaxy. The overall appearance of
  quiescent galaxies evolves from ``naked'' to ``clothed'' cores
  (i.e., larger $r_{e}$ at similar \sigone). Such evolution is likely
  driven by size growth in quiescent galaxies formed at high-z, and
  the formation of new quiescent galaxies from compact SFGs with
  larger sizes.}
\end{figure*} 

The existence of a universal compaction-quenching diagram and the
increasing number density of quiescent galaxies as a function of time
supports the idea that the L-shaped diagram is indeed an evolutionary
sequence, i.e., compact SFGs in the knee of the relation are immediate
progenitors of the quiescent galaxies at lower
redshifts. Figure~\ref{ndensity} shows the evolution in the number
density of compact SFGs, selected using $\Delta$\sigoneq~$>-0.2$, and
all quiescent galaxies, regardless of their structural properties,
since $z\sim3$. The latter grows approximately as a power-law, N$^{\rm
  Q}\sim10^{-0.5(1+z)}$, in good agreement with previous results
(e.g., \citealt{muzzin13smf}). Assuming that quiescent galaxies are
descendants of compact SFGs at higher redshift, the number of
quiescent galaxies is the cumulative distribution of quenching compact
SFGs as a function of time. To quantify the relative numbers of these
two populations we assume that the number density of compact SFGs also
follows a power-law evolution as the result of 2 opposite processes
that cause either an increase (due to compaction) or decrease (due to
quenching) in their total number: $dN_{\rm SFG}/dz=\lambda_{\rm
  C}N_{\rm SFG}+\lambda_{\rm Q}N_{\rm SFG}$.  In this model, the
evolution in the number of quiescent galaxies is proportional to the
number of quenching SFGs, and thus inherit the power-law dependence:
$dN^{\rm Q}/dz=-\lambda_{d}N_{\rm SFG}\sim-\lambda_{\rm
  Q}10^{(\lambda_{\rm C}+\lambda_{\rm Q})(1+z)}$.  There are two
additional parameters which determine the initial number compact SFGs
and quiescent galaxies, N$^{\rm SFG}_{z0}$ and N$^{\rm Q}_{z0}$. We
set N$^{\rm Q}_{z0}=0$ at $z0=4$, and we fit the other 3 parameters to
the observed number density of compact SFGs and quiescent galaxies,
which yields $\lambda_{i}=1.27\pm0.15$, $\lambda_{d}=-0.75\pm0.12$ and
N$^{\rm SFG}_{z0}=5\cdot10^{-6}$~Mpc$^{-3}$. Thus, the characteristic
quenching timescale for a compact SFG in units of redshift is
$z_{1/2,Q}=\ln(10)/\ln(2)/\lambda_{\rm Q}=0.38$, which implies that
the quenching time increases from $t_{\rm Q}=700$~Myr to 1.1~Gyr at
$z=3$ and $z=1$. In spite of its simplicity, this model provides a
more realistic approximation than previous assumptions that all
compact SFGs form and quench in discrete intervals of time. In this
case, the number densities vary continuously while preserving the
evolutionary connection as a characteristic quenching time.

\subsection{Galaxy morphologies in the universal compaction-quenching sequence}\label{cqprops}

As discussed in \S~\ref{2dsequence}, the shape of the
compaction-quenching sequence is independent of redshift. However, the
evolutionary pace along the sequence declines with time as compaction,
and most likely quenching, processes become slower. Furthermore, the
evolution in the normalization of \sige~clearly indicates that the
structural and morphological properties of galaxies in the sequence
depend on the redshift.

Figure~\ref{btdiagrams} illustrates the evolution in the visual
appearances of SFGs and quiescent galaxies as a function of time in
the core-to-total vs. mass diagram. The thumbnail frames are color
coded according to the location of the galaxies in the 3 regions of
the compaction-quenching diagram (Figure~\ref{usr}). Overall, the
evolutionary tracks in this diagram follow decreasing trends in C/T
for both SFGs (i.e., inside-out growth of a disk), and quiescent
galaxies (i.e., stellar halo growth). Only SFGs experiencing a
compaction event (from blue to green) follow a trend of increasing
C/T. Note that, conceptually, C/T is similar to a {\it bulge}-to-total
ratio. However, bulges can typically grow beyond a radius of 1~kpc, in
fact, recent results indicate that both the bulge size and its mass
correlate with the total mass of the galaxy, i.e., B/T increases with
mass, as opposed to C/T (e.g., \citealt{lang14}). Qualitatively, the
visual appearances agree with this evolution. The bulk of SFGs exhibit
larger are more-settled disks as a function of time, while also
increasing their central densities. The steeper slope of
\sigonesf~seems clear in this diagram (dashed blue line), particularly
at $z<1$, where slow compaction processes push most SFGs above
C/T$\gtrsim10\%$.  Quiescent galaxies also show larger ({\it puffed
  up}) appearances with time with values declining from $C/T\sim60\%$
at $z=2.6$ (``naked''-cores) to 35\% and $z=0.75$ (``clothed''-cores).

To further illustrate the change in visual appearances within an
evolutionary path, Figure~\ref{postages} shows images of galaxies with
increasing stellar mass a function of time in the 3 regions of the
compaction-quenching sequence. The increase in stellar mass with
redshift follows the average mass growth in galaxies with
\lmass~$=10.5$ at $z=2.5$. There several methods to estimate the
typical stellar mass growth with redshift (e.g. \citealt{dokkum10};
\citealt{patel13}; \citealt{marchesini14}). Here we adopt the
evolutionary tracks of \citet{moster13} determined from semi-analytic
models, with average mass growth of $\Delta M_{\star}=0.7$~dex from
$z=2.5$ to $z=0.75$.  Since galaxies of a given mass can be in any of
the three stages at a given redshift, we assume that at each redshift
some SFGs remain in the left region (blue) growing along the
\sigone-MS, others move rightward due to compaction (blue to green),
and some compact SFGs quench downward after reaching a maximum central
density (green to red).

The morphologies and visual appearances change strongly as a function
of redshift: 1) at $z\gtrsim1.4$, compaction involves a significant
structural transformation going from irregular, clumpy SFGs to compact
``naked'' spheroids. Compact SFGs quench having almost identical
morphologies than quiescent galaxies at those redshifts (see also
\citealt{barro14a}). Extended SFGs in the $\Sigma$-MS (blue) increase
their size, central density and total mass with time. However, as
total mass grows faster than the core mass, their C/T decreases. By
$z\lesssim2$ these galaxies exhibit clear disk-like structures and
distinct cores with redder colors (see also \citealt{wuyts12};
\citealt{lang14}; \citealt{tacchella15}). 2) at $z<1.4$, extended SFGs
show larger, more settled disks and similarly more massive
cores. Compact SFGs and quiescent galaxies also have denser cores,
but, in contrast with the high-z appearances, they exhibit diffuse
extended components (some clearly star-forming). This suggests that
the compaction process at lower redshift preserves the already
existing disk structure. The evolution towards disk-like morphologies
with progressively bigger cores eventually results in the prominence
of disks with bulges among massive SFGs and quiescent galaxies at
$z\gtrsim1$ reported in previous works (e.g., \citealt{bundy10};
\citealt{bruce12}; \citealt{bruce14b}; \citealt{buitrago13};
\citealt{mclure13}; \citealt{lang14}; Huertas-Company et al 2015). As
noted in previous works, however, the progenitor bias makes it nearly
impossible to distinguish between puffed-up, older galaxies and newly
quenched galaxies with larger sizes attending only to their
morphologies.

\section{Summary}

We analyze the star-formation and structural properties of massive
galaxies in the CANDELS/GOODS-S field to study the relation between
stellar mass and structural growth, and the role of the latter in the
quenching of star-formation since $z\sim3$. We characterize the
structural properties as a function of redshift by studying the
correlations in the mass surface density within the effective radius,
\sige, and within the central 1~kpc, \sigone, vs. stellar
mass. \sigone~traces the stellar mass growth in the galaxy core, and
thus it is close to the concept of a cosmic clock (i.e., it increases
with time). \sige~, however, depends on the relative balance between
stellar mass and size growth, and thus it exhibits positive and
negative fluctuations.

We find that SFGs and quiescent galaxies follow clear and distinct
correlations in \sige~and \sigone vs. stellar mass since $z\sim3$.
These correlations are well-described by linear relations in log-log
space. The slopes and scatter of these relations are relatively
constant with time, while their normalizations decline (see
Table~\ref{powerlaw}).  The scatter in the \sigone~structural
relations is $\sim$2$\times$ tighter than in the \sige~relations for
~for both SFGs and quiescent galaxies. For SFGs, the normalizations in
\sige~and \sigone~decrease by less than factor of $\sim$2 from $z=3$
to $z=0.5$. For quiescent galaxies, the decline in \sige~is 5$\times$
larger than in \sigone~ ($\sim0.3$~dex vs. $\sim1$~dex,
respectively). The differential evolution of the normalization in
\sigeq~and \sigoneq~is inconsistent with a simple minor merger
scenario in which the core mass remains constant while the size
increases. However, the normalizations at redshifts $z=3$ to $z=0.5$
agree well with one another if the mass profiles of quiescent galaxies
follow a single S\'ersic with $n\sim4$. Thus the large increase with
time in the size of quiescent galaxies is consistent with just modest
decline in \sigone~at constant S\'ersic. Such decline can be caused by
the formation of new quiescent galaxies with lower density cores or
mass loss due to passive evolution in the already existing population.

Based on the slow decline in the normalizations of the structural
relations for SFGs, we speculate that these galaxies follow
evolutionary paths {\it along} the $\Sigma_{e,1}$ correlations. We
define these paths as the $\Sigma$ main sequence, following the
paradigm of the SFR-MS as a smooth phase of stellar and structural
growth (e.g., \citealt{elbaz11}; \citealt{rodi11}). The evolution in
the $\Sigma$-MS is consistent with the inside-out growth of an
exponential disk due to in-situ star-formation, i.e., an increase in
both the core mass and the overall size of the galaxy.

At every mass and redshift, quiescent galaxies have steeper mass
profiles (higher S\'ersic) and higher surface densities than SFGs.
This implies that growing a dense stellar core is a pre-requisite for
quenching star-formation (see also, \citealt{cheung12},
\citealt{bell12}; \citealt{fang13}; \citealt{dokkum14}). Thus, the
immediate star-forming progenitors of quiescent galaxies must
experience a phase of stronger core-growth, relative to the
$\Sigma$-MS. We define this phase(s) of fast increase in \sigone,
\sige~and $n$ as {\it compaction}. The compaction phase is typically
associated with dissipational processes ranging from major mergers
(\citealt{hopkins06}; \citealt{hopkins08a}), to violent gravitational
instabilities (\citealt{dekel09a,dekel13b}; \citealt{ceverino10}), and
weaker {\it secular} instabilities (e.g., bars and spiral arms;
\citealt{kor04}). The evolutionary tracks of massive SFGs in recent
hydrodynamical simulations exhibit an excellent agreement with the
$\Sigma$-MS/compaction scenario (\citealt{ceverino15};
\citealt{zolotov14}). The simulations suggest that SFGs depart from
a steady-state evolution along the $\Sigma$-MS as a result of
dissipational compaction events triggered by intense episodes of gas
inflow. The strength and duration of these events declines with time
following a decline in the gas reservoirs of SFGs
(\citealt{dekel13a}).

\begin{figure}[t]
\centering
\includegraphics[width=8cm,angle=0.]{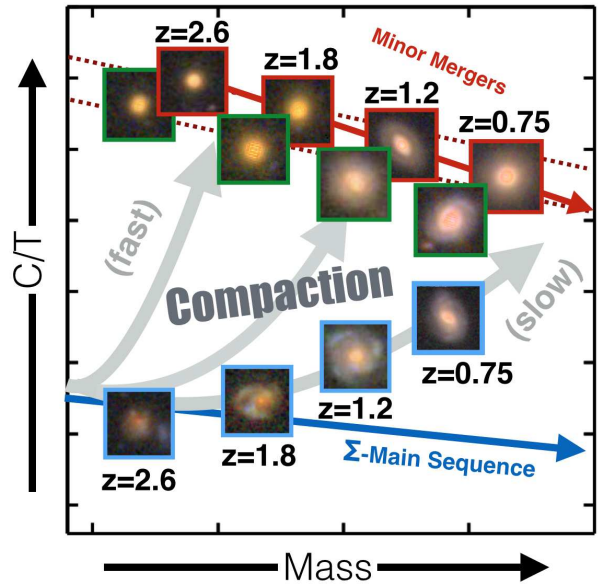}
\caption{\label{cartoon} Structural evolutionary path of massive
  galaxies since $z\sim3$ in terms of their core ($r<1$~kpc) to total
  mass (C/T) vs. stellar mass. The typical SFG follows a structural
  ``main sequence'' that describes the inside-out growth of a disk
  (blue-framed images), i.e., increasing its stellar mass and $r_{e}$
  as does its central density (bulge growth). Prior to quenching, SFGs
  experience a dissipative compaction phase of enhanced core growth
  (green-framed images) fueled by the inward migration of gas into the
  central region. Compaction events are strongest at high-z, due to
  higher gas fractions, forming extremely compact SFGs on very short
  timescales. At low-z the strength of compaction diminishes, the
  core growth becomes slower and it preserves the underlying disk
  structure.  Newly quenched galaxies at each redshift (red-framed
  images) are direct descendants of compact SFGs, which shut down
  star-formation after forming a dense stellar core with no further
  structural changes. Quiescent galaxies grow an extended stellar halo
  due to minor merging, decreasing their C/T. Quenching timescales
  depend on the strenght of the compaction process, which drives a
  major gas consumption event. Thus, qualitatively, quenching follows
  the fast shut-down of a compact core at high-z and the slower fading
  of a star-forming disk at low-z.}
\end{figure} 

We find that the 2D distribution of SFGs and quiescent galaxies
relative to the SFR-MS and the quiescent structural relations, $\Delta
{\rm SFR_{\rm MS}}-\Delta\Sigma_{e,1}^{\rm Q}$, exhibits a universal
L-shape that is independent of redshift (Figure~\ref{deltaseq}). Each
branch of this distribution describes a fundamental transition in the
evolution of SFGs, namely: compaction and quenching. The horizontal
branch describes the growth in size and core mass of a SFG along the
$\Sigma$-MS, followed by a compaction event that causes a strong
increase in surface density and S\'ersic index. Compaction marks the
departure from the $\Sigma$-MS towards the knee of the L-shaped
distribution. The vertical branch describes the quenching of
star-formation at maximum core and effective surface density.  The
latter implies that the formation of a dense core precedes the
quenching of star-formation. Therefore, {\it compact} SFGs in the knee
of the L-diagram at every redshift are the most likely progenitors
quiescent galaxies at lower-z. Owing to the mild decline in the
normalization of the \sigoneq~relation with time, a constant threshold
of $\Sigma_{1}^{\rm
  Q}\equiv\Sigma_{1}-0.65\log(M_{\star}-10.5)\gtrsim9.5$ identifies
compact SFGs at any redshift $z>0.5$.

Although the L-shape of the compaction-quenching sequence is
independent of redshift, the rate of growth along the sequence and the
strenght of the transformation processes declines with time following
the similar decline in the overall SFR (e.g., \citealt{whitaker14};
\citealt{speagle14}). As a result, the morphologies of galaxies in
each region of the sequence are substantially different with
time. Figure~\ref{cartoon} describes qualitatively the change in
morphologies and core-to-total ratios for galaxies following
fast-to-slow evolutionary paths in the compaction-quenching sequence
as a function of time.

At high-z, SFGs have larger gas reservoirs and thus follow a faster
mass growth (\citealt{tacconi10}; \citealt{tacconi13}). Those SFGs are
also prone to stronger gravitational instabilities and expedited
compaction processes capable of transforming SFGs with clumpy,
irregular morphologies into compact SFGs with spheroid-like
morphologies, high S\'ersic and and high core-to-total ratios
(\citealt{dekel13b}; \citealt{wellons14}; \citealt{ceverino15};
\citealt{zolotov14}). I.e., ``naked'' star-forming cores, which appear
to quench fast turning into compact quiescent galaxies
(\citealt{bezanson13}; \citealt{onodera14}; \citealt{belli14b};
\citealt{barro15}) with similar masses and structural properties.

At low-z, SFGs have lower gas fractions and SFRs, which implies a
slower mass growth and weaker gravitational instabilities. This is
consistent with SFGs having more settled, disk-like morphologies,
larger sizes, and progressively denser cores (see also,
\citealt{bruce12}; \citealt{lang14}; \citealt{tacchella15}). As
opposed to high-z, the similar morphologies of compact SFGs
(green-framed images) and SFGs in the $\Sigma$-MS (blue-framed images)
suggests that compaction at low-z causes an enhanced core growth,
increasing C/T within an already existing disk.  Many SFGs at low-z
exhibit blue disks and red cores, suggesting that quenching
star-formation (in the whole galaxy) is a slower process that causes
also minor changes in their appearances due to disk-fading.

\section*{Acknowledgments}

Support for Program number HST-GO-12060 was provided by NASA through a
grant from the Space Telescope Science Institute, which is operated by
the Association of Universities for Research in Astronomy,
Incorporated, under NASA contract NAS5-26555. GB acknowledges support
from NSF grant AST-08-08133. PGP-G acknowledges support from grant
AYA2012-31277.  This work has made use of the Rainbow Cosmological
Surveys Database, which is operated by the Universidad Complutense de
Madrid (UCM), partnered with the University of California
Observatories at Santa Cruz (UCO/Lick,UCSC). This work was partly
supported, by MINECO grant AYA2012-32295. FL acknowledges support from
NSFC grant 11573017

\bibliographystyle{aa}
\bibliography{referencias}
\clearpage
\label{lastpage}

\end{document}

%% file: etab1.tex
\begin{table*}[t]
\centering
\tabletypesize{\scriptsize}
\setlength{\tabcolsep}{0.025in} 

\begin{threeparttable}
    \caption{Power Law Fits}\label{powerlaw}
    \begin{tabular}{|l|cc||cc|cc||cc|cc|}
      \hline 
                     &\multicolumn{2}{c||}{$SFR_{MS}-$~SFGs}&\multicolumn{2}{c|}{\sige~$-$~Quiescent}&\multicolumn{2}{c||}{\sige~$-$~SFGs}& \multicolumn{2}{c|}{\sigone~$-$~Quiescent}&\multicolumn{2}{c|}{\sigone~$-$~SFGs}\\
      \hline
      redshift range & $\mu$ & $\log$C        & $\alpha$ & $\log$A & $\alpha$ & $\log$A & $\beta$ & $\log$B & $\beta$ & $\log$B\\
      \hline
      $0.5<z<1.0$  & $0.19\pm0.08$ & $1.21\pm0.02$ & -$0.42\pm0.13$ & $9.15\pm0.06$  & $0.60\pm0.04$ & $8.46\pm0.04$  & $0.65\pm0.03$ & $9.53\pm0.05$& $0.89\pm0.03$ & $9.12\pm0.03$\\
      $1.0<z<1.4$  & $0.53\pm0.07$ & $1.44\pm0.04$ & -$0.45\pm0.14$ & $9.53\pm0.05$  & $0.60\pm0.05$ & $8.54\pm0.05$  & $0.65\pm0.04$ & $9.64\pm0.04$& $0.88\pm0.03$ & $9.16\pm0.04$\\
      $1.4<z<2.2$  & $0.64\pm0.06$ & $1.75\pm0.05$ & -$0.52\pm0.14$ & $9.91\pm0.07$  & $0.64\pm0.05$ & $8.68\pm0.04$  & $0.64\pm0.03$ & $9.74\pm0.05$& $0.86\pm0.04$ & $9.25\pm0.03$\\
      $2.2<z<3.0$  & $0.68\pm0.06$ & $1.92\pm0.05$ & -$0.48\pm0.17$ & $10.16\pm0.05$ & $0.56\pm0.06$ & $8.80\pm0.04$  & $0.67\pm0.04$ & $9.80\pm0.05$& $0.89\pm0.04$ & $9.33\pm0.05$\\
      \hline
      \noalign{\smallskip}
    \end{tabular}
    \begin{tablenotes}
      \small
    \item \emph{Notes.} power law coefficients parameterizing the
      evolution of the SFR
      ($\log$SFR$=\mu\log(M_{\star}-10.5)+\log{\rm C}$), the effective
      mass surface density
      ($\log$\sige$=\alpha\log(M_{\star}-10.5)+\log{\rm A}$) and the
      mass surface density within the central 1~kpc
      ($\log$\sigone$=\beta\log(M_{\star}-10.5)+\log{\rm B}$)
      relations with stellar mass as a function of redshift.
    \end{tablenotes}
  \end{threeparttable}
\end{table*}

%% file: etab2.tex
\begin{table*}[t]
\centering
\tabletypesize{\scriptsize}
\setlength{\tabcolsep}{0.025in} 
\begin{threeparttable}
    \caption{2nd-order Power Law Fits}\label{2ndorder}
    \begin{tabular}{|l|ccc|ccc|}
      \hline 
       &\multicolumn{3}{c|}{\sige$-$SFGs}&\multicolumn{3}{c|}{\sigone$-$SFGs}\\
      \hline
      redshift range & $\alpha_{1}$ & $\alpha_{2}$ & $\log$A 
                     & $\beta_{1}$ & $\beta_{2}$ & $\log$B\\                      
      \hline
      $0.5<z<1.0$  & $0.09\pm0.08$ & $1.10\pm0.07$  & $9.25\pm0.03$ & $0.08\pm0.04$ & $0.80\pm0.03$  & $8.56\pm0.05$\\
      $1.0<z<1.4$  & $0.15\pm0.07$ & $1.19\pm0.06$  & $9.25\pm0.02$ & $0.08\pm0.03$ & $0.95\pm0.04$  & $8.65\pm0.04$\\
      $1.4<z<2.2$  & $0.16\pm0.06$ & $1.18\pm0.07$  & $9.36\pm0.02$ & $0.09\pm0.03$ & $1.12\pm0.03$  & $8.82\pm0.05$\\
      $2.2<z<3.0$  & $0.14\pm0.06$ & $1.10\pm0.05$  & $9.41\pm0.02$ & $0.11\pm0.04$ & $1.10\pm0.04$  & $8.93\pm0.05$\\
      \hline
      \noalign{\smallskip}
    \end{tabular}
    \begin{tablenotes}
      \small
    \item \emph{Notes.} Power law coefficients parameterizing the
      evolution of \sige~and \sigone~ vs. stellar mass as 2nd order polynomials:
      $\log$\sige$=\alpha_{1}(\log M)^{2}+\alpha_{2}\log M+\log {\rm A}$ and
      $\log$\sigone$=\beta_{1}(\log M)^{2}+\beta_{2}\log M+\log {\rm B}$ 
    \end{tablenotes}
  \end{threeparttable}
\end{table*}

%% file: ms.bbl
\begin{thebibliography}{129}
\expandafter\ifx\csname natexlab\endcsname\relax\def\natexlab#1{#1}\fi

\bibitem[{{Barro} {et~al.}(2015){Barro}, {Faber}, {Dekel}, {Pacifici},
  {Perez-Gonzalez}, {Toloba}, {Koo}, {Trump}, {Inoue}, {Guo}, {Liu}, {Primack},
  {Koekemoer}, {Brammer}, {Cava}, {Cardiel}, {Ceverino}, {Eliche}, {Fang},
  {Finkelstein}, {Kocevski}, {Livermore}, \& {McGrath}}]{barro15}
{Barro}, G., {Faber}, S.~M., {Dekel}, A., {et~al.} 2015, ArXiv e-prints

\bibitem[{{Barro} {et~al.}(2013){Barro}, {Faber}, {P{\'e}rez-Gonz{\'a}lez},
  {Koo}, {Williams}, {Kocevski}, {Trump}, {Mozena}, {McGrath}, {van der Wel},
  {Wuyts}, {Bell}, {Croton}, {Ceverino}, {Dekel}, {Ashby}, {Cheung},
  {Ferguson}, {Fontana}, {Fang}, {Giavalisco}, {Grogin}, {Guo}, {Hathi},
  {Hopkins}, {Huang}, {Koekemoer}, {Kartaltepe}, {Lee}, {Newman}, {Porter},
  {Primack}, {Ryan}, {Rosario}, {Somerville}, {Salvato}, \& {Hsu}}]{barro13}
{Barro}, G., {Faber}, S.~M., {P{\'e}rez-Gonz{\'a}lez}, P.~G., {et~al.} 2013,
  \apj, 765, 104

\bibitem[{{Barro} {et~al.}(2014{\natexlab{a}}){Barro}, {Faber},
  {P{\'e}rez-Gonz{\'a}lez}, {Pacifici}, {Trump}, {Koo}, {Wuyts}, {Guo}, {Bell},
  {Dekel}, {Porter}, {Primack}, {Ferguson}, {Ashby}, {Caputi}, {Ceverino},
  {Croton}, {Fazio}, {Giavalisco}, {Hsu}, {Kocevski}, {Koekemoer},
  {Kurczynski}, {Kollipara}, {Lee}, {McIntosh}, {McGrath}, {Moody},
  {Somerville}, {Papovich}, {Salvato}, {Santini}, {Tal}, {van der Wel},
  {Williams}, {Willner}, \& {Zolotov}}]{barro14a}
---. 2014{\natexlab{a}}, \apj, 791, 52

\bibitem[{{Barro} {et~al.}(2011){Barro}, {P{\'e}rez-Gonz{\'a}lez}, {Gallego},
  {Ashby}, {Kajisawa}, {Miyazaki}, {Villar}, {Yamada}, \&
  {Zamorano}}]{barro11b}
{Barro}, G., {P{\'e}rez-Gonz{\'a}lez}, P.~G., {Gallego}, J., {et~al.} 2011,
  \apjs, 193, 30

\bibitem[{{Barro} {et~al.}(2014{\natexlab{b}}){Barro}, {Trump}, {Koo}, {Dekel},
  {Kassin}, {Kocevski}, {Faber}, {van der Wel}, {Guo}, {Perez-Gonzalez},
  {Toloba}, {Fang}, {Pacifici}, {Simons}, {Campbell}, {Ceverino},
  {Finkelstein}, {Goodrich}, {Kassis}, {Koekemoer}, {Konidaris}, {Livermore},
  {Lyke}, {Mobasher}, {Nayyeri}, {Peth}, {Primack}, {Rizzi}, {Somerville},
  {Wirth}, \& {Zolotov}}]{barro14b}
{Barro}, G., {Trump}, J.~R., {Koo}, D.~C., {et~al.} 2014{\natexlab{b}}, ArXiv
  e-prints

\bibitem[{{Bell}(2008)}]{bell08}
{Bell}, E.~F. 2008, \apj, 682, 355

\bibitem[{{Bell} {et~al.}(2005){Bell}, {Papovich}, {Wolf}, {Le Floc'h},
  {Caldwell}, {Barden}, {Egami}, {McIntosh}, {Meisenheimer},
  {P{\'e}rez-Gonz{\'a}lez}, {Rieke}, {Rieke}, {Rigby}, \& {Rix}}]{bell05}
{Bell}, E.~F., {Papovich}, C., {Wolf}, C., {et~al.} 2005, \apj, 625, 23

\bibitem[{{Bell} {et~al.}(2012){Bell}, {van der Wel}, {Papovich}, {Kocevski},
  {Lotz}, {McIntosh}, {Kartaltepe}, {Faber}, {Ferguson}, {Koekemoer}, {Grogin},
  {Wuyts}, {Cheung}, {Conselice}, {Dekel}, {Dunlop}, {Giavalisco},
  {Herrington}, {Koo}, {McGrath}, {de Mello}, {Rix}, {Robaina}, \&
  {Williams}}]{bell12}
{Bell}, E.~F., {van der Wel}, A., {Papovich}, C., {et~al.} 2012, \apj, 753, 167

\bibitem[{{Belli} {et~al.}(2014{\natexlab{a}}){Belli}, {Newman}, \&
  {Ellis}}]{belli14a}
{Belli}, S., {Newman}, A.~B., \& {Ellis}, R.~S. 2014{\natexlab{a}}, \apj, 783,
  117

\bibitem[{{Belli} {et~al.}(2014{\natexlab{b}}){Belli}, {Newman}, {Ellis}, \&
  {Konidaris}}]{belli14b}
{Belli}, S., {Newman}, A.~B., {Ellis}, R.~S., \& {Konidaris}, N.~P.
  2014{\natexlab{b}}, \apjl, 788, L29

\bibitem[{{Bezanson} {et~al.}(2013){Bezanson}, {van Dokkum}, {van de Sande},
  {Franx}, \& {Kriek}}]{bezanson13}
{Bezanson}, R., {van Dokkum}, P., {van de Sande}, J., {Franx}, M., \& {Kriek},
  M. 2013, \apjl, 764, L8

\bibitem[{{Bezanson} {et~al.}(2009){Bezanson}, {van Dokkum}, {Tal},
  {Marchesini}, {Kriek}, {Franx}, \& {Coppi}}]{bezanson09}
{Bezanson}, R., {van Dokkum}, P.~G., {Tal}, T., {et~al.} 2009, \apj, 697, 1290

\bibitem[{{Bournaud} {et~al.}(2011){Bournaud}, {Chapon}, {Teyssier}, {Powell},
  {Elmegreen}, {Elmegreen}, {Duc}, {Contini}, {Epinat}, \&
  {Shapiro}}]{bournaud11a}
{Bournaud}, F., {Chapon}, D., {Teyssier}, R., {et~al.} 2011, \apj, 730, 4

\bibitem[{{Bouwens} {et~al.}(2010){Bouwens}, {Illingworth}, {Oesch}, {Trenti},
  {Stiavelli}, {Carollo}, {Franx}, {van Dokkum}, {Labb{\'e}}, \&
  {Magee}}]{bouwens10}
{Bouwens}, R.~J., {Illingworth}, G.~D., {Oesch}, P.~A., {et~al.} 2010, \apjl,
  708, L69

\bibitem[{{Brammer} {et~al.}(2008){Brammer}, {van Dokkum}, \& {Coppi}}]{eazy}
{Brammer}, G.~B., {van Dokkum}, P.~G., \& {Coppi}, P. 2008, \apj, 686, 1503

\bibitem[{{Brammer} {et~al.}(2011){Brammer}, {Whitaker}, {van Dokkum},
  {Marchesini}, {Franx}, {Kriek}, {Labb{\'e}}, {Lee}, {Muzzin}, {Quadri},
  {Rudnick}, \& {Williams}}]{brammer11}
{Brammer}, G.~B., {Whitaker}, K.~E., {van Dokkum}, P.~G., {et~al.} 2011, \apj,
  739, 24

\bibitem[{{Bruce} {et~al.}(2012){Bruce}, {Dunlop}, {Cirasuolo}, {McLure},
  {Targett}, {Bell}, {Croton}, {Dekel}, {Faber}, {Ferguson}, {Grogin},
  {Kocevski}, {Koekemoer}, {Koo}, {Lai}, {Lotz}, {McGrath}, {Newman}, \& {van
  der Wel}}]{bruce12}
{Bruce}, V.~A., {Dunlop}, J.~S., {Cirasuolo}, M., {et~al.} 2012, ArXiv e-prints

\bibitem[{{Bruce} {et~al.}(2014{\natexlab{a}}){Bruce}, {Dunlop}, {McLure},
  {Cirasuolo}, {Buitrago}, {Bowler}, {Targett}, {Bell}, {McIntosh}, {Dekel},
  {Faber}, {Ferguson}, {Grogin}, {Hartley}, {Kocevski}, {Koekemoer}, {Koo}, \&
  {McGrath}}]{bruce14b}
{Bruce}, V.~A., {Dunlop}, J.~S., {McLure}, R.~J., {et~al.} 2014{\natexlab{a}},
  \mnras, 444, 1660

\bibitem[{{Bruce} {et~al.}(2014{\natexlab{b}}){Bruce}, {Dunlop}, {McLure},
  {Cirasuolo}, {Buitrago}, {Bowler}, {Targett}, {Bell}, {McIntosh}, {Dekel},
  {Faber}, {Ferguson}, {Grogin}, {Hartley}, {Kocevski}, {Koekemoer}, {Koo}, \&
  {McGrath}}]{bruce14a}
---. 2014{\natexlab{b}}, \mnras, 444, 1001

\bibitem[{{Bruzual} \& {Charlot}(2003)}]{bc03}
{Bruzual}, G. \& {Charlot}, S. 2003, \mnras, 344, 1000

\bibitem[{{Buitrago} {et~al.}(2008){Buitrago}, {Trujillo}, {Conselice},
  {Bouwens}, {Dickinson}, \& {Yan}}]{buitrago08}
{Buitrago}, F., {Trujillo}, I., {Conselice}, C.~J., {et~al.} 2008, ArXiv
  e-prints

\bibitem[{{Buitrago} {et~al.}(2013){Buitrago}, {Trujillo}, {Conselice}, \&
  {H{\"a}u{\ss}ler}}]{buitrago13}
{Buitrago}, F., {Trujillo}, I., {Conselice}, C.~J., \& {H{\"a}u{\ss}ler}, B.
  2013, \mnras, 428, 1460

\bibitem[{{Bundy} {et~al.}(2010){Bundy}, {Scarlata}, {Carollo}, {Ellis},
  {Drory}, {Hopkins}, {Salvato}, {Leauthaud}, {Koekemoer}, {Murray}, {Ilbert},
  {Oesch}, {Ma}, {Capak}, {Pozzetti}, \& {Scoville}}]{bundy10}
{Bundy}, K., {Scarlata}, C., {Carollo}, C.~M., {et~al.} 2010, \apj, 719, 1969

\bibitem[{{Calzetti} {et~al.}(2000){Calzetti}, {Armus}, {Bohlin}, {Kinney},
  {Koornneef}, \& {Storchi-Bergmann}}]{calzetti}
{Calzetti}, D., {Armus}, L., {Bohlin}, R.~C., {et~al.} 2000, \apj, 533, 682

\bibitem[{{Carollo} {et~al.}(2013){Carollo}, {Bschorr}, {Renzini}, {Lilly},
  {Capak}, {Cibinel}, {Ilbert}, {Onodera}, {Scoville}, {Cameron}, {Mobasher},
  {Sanders}, \& {Taniguchi}}]{carollo13}
{Carollo}, C.~M., {Bschorr}, T.~J., {Renzini}, A., {et~al.} 2013, \apj, 773,
  112

\bibitem[{{Cassata} {et~al.}(2011){Cassata}, {Giavalisco}, {Guo}, {Renzini},
  {Ferguson}, {Koekemoer}, {Salimbeni}, {Scarlata}, {Grogin}, {Conselice},
  {Dahlen}, {Lotz}, {Dickinson}, \& {Lin}}]{cassata11}
{Cassata}, P., {Giavalisco}, M., {Guo}, Y., {et~al.} 2011, \apj, 743, 96

\bibitem[{{Cassata} {et~al.}(2013){Cassata}, {Giavalisco}, {Williams}, {Guo},
  {Lee}, {Renzini}, {Ferguson}, {Faber}, {Barro}, {McIntosh}, {Lu}, {Bell},
  {Koo}, {Papovich}, {Ryan}, {Conselice}, {Grogin}, {Koekemoer}, \&
  {Hathi}}]{cassata13}
{Cassata}, P., {Giavalisco}, M., {Williams}, C.~C., {et~al.} 2013, \apj, 775,
  106

\bibitem[{{Ceverino} {et~al.}(2010){Ceverino}, {Dekel}, \&
  {Bournaud}}]{ceverino10}
{Ceverino}, D., {Dekel}, A., \& {Bournaud}, F. 2010, \mnras, 404, 2151

\bibitem[{{Ceverino} {et~al.}(2015){Ceverino}, {Dekel}, {Tweed}, \&
  {Primack}}]{ceverino15}
{Ceverino}, D., {Dekel}, A., {Tweed}, D., \& {Primack}, J. 2015, \mnras, 447,
  3291

\bibitem[{{Ceverino} {et~al.}(2014){Ceverino}, {Klypin}, {Klimek},
  {Trujillo-Gomez}, {Churchill}, {Primack}, \& {Dekel}}]{ceverino14}
{Ceverino}, D., {Klypin}, A., {Klimek}, E.~S., {et~al.} 2014, \mnras, 442, 1545

\bibitem[{{Chabrier}(2003)}]{chabrier}
{Chabrier}, G. 2003, \pasp, 115, 763

\bibitem[{{Chary} \& {Elbaz}(2001)}]{ce01}
{Chary}, R. \& {Elbaz}, D. 2001, \apj, 556, 562

\bibitem[{{Cheung} {et~al.}(2012){Cheung}, {Faber}, {Koo}, {Dutton}, {Simard},
  {McGrath}, {Huang}, {Bell}, {Dekel}, {Fang}, {Salim}, {Barro}, {Bundy},
  {Coil}, {Cooper}, {Conselice}, {Davis}, {Dominguez}, {Kassin}, {Kocevski},
  {Koekemoer}, {Lin}, {Lotz}, {Newman}, {Phillips}, {Rosario}, {Weiner}, \&
  {Willmer}}]{cheung12}
{Cheung}, E., {Faber}, S.~M., {Koo}, D.~C., {et~al.} 2012, ArXiv e-prints

\bibitem[{{Croton} {et~al.}(2006){Croton}, {Springel}, {White}, {De Lucia},
  {Frenk}, {Gao}, {Jenkins}, {Kauffmann}, {Navarro}, \& {Yoshida}}]{croton06}
{Croton}, D.~J., {Springel}, V., {White}, S.~D.~M., {et~al.} 2006, \mnras, 365,
  11

\bibitem[{{Daddi} {et~al.}(2010){Daddi}, {Bournaud}, {Walter}, {Dannerbauer},
  {Carilli}, {Dickinson}, {Elbaz}, {Morrison}, {Riechers}, {Onodera}, {Salmi},
  {Krips}, \& {Stern}}]{daddi10a}
{Daddi}, E., {Bournaud}, F., {Walter}, F., {et~al.} 2010, \apj, 713, 686

\bibitem[{{Damjanov} {et~al.}(2011){Damjanov}, {Abraham}, {Glazebrook},
  {McCarthy}, {Caris}, {Carlberg}, {Chen}, {Crampton}, {Green}, {J{\o}rgensen},
  {Juneau}, {Le Borgne}, {Marzke}, {Mentuch}, {Murowinski}, {Roth}, {Savaglio},
  \& {Yan}}]{damjanov11}
{Damjanov}, I., {Abraham}, R.~G., {Glazebrook}, K., {et~al.} 2011, \apjl, 739,
  L44

\bibitem[{{Damjanov} {et~al.}(2009){Damjanov}, {McCarthy}, {Abraham},
  {Glazebrook}, {Yan}, {Mentuch}, {Le Borgne}, {Savaglio}, {Crampton},
  {Murowinski}, {Juneau}, {Carlberg}, {J{\o}rgensen}, {Roth}, {Chen}, \&
  {Marzke}}]{damjanov09}
{Damjanov}, I., {McCarthy}, P.~J., {Abraham}, R.~G., {et~al.} 2009, \apj, 695,
  101

\bibitem[{{Dekel} \& {Birnboim}(2006)}]{dekel06}
{Dekel}, A. \& {Birnboim}, Y. 2006, \mnras, 368, 2

\bibitem[{{Dekel} {et~al.}(2009{\natexlab{a}}){Dekel}, {Birnboim}, {Engel},
  {Freundlich}, {Goerdt}, {Mumcuoglu}, {Neistein}, {Pichon}, {Teyssier}, \&
  {Zinger}}]{dekel09a}
{Dekel}, A., {Birnboim}, Y., {Engel}, G., {et~al.} 2009{\natexlab{a}}, \nat,
  457, 451

\bibitem[{{Dekel} \& {Burkert}(2014)}]{dekel13b}
{Dekel}, A. \& {Burkert}, A. 2014, \mnras, 438, 1870

\bibitem[{{Dekel} {et~al.}(2009{\natexlab{b}}){Dekel}, {Sari}, \&
  {Ceverino}}]{dekel09b}
{Dekel}, A., {Sari}, R., \& {Ceverino}, D. 2009{\natexlab{b}}, \apj, 703, 785

\bibitem[{{Dekel} {et~al.}(2013){Dekel}, {Zolotov}, {Tweed}, {Cacciato},
  {Ceverino}, \& {Primack}}]{dekel13a}
{Dekel}, A., {Zolotov}, A., {Tweed}, D., {et~al.} 2013, \mnras, 435, 999

\bibitem[{{Elbaz} {et~al.}(2007){Elbaz}, {Daddi}, {Le Borgne}, {Dickinson},
  {Alexander}, {Chary}, {Starck}, {Brandt}, {Kitzbichler}, {MacDonald},
  {Nonino}, {Popesso}, {Stern}, \& {Vanzella}}]{elbaz07}
{Elbaz}, D., {Daddi}, E., {Le Borgne}, D., {et~al.} 2007, \aap, 468, 33

\bibitem[{{Elbaz} {et~al.}(2011){Elbaz}, {Dickinson}, {Hwang},
  {D{\'{\i}}az-Santos}, {Magdis}, {Magnelli}, {Le Borgne}, {Galliano},
  {Pannella}, {Chanial}, {Armus}, {Charmandaris}, {Daddi}, {Aussel}, {Popesso},
  {Kartaltepe}, {Altieri}, {Valtchanov}, {Coia}, {Dannerbauer}, {Dasyra},
  {Leiton}, {Mazzarella}, {Alexander}, {Buat}, {Burgarella}, {Chary}, {Gilli},
  {Ivison}, {Juneau}, {Le Floc'h}, {Lutz}, {Morrison}, {Mullaney}, {Murphy},
  {Pope}, {Scott}, {Brodwin}, {Calzetti}, {Cesarsky}, {Charlot}, {Dole},
  {Eisenhardt}, {Ferguson}, {F{\"o}rster Schreiber}, {Frayer}, {Giavalisco},
  {Huynh}, {Koekemoer}, {Papovich}, {Reddy}, {Surace}, {Teplitz}, {Yun}, \&
  {Wilson}}]{elbaz11}
{Elbaz}, D., {Dickinson}, M., {Hwang}, H.~S., {et~al.} 2011, \aap, 533, A119

\bibitem[{{Elmegreen} {et~al.}(2008){Elmegreen}, {Bournaud}, \&
  {Elmegreen}}]{elme08}
{Elmegreen}, B.~G., {Bournaud}, F., \& {Elmegreen}, D.~M. 2008, \apj, 688, 67

\bibitem[{{Elmegreen} {et~al.}(2004){Elmegreen}, {Elmegreen}, \&
  {Sheets}}]{elmegreen04}
{Elmegreen}, D.~M., {Elmegreen}, B.~G., \& {Sheets}, C.~M. 2004, \apj, 603, 74

\bibitem[{{Faber} \& {Jackson}(1976)}]{faber76}
{Faber}, S.~M. \& {Jackson}, R.~E. 1976, \apj, 204, 668

\bibitem[{{Fang} {et~al.}(2013){Fang}, {Faber}, {Koo}, \& {Dekel}}]{fang13}
{Fang}, J.~J., {Faber}, S.~M., {Koo}, D.~C., \& {Dekel}, A. 2013, \apj, 776, 63

\bibitem[{{Feldmann} \& {Mayer}(2015)}]{feldmann15}
{Feldmann}, R. \& {Mayer}, L. 2015, \mnras, 446, 1939

\bibitem[{{Franx} {et~al.}(2008){Franx}, {van Dokkum}, {Schreiber}, {Wuyts},
  {Labb{\'e}}, \& {Toft}}]{franx08}
{Franx}, M., {van Dokkum}, P.~G., {Schreiber}, N.~M.~F., {et~al.} 2008, \apj,
  688, 770

\bibitem[{{Genel} {et~al.}(2012){Genel}, {Naab}, {Genzel}, {F{\"o}rster
  Schreiber}, {Sternberg}, {Oser}, {Johansson}, {Dav{\'e}}, {Oppenheimer}, \&
  {Burkert}}]{genel12}
{Genel}, S., {Naab}, T., {Genzel}, R., {et~al.} 2012, \apj, 745, 11

\bibitem[{{Genel} {et~al.}(2014){Genel}, {Vogelsberger}, {Springel}, {Sijacki},
  {Nelson}, {Snyder}, {Rodriguez-Gomez}, {Torrey}, \& {Hernquist}}]{genel14}
{Genel}, S., {Vogelsberger}, M., {Springel}, V., {et~al.} 2014, \mnras, 445,
  175

\bibitem[{{Genzel} {et~al.}(2008){Genzel}, {Burkert}, {Bouch{\'e}}, {Cresci},
  {F{\"o}rster Schreiber}, {Shapley}, {Shapiro}, {Tacconi}, {Buschkamp},
  {Cimatti}, {Daddi}, {Davies}, {Eisenhauer}, {Erb}, {Genel}, {Gerhard},
  {Hicks}, {Lutz}, {Naab}, {Ott}, {Rabien}, {Renzini}, {Steidel}, {Sternberg},
  \& {Lilly}}]{genzel08}
{Genzel}, R., {Burkert}, A., {Bouch{\'e}}, N., {et~al.} 2008, \apj, 687, 59

\bibitem[{{Giavalisco} {et~al.}(2004){Giavalisco}, {Ferguson}, {Koekemoer},
  {Dickinson}, {Alexander}, {Bauer}, {Bergeron}, {Biagetti}, {Brandt},
  {Casertano}, {Cesarsky}, {Chatzichristou}, {Conselice}, {Cristiani}, {Da
  Costa}, {Dahlen}, {de Mello}, {Eisenhardt}, {Erben}, {Fall}, {Fassnacht},
  {Fosbury}, {Fruchter}, {Gardner}, {Grogin}, {Hook}, {Hornschemeier}, {Idzi},
  {Jogee}, {Kretchmer}, {Laidler}, {Lee}, {Livio}, {Lucas}, {Madau},
  {Mobasher}, {Moustakas}, {Nonino}, {Padovani}, {Papovich}, {Park},
  {Ravindranath}, {Renzini}, {Richardson}, {Riess}, {Rosati}, {Schirmer},
  {Schreier}, {Somerville}, {Spinrad}, {Stern}, {Stiavelli}, {Strolger},
  {Urry}, {Vandame}, {Williams}, \& {Wolf}}]{goods}
{Giavalisco}, M., {Ferguson}, H.~C., {Koekemoer}, A.~M., {et~al.} 2004, \apjl,
  600, L93

\bibitem[{{Graham} {et~al.}(2005){Graham}, {Driver}, {Petrosian}, {Conselice},
  {Bershady}, {Crawford}, \& {Goto}}]{graham05}
{Graham}, A.~W., {Driver}, S.~P., {Petrosian}, V., {et~al.} 2005, \aj, 130,
  1535

\bibitem[{{Grogin} {et~al.}(2011){Grogin}, {Kocevski}, {Faber}, {Ferguson},
  {Koekemoer}, {Riess}, {Acquaviva}, {Alexander}, {Almaini}, {Ashby}, {Barden},
  {Bell}, {Bournaud}, {Brown}, {Caputi}, {Casertano}, {Cassata}, {Castellano},
  {Challis}, {Chary}, {Cheung}, {Cirasuolo}, {Conselice}, {Roshan Cooray},
  {Croton}, {Daddi}, {Dahlen}, {Dav{\'e}}, {de Mello}, {Dekel}, {Dickinson},
  {Dolch}, {Donley}, {Dunlop}, {Dutton}, {Elbaz}, {Fazio}, {Filippenko},
  {Finkelstein}, {Fontana}, {Gardner}, {Garnavich}, {Gawiser}, {Giavalisco},
  {Grazian}, {Guo}, {Hathi}, {H{\"a}ussler}, {Hopkins}, {Huang}, {Huang},
  {Jha}, {Kartaltepe}, {Kirshner}, {Koo}, {Lai}, {Lee}, {Li}, {Lotz}, {Lucas},
  {Madau}, {McCarthy}, {McGrath}, {McIntosh}, {McLure}, {Mobasher},
  {Moustakas}, {Mozena}, {Nandra}, {Newman}, {Niemi}, {Noeske}, {Papovich},
  {Pentericci}, {Pope}, {Primack}, {Rajan}, {Ravindranath}, {Reddy}, {Renzini},
  {Rix}, {Robaina}, {Rodney}, {Rosario}, {Rosati}, {Salimbeni}, {Scarlata},
  {Siana}, {Simard}, {Smidt}, {Somerville}, {Spinrad}, {Straughn}, {Strolger},
  {Telford}, {Teplitz}, {Trump}, {van der Wel}, {Villforth}, {Wechsler},
  {Weiner}, {Wiklind}, {Wild}, {Wilson}, {Wuyts}, {Yan}, \& {Yun}}]{candelsgro}
{Grogin}, N.~A., {Kocevski}, D.~D., {Faber}, S.~M., {et~al.} 2011, \apjs, 197,
  35

\bibitem[{{Guo} {et~al.}(2015){Guo}, {Ferguson}, {Bell}, {Koo}, {Conselice},
  {Giavalisco}, {Kassin}, {Lu}, {Lucas}, {Mandelker}, {McIntosh}, {Primack},
  {Ravindranath}, {Barro}, {Ceverino}, {Dekel}, {Faber}, {Fang}, {Koekemoer},
  {Noeske}, {Rafelski}, \& {Straughn}}]{guo15}
{Guo}, Y., {Ferguson}, H.~C., {Bell}, E.~F., {et~al.} 2015, \apj, 800, 39

\bibitem[{{Guo} {et~al.}(2013){Guo}, {Ferguson}, {Giavalisco}, {Barro},
  {Willner}, {Ashby}, {Dahlen}, {Donley}, {Faber}, {Fontana}, {Galametz},
  {Grazian}, {Huang}, {Kocevski}, {Koekemoer}, {Koo}, {McGrath}, {Peth},
  {Salvato}, {Wuyts}, {Castellano}, {Cooray}, {Dickinson}, {Dunlop}, {Fazio},
  {Gardner}, {Gawiser}, {Grogin}, {Hathi}, {Hsu}, {Lee}, {Lucas}, {Mobasher},
  {Nandra}, {Newman}, \& {van der Wel}}]{guo13}
{Guo}, Y., {Ferguson}, H.~C., {Giavalisco}, M., {et~al.} 2013, \apjs, 207, 24

\bibitem[{{Guo} {et~al.}(2012){Guo}, {Giavalisco}, {Cassata}, {Ferguson},
  {Williams}, {Dickinson}, {Koekemoer}, {Grogin}, {Chary}, {Messias}, {Tundo},
  {Lin}, {Lee}, {Salimbeni}, {Fontana}, {Grazian}, {Kocevski}, {Lee},
  {Villanueva}, \& {van der Wel}}]{guo12}
{Guo}, Y., {Giavalisco}, M., {Cassata}, P., {et~al.} 2012, \apj, 749, 149

\bibitem[{{Hopkins} {et~al.}(2009){Hopkins}, {Bundy}, {Murray}, {Quataert},
  {Lauer}, \& {Ma}}]{hopkins09cores}
{Hopkins}, P.~F., {Bundy}, K., {Murray}, N., {et~al.} 2009, \mnras, 398, 898

\bibitem[{{Hopkins} {et~al.}(2006){Hopkins}, {Hernquist}, {Cox}, {Di Matteo},
  {Robertson}, \& {Springel}}]{hopkins06}
{Hopkins}, P.~F., {Hernquist}, L., {Cox}, T.~J., {et~al.} 2006, \apjs, 163, 1

\bibitem[{{Hopkins} {et~al.}(2008){Hopkins}, {Hernquist}, {Cox}, \& {Kere{\v
  s}}}]{hopkins08a}
{Hopkins}, P.~F., {Hernquist}, L., {Cox}, T.~J., \& {Kere{\v s}}, D. 2008,
  \apjs, 175, 356

\bibitem[{{Kauffmann} {et~al.}(2003){Kauffmann}, {Heckman}, {Tremonti},
  {Brinchmann}, {Charlot}, {White}, {Ridgway}, {Brinkmann}, {Fukugita}, {Hall},
  {Ivezi{\'c}}, {Richards}, \& {Schneider}}]{kauffmann03}
{Kauffmann}, G., {Heckman}, T.~M., {Tremonti}, C., {et~al.} 2003, \mnras, 346,
  1055

\bibitem[{{Kennicutt}(1998)}]{ken98}
{Kennicutt}, Jr., R.~C. 1998, \araa, 36, 189

\bibitem[{{Koekemoer} {et~al.}(2011){Koekemoer}, {Faber}, {Ferguson}, {Grogin},
  {Kocevski}, {Koo}, {Lai}, {Lotz}, {Lucas}, {McGrath}, {Ogaz}, {Rajan},
  {Riess}, {Rodney}, {Strolger}, {Casertano}, {Castellano}, {Dahlen},
  {Dickinson}, {Dolch}, {Fontana}, {Giavalisco}, {Grazian}, {Guo}, {Hathi},
  {Huang}, {van der Wel}, {Yan}, {Acquaviva}, {Alexander}, {Almaini}, {Ashby},
  {Barden}, {Bell}, {Bournaud}, {Brown}, {Caputi}, {Cassata}, {Challis},
  {Chary}, {Cheung}, {Cirasuolo}, {Conselice}, {Roshan Cooray}, {Croton},
  {Daddi}, {Dav{\'e}}, {de Mello}, {de Ravel}, {Dekel}, {Donley}, {Dunlop},
  {Dutton}, {Elbaz}, {Fazio}, {Filippenko}, {Finkelstein}, {Frazer}, {Gardner},
  {Garnavich}, {Gawiser}, {Gruetzbauch}, {Hartley}, {H{\"a}ussler},
  {Herrington}, {Hopkins}, {Huang}, {Jha}, {Johnson}, {Kartaltepe},
  {Khostovan}, {Kirshner}, {Lani}, {Lee}, {Li}, {Madau}, {McCarthy},
  {McIntosh}, {McLure}, {McPartland}, {Mobasher}, {Moreira}, {Mortlock},
  {Moustakas}, {Mozena}, {Nandra}, {Newman}, {Nielsen}, {Niemi}, {Noeske},
  {Papovich}, {Pentericci}, {Pope}, {Primack}, {Ravindranath}, {Reddy},
  {Renzini}, {Rix}, {Robaina}, {Rosario}, {Rosati}, {Salimbeni}, {Scarlata},
  {Siana}, {Simard}, {Smidt}, {Snyder}, {Somerville}, {Spinrad}, {Straughn},
  {Telford}, {Teplitz}, {Trump}, {Vargas}, {Villforth}, {Wagner}, {Wandro},
  {Wechsler}, {Weiner}, {Wiklind}, {Wild}, {Wilson}, {Wuyts}, \&
  {Yun}}]{candelskoe}
{Koekemoer}, A.~M., {Faber}, S.~M., {Ferguson}, H.~C., {et~al.} 2011, \apjs,
  197, 36

\bibitem[{{Kormendy} \& {Kennicutt}(2004)}]{kor04}
{Kormendy}, J. \& {Kennicutt}, Jr., R.~C. 2004, \araa, 42, 603

\bibitem[{{Kriek} {et~al.}(2009{\natexlab{a}}){Kriek}, {van Dokkum}, {Franx},
  {Illingworth}, \& {Magee}}]{kriek09}
{Kriek}, M., {van Dokkum}, P.~G., {Franx}, M., {Illingworth}, G.~D., \&
  {Magee}, D.~K. 2009{\natexlab{a}}, \apjl, 705, L71

\bibitem[{{Kriek} {et~al.}(2009{\natexlab{b}}){Kriek}, {van Dokkum},
  {Labb{\'e}}, {Franx}, {Illingworth}, {Marchesini}, \& {Quadri}}]{fast}
{Kriek}, M., {van Dokkum}, P.~G., {Labb{\'e}}, I., {et~al.} 2009{\natexlab{b}},
  \apj, 700, 221

\bibitem[{{Laidler} {et~al.}(2006){Laidler}, {Grogin}, {Clubb}, {Ferguson},
  {Papovich}, {Dickinson}, {Idzi}, {MacDonald}, {Ouchi}, \& {Mobasher}}]{tfit}
{Laidler}, V.~G., {Grogin}, N., {Clubb}, K., {et~al.} 2006, in Astronomical
  Society of the Pacific Conference Series, Vol. 351, Astronomical Data
  Analysis Software and Systems XV, ed. C.~{Gabriel}, C.~{Arviset}, D.~{Ponz},
  \& S.~{Enrique}, 228

\bibitem[{{Lang} {et~al.}(2014){Lang}, {Wuyts}, {Somerville}, {F{\"o}rster
  Schreiber}, {Genzel}, {Bell}, {Brammer}, {Dekel}, {Faber}, {Ferguson},
  {Grogin}, {Kocevski}, {Koekemoer}, {Lutz}, {McGrath}, {Momcheva}, {Nelson},
  {Primack}, {Rosario}, {Skelton}, {Tacconi}, {van Dokkum}, \&
  {Whitaker}}]{lang14}
{Lang}, P., {Wuyts}, S., {Somerville}, R.~S., {et~al.} 2014, \apj, 788, 11

\bibitem[{{Law} {et~al.}(2012){Law}, {Steidel}, {Shapley}, {Nagy}, {Reddy}, \&
  {Erb}}]{law12a}
{Law}, D.~R., {Steidel}, C.~C., {Shapley}, A.~E., {et~al.} 2012, \apj, 745, 85

\bibitem[{{Liu} {et~al.}(2013){Liu}, {Guo}, {Koo}, {Trump}, {Barro}, {Yesuf},
  {Faber}, {Giavalisco}, {Cassata}, {Koekemoer}, {Pentericci}, {Castellano},
  {Cheung}, {Mao}, {Xia}, {Grogin}, {Hathi}, {Huang}, {Kocevski}, {McGrath}, \&
  {Wuyts}}]{liu13}
{Liu}, F.~S., {Guo}, Y., {Koo}, D.~C., {et~al.} 2013, \apj, 769, 147

\bibitem[{{L{\'o}pez-Sanjuan} {et~al.}(2012){L{\'o}pez-Sanjuan}, {Le
  F{\`e}vre}, {Ilbert}, {Tasca}, {Bridge}, {Cucciati}, {Kampczyk}, {Pozzetti},
  {Xu}, {Carollo}, {Contini}, {Kneib}, {Lilly}, {Mainieri}, {Renzini},
  {Sanders}, {Scodeggio}, {Scoville}, {Taniguchi}, {Zamorani}, {Aussel},
  {Bardelli}, {Bolzonella}, {Bongiorno}, {Capak}, {Caputi}, {de la Torre}, {de
  Ravel}, {Franzetti}, {Garilli}, {Iovino}, {Knobel}, {Kova{\v c}},
  {Lamareille}, {Le Borgne}, {Le Brun}, {Le Floc'h}, {Maier}, {McCracken},
  {Mignoli}, {Pell{\'o}}, {Peng}, {P{\'e}rez-Montero}, {Presotto},
  {Ricciardelli}, {Salvato}, {Silverman}, {Tanaka}, {Tresse}, {Vergani},
  {Zucca}, {Barnes}, {Bordoloi}, {Cappi}, {Cimatti}, {Coppa}, {Koekoemoer},
  {Liu}, {Moresco}, {Nair}, {Oesch}, {Schawinski}, \& {Welikala}}]{carlos12}
{L{\'o}pez-Sanjuan}, C., {Le F{\`e}vre}, O., {Ilbert}, O., {et~al.} 2012, ArXiv
  e-prints

\bibitem[{{Magdis} {et~al.}(2010){Magdis}, {Rigopoulou}, {Huang}, \&
  {Fazio}}]{magdis10}
{Magdis}, G.~E., {Rigopoulou}, D., {Huang}, J.-S., \& {Fazio}, G.~G. 2010,
  \mnras, 401, 1521

\bibitem[{{Marchesini} {et~al.}(2014){Marchesini}, {Muzzin}, {Stefanon},
  {Franx}, {Brammer}, {Marsan}, {Vulcani}, {Fynbo}, {Milvang-Jensen}, {Dunlop},
  \& {Buitrago}}]{marchesini14}
{Marchesini}, D., {Muzzin}, A., {Stefanon}, M., {et~al.} 2014, \apj, 794, 65

\bibitem[{{Martig} {et~al.}(2009){Martig}, {Bournaud}, {Teyssier}, \&
  {Dekel}}]{martig09}
{Martig}, M., {Bournaud}, F., {Teyssier}, R., \& {Dekel}, A. 2009, \apj, 707,
  250

\bibitem[{{McGrath} {et~al.}(2008){McGrath}, {Stockton}, {Canalizo}, {Iye}, \&
  {Maihara}}]{mcgrath08}
{McGrath}, E.~J., {Stockton}, A., {Canalizo}, G., {Iye}, M., \& {Maihara}, T.
  2008, \apj, 682, 303

\bibitem[{{McLure} {et~al.}(2013){McLure}, {Pearce}, {Dunlop}, {Cirasuolo},
  {Curtis-Lake}, {Bruce}, {Caputi}, {Almaini}, {Bonfield}, {Bradshaw},
  {Buitrago}, {Chuter}, {Foucaud}, {Hartley}, \& {Jarvis}}]{mclure13}
{McLure}, R.~J., {Pearce}, H.~J., {Dunlop}, J.~S., {et~al.} 2013, \mnras, 428,
  1088

\bibitem[{{Mosleh} {et~al.}(2012){Mosleh}, {Williams}, {Franx}, {Gonzalez},
  {Bouwens}, {Oesch}, {Labbe}, {Illingworth}, \& {Trenti}}]{mosleh12}
{Mosleh}, M., {Williams}, R.~J., {Franx}, M., {et~al.} 2012, \apjl, 756, L12

\bibitem[{{Moster} {et~al.}(2013){Moster}, {Naab}, \& {White}}]{moster13}
{Moster}, B.~P., {Naab}, T., \& {White}, S.~D.~M. 2013, \mnras, 428, 3121

\bibitem[{{Muzzin} {et~al.}(2013){Muzzin}, {Marchesini}, {Stefanon}, {Franx},
  {McCracken}, {Milvang-Jensen}, {Dunlop}, {Fynbo}, {Le Fevre}, {Brammer}, \&
  {Labbe}}]{muzzin13smf}
{Muzzin}, A., {Marchesini}, D., {Stefanon}, M., {et~al.} 2013, ArXiv e-prints

\bibitem[{{Nelson} {et~al.}(2014){Nelson}, {van Dokkum}, {Franx}, {Brammer},
  {Momcheva}, {Schreiber}, {da Cunha}, {Tacconi}, {Bezanson}, {Kirkpatrick},
  {Leja}, {Rix}, {Skelton}, {van der Wel}, {Whitaker}, \& {Wuyts}}]{nelson14}
{Nelson}, E., {van Dokkum}, P., {Franx}, M., {et~al.} 2014, \nat, 513, 394

\bibitem[{{Nelson} {et~al.}(2012){Nelson}, {van Dokkum}, {Brammer},
  {F{\"o}rster Schreiber}, {Franx}, {Fumagalli}, {Patel}, {Rix}, {Skelton},
  {Bezanson}, {Da Cunha}, {Kriek}, {Labbe}, {Lundgren}, {Quadri}, \&
  {Schmidt}}]{nelson12}
{Nelson}, E.~J., {van Dokkum}, P.~G., {Brammer}, G., {et~al.} 2012, \apjl, 747,
  L28

\bibitem[{{Nelson} {et~al.}(2013){Nelson}, {van Dokkum}, {Momcheva}, {Brammer},
  {Lundgren}, {Skelton}, {Whitaker}, {Da Cunha}, {F{\"o}rster Schreiber},
  {Franx}, {Fumagalli}, {Kriek}, {Labbe}, {Leja}, {Patel}, {Rix}, {Schmidt},
  {van der Wel}, \& {Wuyts}}]{nelson13}
{Nelson}, E.~J., {van Dokkum}, P.~G., {Momcheva}, I., {et~al.} 2013, \apjl,
  763, L16

\bibitem[{{Newman} {et~al.}(2012){Newman}, {Ellis}, {Bundy}, \&
  {Treu}}]{newman12}
{Newman}, A.~B., {Ellis}, R.~S., {Bundy}, K., \& {Treu}, T. 2012, \apj, 746,
  162

\bibitem[{{Noeske} {et~al.}(2007){Noeske}, {Weiner}, {Faber}, {Papovich},
  {Koo}, {Somerville}, {Bundy}, {Conselice}, {Newman}, {Schiminovich}, {Le
  Floc'h}, {Coil}, {Rieke}, {Lotz}, {Primack}, {Barmby}, {Cooper}, {Davis},
  {Ellis}, {Fazio}, {Guhathakurta}, {Huang}, {Kassin}, {Martin}, {Phillips},
  {Rich}, {Small}, {Willmer}, \& {Wilson}}]{mainseq}
{Noeske}, K.~G., {Weiner}, B.~J., {Faber}, S.~M., {et~al.} 2007, \apjl, 660,
  L43

\bibitem[{{Onodera} {et~al.}(2014){Onodera}, {Carollo}, {Renzini},
  {Cappellari}, {Mancini}, {Arimoto}, {Daddi}, {Gobat}, {Strazzullo},
  {Tacchella}, \& {Yamada}}]{onodera14}
{Onodera}, M., {Carollo}, C.~M., {Renzini}, A., {et~al.} 2014, ArXiv e-prints

\bibitem[{{Oser} {et~al.}(2012){Oser}, {Naab}, {Ostriker}, \&
  {Johansson}}]{oser12}
{Oser}, L., {Naab}, T., {Ostriker}, J.~P., \& {Johansson}, P.~H. 2012, \apj,
  744, 63

\bibitem[{{Pannella} {et~al.}(2009){Pannella}, {Carilli}, {Daddi}, {McCracken},
  {Owen}, {Renzini}, {Strazzullo}, {Civano}, {Koekemoer}, {Schinnerer},
  {Scoville}, {Smol{\v c}i{\'c}}, {Taniguchi}, {Aussel}, {Kneib}, {Ilbert},
  {Mellier}, {Salvato}, {Thompson}, \& {Willott}}]{pannella09}
{Pannella}, M., {Carilli}, C.~L., {Daddi}, E., {et~al.} 2009, \apjl, 698, L116

\bibitem[{{Pannella} {et~al.}(2014){Pannella}, {Elbaz}, {Daddi}, {Dickinson},
  {Hwang}, {Schreiber}, {Strazzullo}, {Aussel}, {Bethermin}, {Buat},
  {Charmandaris}, {Cibinel}, {Juneau}, {Ivison}, {Le Borgne}, {Le Floc'h},
  {Leiton}, {Lin}, {Magdis}, {Morrison}, {Mullaney}, {Onodera}, {Renzini},
  {Salim}, {Sargent}, {Scott}, {Shu}, \& {Wang}}]{pannella14}
{Pannella}, M., {Elbaz}, D., {Daddi}, E., {et~al.} 2014, ArXiv e-prints

\bibitem[{{Patel} {et~al.}(2013){Patel}, {van Dokkum}, {Franx}, {Quadri},
  {Muzzin}, {Marchesini}, {Williams}, {Holden}, \& {Stefanon}}]{patel13}
{Patel}, S.~G., {van Dokkum}, P.~G., {Franx}, M., {et~al.} 2013, ArXiv e-prints

\bibitem[{{Peng} {et~al.}(2002){Peng}, {Ho}, {Impey}, \& {Rix}}]{galfit}
{Peng}, C.~Y., {Ho}, L.~C., {Impey}, C.~D., \& {Rix}, H.-W. 2002, \aj, 124, 266

\bibitem[{{Peng} {et~al.}(2010){Peng}, {Lilly}, {Kova{\v c}}, {Bolzonella},
  {Pozzetti}, {Renzini}, {Zamorani}, {Ilbert}, {Knobel}, {Iovino}, {Maier},
  {Cucciati}, {Tasca}, {Carollo}, {Silverman}, {Kampczyk}, {de Ravel},
  {Sanders}, {Scoville}, {Contini}, {Mainieri}, {Scodeggio}, {Kneib}, {Le
  F{\`e}vre}, {Bardelli}, {Bongiorno}, {Caputi}, {Coppa}, {de la Torre},
  {Franzetti}, {Garilli}, {Lamareille}, {Le Borgne}, {Le Brun}, {Mignoli},
  {Perez Montero}, {Pello}, {Ricciardelli}, {Tanaka}, {Tresse}, {Vergani},
  {Welikala}, {Zucca}, {Oesch}, {Abbas}, {Barnes}, {Bordoloi}, {Bottini},
  {Cappi}, {Cassata}, {Cimatti}, {Fumana}, {Hasinger}, {Koekemoer},
  {Leauthaud}, {Maccagni}, {Marinoni}, {McCracken}, {Memeo}, {Meneux}, {Nair},
  {Porciani}, {Presotto}, \& {Scaramella}}]{peng10}
{Peng}, Y.-j., {Lilly}, S.~J., {Kova{\v c}}, K., {et~al.} 2010, \apj, 721, 193

\bibitem[{{Poggianti} {et~al.}(2013){Poggianti}, {Moretti}, {Calvi},
  {D'Onofrio}, {Valentinuzzi}, {Fritz}, \& {Renzini}}]{poggianti13}
{Poggianti}, B.~M., {Moretti}, A., {Calvi}, R., {et~al.} 2013, \apj, 777, 125

\bibitem[{{Porter} {et~al.}(2014){Porter}, {Somerville}, {Primack}, \&
  {Johansson}}]{porter14}
{Porter}, L.~A., {Somerville}, R.~S., {Primack}, J.~R., \& {Johansson}, P.~H.
  2014, \mnras, 444, 942

\bibitem[{{Rodighiero} {et~al.}(2010){Rodighiero}, {Cimatti}, {Gruppioni},
  {Popesso}, {Andreani}, {Altieri}, {Aussel}, {Berta}, {Bongiovanni},
  {Brisbin}, {Cava}, {Cepa}, {Daddi}, {Dominguez-Sanchez}, {Elbaz}, {Fontana},
  {F{\"o}rster Schreiber}, {Franceschini}, {Genzel}, {Grazian}, {Lutz},
  {Magdis}, {Magliocchetti}, {Magnelli}, {Maiolino}, {Mancini}, {Nordon},
  {Perez Garcia}, {Poglitsch}, {Santini}, {Sanchez-Portal}, {Pozzi},
  {Riguccini}, {Saintonge}, {Shao}, {Sturm}, {Tacconi}, {Valtchanov},
  {Wetzstein}, \& {Wieprecht}}]{rodi10b}
{Rodighiero}, G., {Cimatti}, A., {Gruppioni}, C., {et~al.} 2010, \aap, 518, L25

\bibitem[{{Rodighiero} {et~al.}(2011){Rodighiero}, {Daddi}, {Baronchelli},
  {Cimatti}, {Renzini}, {Aussel}, {Popesso}, {Lutz}, {Andreani}, {Berta},
  {Cava}, {Elbaz}, {Feltre}, {Fontana}, {F{\"o}rster Schreiber},
  {Franceschini}, {Genzel}, {Grazian}, {Gruppioni}, {Ilbert}, {Le Floch},
  {Magdis}, {Magliocchetti}, {Magnelli}, {Maiolino}, {McCracken}, {Nordon},
  {Poglitsch}, {Santini}, {Pozzi}, {Riguccini}, {Tacconi}, {Wuyts}, \&
  {Zamorani}}]{rodi11}
{Rodighiero}, G., {Daddi}, E., {Baronchelli}, I., {et~al.} 2011, \apjl, 739,
  L40

\bibitem[{{Salim} {et~al.}(2007){Salim}, {Rich}, {Charlot}, {Brinchmann},
  {Johnson}, {Schiminovich}, {Seibert}, {Mallery}, {Heckman}, {Forster},
  {Friedman}, {Martin}, {Morrissey}, {Neff}, {Small}, {Wyder}, {Bianchi},
  {Donas}, {Lee}, {Madore}, {Milliard}, {Szalay}, {Welsh}, \& {Yi}}]{salim07}
{Salim}, S., {Rich}, R.~M., {Charlot}, S., {et~al.} 2007, \apjs, 173, 267

\bibitem[{{Santini} {et~al.}(2015){Santini}, {Ferguson}, {Fontana}, {Mobasher},
  {Barro}, {Castellano}, {Finkelstein}, {Grazian}, {Hsu}, {Lee}, {Lee},
  {Pforr}, {Salvato}, {Wiklind}, {Wuyts}, {Almaini}, {Cooper}, {Galametz},
  {Weiner}, {Amorin}, {Boutsia}, {Conselice}, {Dahlen}, {Dickinson},
  {Giavalisco}, {Grogin}, {Guo}, {Hathi}, {Kocevski}, {Koekemoer},
  {Kurczynski}, {Merlin}, {Mortlock}, {Newman}, {Paris}, {Pentericci},
  {Simons}, \& {Willner}}]{santini15}
{Santini}, P., {Ferguson}, H.~C., {Fontana}, A., {et~al.} 2015, \apj, 801, 97

\bibitem[{{Schiminovich} {et~al.}(2007){Schiminovich}, {Wyder}, {Martin},
  {Johnson}, {Salim}, {Seibert}, {Treyer}, {Budav{\'a}ri}, {Hoopes},
  {Zamojski}, {Barlow}, {Forster}, {Friedman}, {Morrissey}, {Neff}, {Small},
  {Bianchi}, {Donas}, {Heckman}, {Lee}, {Madore}, {Milliard}, {Rich}, {Szalay},
  {Welsh}, \& {Yi}}]{schiminovich07}
{Schiminovich}, D., {Wyder}, T.~K., {Martin}, D.~C., {et~al.} 2007, \apjs, 173,
  315

\bibitem[{{Schreiber} {et~al.}(2015){Schreiber}, {Pannella}, {Elbaz},
  {B{\'e}thermin}, {Inami}, {Dickinson}, {Magnelli}, {Wang}, {Aussel}, {Daddi},
  {Juneau}, {Shu}, {Sargent}, {Buat}, {Faber}, {Ferguson}, {Giavalisco},
  {Koekemoer}, {Magdis}, {Morrison}, {Papovich}, {Santini}, \&
  {Scott}}]{coren15}
{Schreiber}, C., {Pannella}, M., {Elbaz}, D., {et~al.} 2015, \aap, 575, A74

\bibitem[{{S{\'e}rsic}(1963)}]{sersic}
{S{\'e}rsic}, J.~L. 1963, Boletin de la Asociacion Argentina de Astronomia La
  Plata Argentina, 6, 41

\bibitem[{{Speagle} {et~al.}(2014){Speagle}, {Steinhardt}, {Capak}, \&
  {Silverman}}]{speagle14}
{Speagle}, J.~S., {Steinhardt}, C.~L., {Capak}, P.~L., \& {Silverman}, J.~D.
  2014, \apjs, 214, 15

\bibitem[{{Stefanon} {et~al.}(2013){Stefanon}, {Marchesini}, {Rudnick},
  {Brammer}, \& {Whitaker}}]{stefanon13}
{Stefanon}, M., {Marchesini}, D., {Rudnick}, G.~H., {Brammer}, G.~B., \&
  {Whitaker}, K.~E. 2013, \apj, 768, 92

\bibitem[{{Szomoru} {et~al.}(2011){Szomoru}, {Franx}, {Bouwens}, {van Dokkum},
  {Labb{\'e}}, {Illingworth}, \& {Trenti}}]{szo11}
{Szomoru}, D., {Franx}, M., {Bouwens}, R.~J., {et~al.} 2011, \apjl, 735, L22

\bibitem[{{Szomoru} {et~al.}(2012){Szomoru}, {Franx}, \& {van Dokkum}}]{szo12}
{Szomoru}, D., {Franx}, M., \& {van Dokkum}, P.~G. 2012, \apj, 749, 121

\bibitem[{{Tacchella} {et~al.}(2015){Tacchella}, {Carollo}, {Renzini},
  {Schreiber}, {Lang}, {Wuyts}, {Cresci}, {Dekel}, {Genzel}, {Lilly},
  {Mancini}, {Newman}, {Onodera}, {Shapley}, {Tacconi}, {Woo}, \&
  {Zamorani}}]{tacchella15}
{Tacchella}, S., {Carollo}, C.~M., {Renzini}, A., {et~al.} 2015, Science, 348,
  314

\bibitem[{{Tacconi} {et~al.}(2010){Tacconi}, {Genzel}, {Neri}, {Cox}, {Cooper},
  {Shapiro}, {Bolatto}, {Bouch{\'e}}, {Bournaud}, {Burkert}, {Combes},
  {Comerford}, {Davis}, {Schreiber}, {Garcia-Burillo}, {Gracia-Carpio}, {Lutz},
  {Naab}, {Omont}, {Shapley}, {Sternberg}, \& {Weiner}}]{tacconi10}
{Tacconi}, L.~J., {Genzel}, R., {Neri}, R., {et~al.} 2010, \nat, 463, 781

\bibitem[{{Tacconi} {et~al.}(2013){Tacconi}, {Neri}, {Genzel}, {Combes},
  {Bolatto}, {Cooper}, {Wuyts}, {Bournaud}, {Burkert}, {Comerford}, {Cox},
  {Davis}, {F{\"o}rster Schreiber}, {Garc{\'{\i}}a-Burillo}, {Gracia-Carpio},
  {Lutz}, {Naab}, {Newman}, {Omont}, {Saintonge}, {Shapiro Griffin}, {Shapley},
  {Sternberg}, \& {Weiner}}]{tacconi13}
{Tacconi}, L.~J., {Neri}, R., {Genzel}, R., {et~al.} 2013, \apj, 768, 74

\bibitem[{{Tomczak} {et~al.}(2013){Tomczak}, {Quadri}, {Tran}, {Labbe},
  {Straatman}, {Papovich}, {Glazebrook}, {Allen}, {Kacprzak},
  {Kawinwanichakij}, {Kelson}, {McCarthy}, {Mehrtens}, {Monson}, {Persson},
  {Spitler}, {Tilvi}, \& {van Dokkum}}]{tomczak13}
{Tomczak}, A.~R., {Quadri}, R.~F., {Tran}, K.-V.~H., {et~al.} 2013, ArXiv
  e-prints

\bibitem[{{van de Sande} {et~al.}(2013){van de Sande}, {Kriek}, {Franx}, {van
  Dokkum}, {Bezanson}, {Bouwens}, {Quadri}, {Rix}, \& {Skelton}}]{vandesande13}
{van de Sande}, J., {Kriek}, M., {Franx}, M., {et~al.} 2013, \apj, 771, 85

\bibitem[{{van der Wel} {et~al.}(2012){van der Wel}, {Bell}, {H{\"a}ussler},
  {McGrath}, {Chang}, {Guo}, {McIntosh}, {Rix}, {Barden}, {Cheung}, {Faber},
  {Ferguson}, {Galametz}, {Grogin}, {Hartley}, {Kartaltepe}, {Kocevski},
  {Koekemoer}, {Lotz}, {Mozena}, {Peth}, \& {Peng}}]{vdw12}
{van der Wel}, A., {Bell}, E.~F., {H{\"a}ussler}, B., {et~al.} 2012, \apjs,
  203, 24

\bibitem[{{van der Wel} {et~al.}(2014){van der Wel}, {Franx}, {van Dokkum},
  {Skelton}, {Momcheva}, {Whitaker}, {Brammer}, {Bell}, {Rix}, {Wuyts},
  {Ferguson}, {Holden}, {Barro}, {Koekemoer}, {Chang}, {McGrath},
  {H{\"a}ussler}, {Dekel}, {Behroozi}, {Fumagalli}, {Leja}, {Lundgren},
  {Maseda}, {Nelson}, {Wake}, {Patel}, {Labb{\'e}}, {Faber}, {Grogin}, \&
  {Kocevski}}]{vdw14}
{van der Wel}, A., {Franx}, M., {van Dokkum}, P.~G., {et~al.} 2014, \apj, 788,
  28

\bibitem[{{van der Wel} {et~al.}(2011){van der Wel}, {Rix}, {Wuyts}, {McGrath},
  {Koekemoer}, {Bell}, {Holden}, {Robaina}, \& {McIntosh}}]{vdw11a}
{van der Wel}, A., {Rix}, H.-W., {Wuyts}, S., {et~al.} 2011, \apj, 730, 38

\bibitem[{{van Dokkum} {et~al.}(2014){van Dokkum}, {Bezanson}, {van der Wel},
  {Nelson}, {Momcheva}, {Skelton}, {Whitaker}, {Brammer}, {Conroy},
  {F{\"o}rster Schreiber}, {Fumagalli}, {Kriek}, {Labb{\'e}}, {Leja},
  {Marchesini}, {Muzzin}, {Oesch}, \& {Wuyts}}]{dokkum14}
{van Dokkum}, P.~G., {Bezanson}, R., {van der Wel}, A., {et~al.} 2014, \apj,
  791, 45

\bibitem[{{van Dokkum} {et~al.}(2015){van Dokkum}, {Nelson}, {Franx},
  {Momcheva}, {Brammer}, {Forster Schreiber}, {Skelton}, {Whitaker}, {van der
  Wel}, {Bezanson}, {Fumagalli}, {Kriek}, {Leja}, \& {Wuyts}}]{dokkum15}
{van Dokkum}, P.~G., {Nelson}, E.~J., {Franx}, M., {et~al.} 2015, ArXiv
  e-prints

\bibitem[{{van Dokkum} {et~al.}(2010){van Dokkum}, {Whitaker}, {Brammer},
  {Franx}, {Kriek}, {Labb{\'e}}, {Marchesini}, {Quadri}, {Bezanson},
  {Illingworth}, {Muzzin}, {Rudnick}, {Tal}, \& {Wake}}]{dokkum10}
{van Dokkum}, P.~G., {Whitaker}, K.~E., {Brammer}, G., {et~al.} 2010, \apj,
  709, 1018

\bibitem[{{Wellons} {et~al.}(2014){Wellons}, {Torrey}, {Ma}, {Rodriguez-Gomez},
  {Vogelsberger}, {Kriek}, {van Dokkum}, {Nelson}, {Genel}, {Pillepich},
  {Springel}, {Sijacki}, {Snyder}, {Nelson}, {Sales}, \&
  {Hernquist}}]{wellons14}
{Wellons}, S., {Torrey}, P., {Ma}, C.-P., {et~al.} 2014, ArXiv e-prints

\bibitem[{{Whitaker} {et~al.}(2014){Whitaker}, {Franx}, {Leja}, {van Dokkum},
  {Henry}, {Skelton}, {Fumagalli}, {Momcheva}, {Brammer}, {Labb{\'e}},
  {Nelson}, \& {Rigby}}]{whitaker14}
{Whitaker}, K.~E., {Franx}, M., {Leja}, J., {et~al.} 2014, \apj, 795, 104

\bibitem[{{Whitaker} {et~al.}(2012){Whitaker}, {van Dokkum}, {Brammer}, \&
  {Franx}}]{whitaker12b}
{Whitaker}, K.~E., {van Dokkum}, P.~G., {Brammer}, G., \& {Franx}, M. 2012,
  \apjl, 754, L29

\bibitem[{{Whitaker} {et~al.}(2013){Whitaker}, {van Dokkum}, {Brammer},
  {Momcheva}, {Skelton}, {Franx}, {Kriek}, {Labbe}, {Fumagalli}, {Lundgren},
  {Nelson}, {Patel}, \& {Rix}}]{whitaker13}
{Whitaker}, K.~E., {van Dokkum}, P.~G., {Brammer}, G., {et~al.} 2013, ArXiv
  e-prints

\bibitem[{{Williams} {et~al.}(2013){Williams}, {Giavalisco}, {Cassata},
  {Tundo}, {Wiklind}, {Guo}, {Lee}, {Barro}, {Wuyts}, {Bell}, {Conselice},
  {Dekel}, {Faber}, {Ferguson}, {Grogin}, {Hathi}, {Huang}, {Kocevski},
  {Koekemoer}, {Koo}, {Ravindranath}, \& {Salimbeni}}]{williams13}
{Williams}, C.~C., {Giavalisco}, M., {Cassata}, P., {et~al.} 2013, ArXiv
  e-prints

\bibitem[{{Williams} {et~al.}(2010){Williams}, {Quadri}, {Franx}, {van Dokkum},
  {Toft}, {Kriek}, \& {Labb{\'e}}}]{williams10}
{Williams}, R.~J., {Quadri}, R.~F., {Franx}, M., {et~al.} 2010, \apj, 713, 738

\bibitem[{{Woo} {et~al.}(2015){Woo}, {Dekel}, {Faber}, \& {Koo}}]{woo15}
{Woo}, J., {Dekel}, A., {Faber}, S.~M., \& {Koo}, D.~C. 2015, \mnras, 448, 237

\bibitem[{{Wuyts} {et~al.}(2012){Wuyts}, {F{\"o}rster Schreiber}, {Genzel},
  {Guo}, {Barro}, {Bell}, {Dekel}, {Faber}, {Ferguson}, {Giavalisco}, {Grogin},
  {Hathi}, {Huang}, {Kocevski}, {Koekemoer}, {Koo}, {Lotz}, {Lutz}, {McGrath},
  {Newman}, {Rosario}, {Saintonge}, {Tacconi}, {Weiner}, \& {van der
  Wel}}]{wuyts12}
{Wuyts}, S., {F{\"o}rster Schreiber}, N.~M., {Genzel}, R., {et~al.} 2012, \apj,
  753, 114

\bibitem[{{Wuyts} {et~al.}(2011{\natexlab{a}}){Wuyts}, {F{\"o}rster Schreiber},
  {Lutz}, {Nordon}, {Berta}, {Altieri}, {Andreani}, {Aussel}, {Bongiovanni},
  {Cepa}, {Cimatti}, {Daddi}, {Elbaz}, {Genzel}, {Koekemoer}, {Magnelli},
  {Maiolino}, {McGrath}, {P{\'e}rez Garc{\'{\i}}a}, {Poglitsch}, {Popesso},
  {Pozzi}, {Sanchez-Portal}, {Sturm}, {Tacconi}, \& {Valtchanov}}]{wuyts11a}
{Wuyts}, S., {F{\"o}rster Schreiber}, N.~M., {Lutz}, D., {et~al.}
  2011{\natexlab{a}}, \apj, 738, 106

\bibitem[{{Wuyts} {et~al.}(2013){Wuyts}, {F{\"o}rster Schreiber}, {Nelson},
  {van Dokkum}, {Brammer}, {Chang}, {Faber}, {Ferguson}, {Franx}, {Fumagalli},
  {Genzel}, {Grogin}, {Kocevski}, {Koekemoer}, {Lundgren}, {Lutz}, {McGrath},
  {Momcheva}, {Rosario}, {Skelton}, {Tacconi}, {van der Wel}, \&
  {Whitaker}}]{wuyts13}
{Wuyts}, S., {F{\"o}rster Schreiber}, N.~M., {Nelson}, E.~J., {et~al.} 2013,
  \apj, 779, 135

\bibitem[{{Wuyts} {et~al.}(2011{\natexlab{b}}){Wuyts}, {F{\"o}rster Schreiber},
  {van der Wel}, {Magnelli}, {Guo}, {Genzel}, {Lutz}, {Aussel}, {Barro},
  {Berta}, {Cava}, {Graci{\'a}-Carpio}, {Hathi}, {Huang}, {Kocevski},
  {Koekemoer}, {Lee}, {Le Floc'h}, {McGrath}, {Nordon}, {Popesso}, {Pozzi},
  {Riguccini}, {Rodighiero}, {Saintonge}, \& {Tacconi}}]{wuyts11b}
{Wuyts}, S., {F{\"o}rster Schreiber}, N.~M., {van der Wel}, A., {et~al.}
  2011{\natexlab{b}}, \apj, 742, 96

\bibitem[{{Zolotov} {et~al.}(2015){Zolotov}, {Dekel}, {Mandelker}, {Tweed},
  {Inoue}, {DeGraf}, {Ceverino}, \& {Primack}}]{zolotov14}
{Zolotov}, A., {Dekel}, A., {Mandelker}, N., {et~al.} 2015, ArXiv e-prints

\end{thebibliography}
